\documentclass[a4paper,11pt]{article}
\pdfoutput=1

\usepackage{jcappub} 
\usepackage{xcolor}
\usepackage[T1]{fontenc} 
 
\usepackage{graphicx}
\usepackage{hyperref}
\usepackage{siunitx}
\usepackage{multirow}
\usepackage{txfonts}
\usepackage{subfigure}
\usepackage{float}
\usepackage{bm,array}
\usepackage{amsmath}

\title{\boldmath Kinetic Inductance Detectors for the OLIMPO experiment: design and pre--flight characterization}

\author[a,b,1]{A. Paiella,\note{Corresponding author.}}
\author[a,b]{A. Coppolecchia,}
\author[a,b]{L. Lamagna,}
\author[c]{P.~A.~R. Ade,}
\author[a,b]{\\E.~S. Battistelli,}
\author[d]{M.~G. Castellano,}
\author[d,e]{I. Colantoni,}
\author[a,b]{F. Columbro,}
\author[a,b]{G. D'Alessandro,}
\author[a,b]{P. de Bernardis,}
\author[f]{S. Gordon,}
\author[a,b]{S. Masi,}
\author[f,g]{\\P. Mauskopf,}
\author[d]{G. Pettinari,}
\author[a,b]{F. Piacentini,}
\author[c]{G. Pisano,}
\author[a,b]{G. Presta}
\author[c]{and C. Tucker}

\affiliation[a]{Dipartimento di Fisica, \emph{Sapienza} Universit\`a di Roma,\\P.le A. Moro 2, 00185 Roma, Italy}
\affiliation[b]{Istituto Nazionale di Fisica Nucleare, Sezione di Roma,\\P.le A. Moro 2, 00185 Roma, Italy}
\affiliation[c]{School of Physics and Astronomy, Cardiff University,\\ Cardiff CF24 3YB, UK}
\affiliation[d]{Istituto di Fotonica e Nanotecnologie -- CNR,\\Via Cineto Romano 42, 00156 Roma, Italy}
\affiliation[e]{{\sl current address}: School of Cosmic Physics, Dublin Institute for Advanced Studies,\\31 Fitzwilliam Place, D02 XF86, Dublin, Ireland}
\affiliation[f]{School of Earth and Space Exploration, Arizona State University,\\Tempe, AZ 85287, USA}
\affiliation[g]{Department of Physics, Arizona State University,\\Tempe, AZ 85257, USA}

\emailAdd{alessandro.paiella@roma1.infn.it}

\abstract{%The study of the Cosmic Microwave Background (CMB) has motivated the development of sensitive low temperature detectors scalable to form large arrays. Kinetic Inductance Detectors (KIDs) satisfy the requirements for CMB spectrum, anisotropy and polarization measurements. They are intrinsically multiplexable, fast, robust and compact, and do not require complex microfabrication steps. The principle of operation of a KID is based on the electrical properties of superconducting films. Incoming mm--wave radiation, incident on a superconducting film, breaks Cooper pairs and changes the relative number density of the paired and unpaired charge carriers, inducing a change in the electrical properties of the superconductor. In a superconducting film shaped as a high--Q microwave resonator, the incoming radiation intensity induces changes in the resonant frequency and quality factor, $Q$, of the KID. This technology has been already demonstrated in the field for ground--based instrumentation, and represents a viable solution for the next generation of sensitive large--format mm--wave cameras. Nonetheless, space--borne applications of KIDs still need to be proven in representative conditions.
We designed, fabricated, and characterized four arrays of horn--coupled, lumped element kinetic inductance detectors (LEKIDs), optimized to work in the spectral bands of the balloon--borne OLIMPO experiment. OLIMPO is a \SI{2.6}{m} aperture telescope, aimed at spectroscopic measurements of the Sunyaev--Zel'dovich (SZ) effect. OLIMPO will also validate the LEKID technology in a representative space environment. The corrected focal plane is filled with diffraction limited horn-coupled KID arrays, with 19, 37, 23, 41 active pixels respectively at 150, 250, 350, and \SI{460}{GHz}.

Here we report on the full electrical and optical characterization performed on these detector arrays before the flight. In a dark laboratory cryostat, we measured the resonator electrical parameters, such as the quality factors and the electrical responsivities, at a base temperature of \SI{300}{mK}. The measured average resonator $Q$s are \SI{1.7e4}{}, \SI{7.0e3}{}, \SI{1.0e4}{}, and \SI{1.0e4} for the 150, 250, 350, and \SI{460}{GHz} arrays, respectively. The average electrical phase responsivities on resonance are \SI{1.4}{rad/pW}, \SI{1.5}{rad/pW}, \SI{2.1}{rad/pW}, and \SI{2.1}{rad/pW}; the electrical noise equivalent powers are $\SI{45}{aW/\sqrt{\rm Hz}}$, \SI{160}{aW/\sqrt{\rm Hz}}, \SI{80}{aW/\sqrt{\rm Hz}}, and \SI{140}{aW/\sqrt{\rm Hz}}, at \SI{12}{Hz}. In the OLIMPO cryostat, we measured the optical properties, such as the noise equivalent temperatures (NET) and the spectral responses. The measured NET$_{\rm RJ}$s are $\SI{200}{\mu K.\sqrt{\rm s}}$, \SI{240}{\mu K.\sqrt{\rm s}}, \SI{240}{\mu K.\sqrt{\rm s}}, and \SI{340}{\mu K.\sqrt{\rm s}}, at \SI{12}{Hz}; under 78, 88, 92, and \SI{90}{mK} Rayleigh--Jeans blackbody load changes respectively for the 150, 250, 350, and \SI{460}{GHz} arrays. The spectral responses were characterized with the OLIMPO differential Fourier transform spectrometer (DFTS) up to THz frequencies, with a resolution of \SI{1.8}{GHz}.}
\keywords{CMBR detectors -- CMBR experiments -- Sunyaev--Zel'dovich effect}

\begin{document}
\maketitle
\flushbottom

\section{Introduction}

Precision measurements of the cosmic microwave background anisotropy, polarization and spectrum, require the development of sensitive low temperature detectors, scalable to form large arrays. Kinetic Inductance Detectors are easily replicable in large (thousands of pixels) arrays, intrinsically multiplexable, and represent a very promising technology for this sector.

\begin{figure}[htb]
\centering
\includegraphics[scale=0.8]{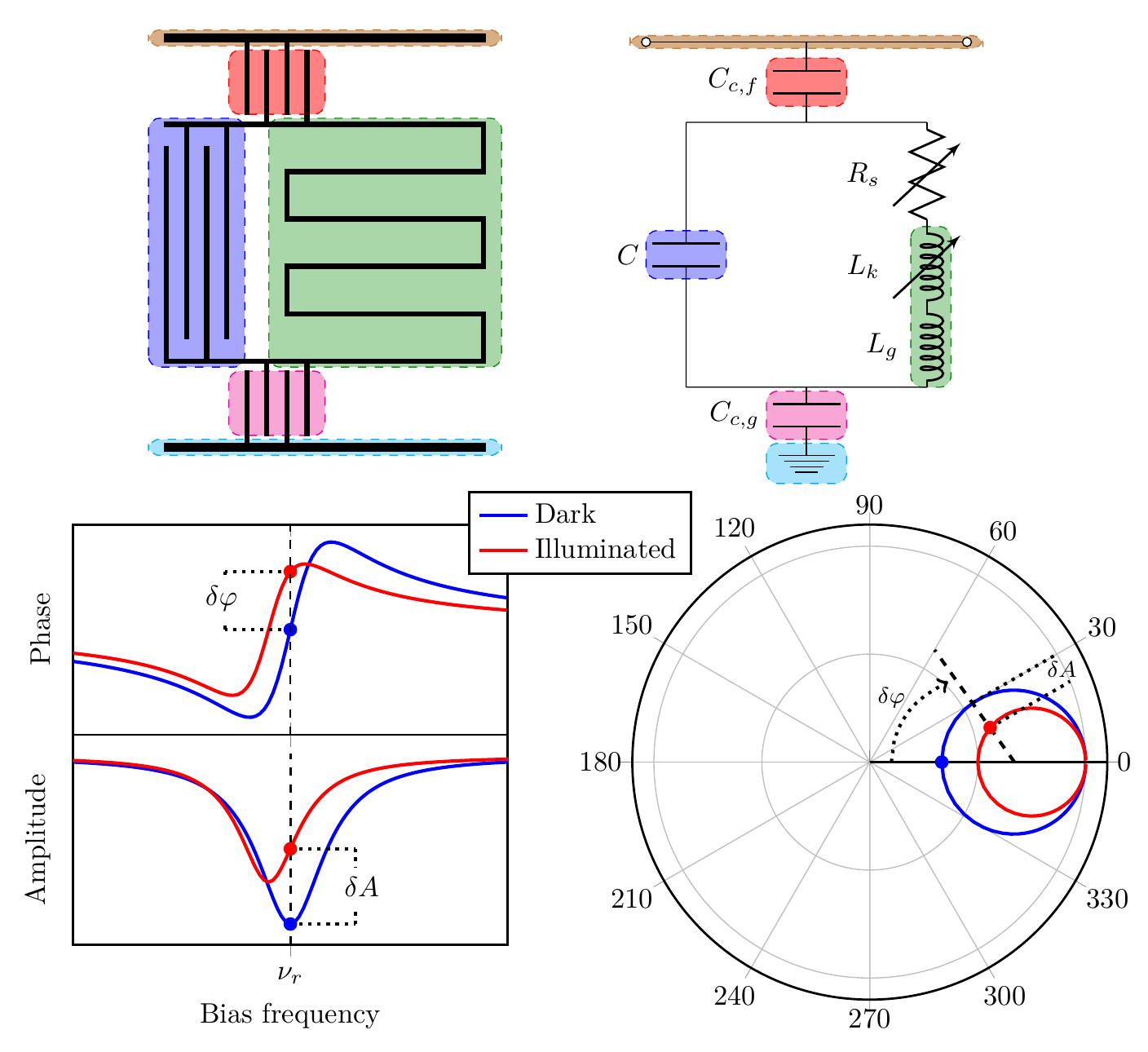}
\caption{\small \emph{Top panel}: Design (\emph{left panel}) and equivalent circuit (\emph{right panel}) of a kinetic inductance detector, capacitively coupled to a feedline and to the ground. $C_{c,f}$ (highlighted in \emph{red}) is the coupling capacitor between the KID and the feedline (highlighted in \emph{brown}), while $C_{c,g}$ (highlighted in \emph{magenta}) is the coupling capacitor between the KID and the ground (highlighted in \emph{cyan}). The KID is composed of a capacitor, $C$ (highlighted in \emph{blue}), a geometric and a kinetic inductance, $L_{g}$ and $L_{k}$ (highlighted in \emph{green}), and a residual resistance, $R_{s}$, due to the non--zero detector temperature, namely to the residual quasiparticles. Both $R_{s}$ and $L_{k}$ depend on the superconducting material, and the geometric parameters of the detector design. \emph{Bottom--left panel}: Bias frequency dependence of the resonance amplitude and phase. The RLC circuit loads the feedline, producing a dip in its transmission (\emph{blue lines}). The quasiparticles, produced by photons, increase both $L_{k}$ and $R_{s}$. This shifts the resonance to lower frequencies, due to $L_{k}$, and makes it broader and shallower, due to $R_{s}$ (\emph{red lines}). \emph{Bottom--right panel}: Polar representation of the resonance. In the polar plane the resonance is a circle (\emph{blue line}), which changes its center and decreases its radius when quasiparticles are produced (\emph{red line}).}
\phantomsection\label{fig:scheme_KIDs}
\end{figure}

A lumped element kinetic inductance detector consists of a high--Q LC resonant circuit, where the inductor acts also as the radiation absorber \citep{Day, Doyle2008}. The principle of operation of KIDs is based on the kinetic inductance, $L_{k}$, dependence on the relative density of paired (Cooper pairs) and unpaired (quasiparticles) charge carriers. Photons with energy greater than the binding energy ($h\nu>2\Delta_{0}$) break the Cooper pairs, inducing an increase of quasiparticle density $n_{qp}$, and, consequently, an increase of the kinetic inductance. This produces a shift of the resonant frequency $\nu_{r}$, and a change of the quality factor $Q$ of the resonator. These are measured by monitoring the amplitude and the phase of a microwave bias signal, transmitted through a feedline coupled to the resonator (see fig.~\ref{fig:scheme_KIDs}).

KIDs, therefore, exploit the phenomenon of superconductivity not only in the detection mechanism, but also in the readout scheme. High quality factor values can be obtained since the superconductor film has very low residual resistance, and allow thousands of kinetic inductance detectors, each with slightly different resonant frequency, to be read out using the same feedline.

In this paper, we describe the design, the fabrication process, and the electrical and optical characterization of the four horn--coupled LEKID arrays of the OLIMPO experiment. OLIMPO (\emph{Osservatorio nel Lontano Infrarosso Montato su Pallone Orientabile}, Far Infrared Observatory Mounted on a Pointed Balloon) is a balloon--borne telescope designed to study the sky in the mm and sub--mm regions of the electromagnetic spectrum, with high angular resolution (matched to the typical angular scales of rich and nearby galaxy clusters) and sensitivity \citep{Coppolecchia2013}. This experiment uses a \SI{2.6}{m} aperture primary mirror and four horn--coupled LEKID arrays, sensitive to four bands centered at 150, 250, 350, and \SI{460}{GHz}. These are cooled to about \SI{300}{mK} through a wet N$_2$ plus $^{4}$He cryostat, and a $^{3}$He sub--K refrigerator. OLIMPO is equipped with a differential Fourier transform spectrometer (DFTS) \citep{schillaci2014, DAlessandro2015}, with a maximum resolution $\Delta\nu=\SI{1.8}{GHz}$, and a custom attitude control system with arcmin pointing accuracy. OLIMPO is designed for a long--duration polar stratospheric balloon flight. This allows the telescope to operate above most of the atmospheric absorption and noise, which are particularly severe at sub--mm frequencies, at a fraction of the cost of a satellite platform.

The main goal of OLIMPO is to observe a sample of galaxy clusters through the Sunyaev--Zel'dovich (SZ) effect \citep{Sunyaev1970}. For this reason its detectors cover four frequency bands matching the negative, zero, and positive regions of the SZ spectrum. Apart from the lowest frequency bands, which have been used extensively by ground--based telescopes in the best observation sites, see e.g. \citep{Staniszewski2009, Hincks2010,Adam2014}, in the other bands atmospheric transmission is very low at ground level, and only space--based experiments can carry out sensitive measurements of the SZ effect \cite{1475-7516-2018-04-019, 1475-7516-2018-04-020}, as demonstrated by the Planck satellite \citep{Planck1, Planck2, Planck3}. The strength of OLIMPO consists in the possibility to perform low resolution spectroscopy ($\Delta\nu=\SI{5}{GHz}$) of the SZ effect along with broad band photometric measurements. These features allow to constrain all the main parameters of the intracluster plasma with low degeneracy and optimal control of the foreground contamination \citep{deBernardis}.

Moreover, OLIMPO will offer the opportunity to qualify the KID technology in a representative near--space environment, in view of future space missions \citep{Delabrouille_CORE, deBernardis_CORE}. 

This paper is structured as follows: Section~\ref{sec:Design} describes the detector design, starting from the requirements (\ref{subsec:Detector_Requirements}), and presenting the results of the optical and the electrical simulations (\ref{subsec:Optical_Simulations} and \ref{subsec:Electrical_Simulations}); Section~\ref{sec:Fabrication_process} summarizes the fabrication process of the LEKID arrays; Section~\ref{sec:Experimental_Setup} describes the experimental setup composed of the dark cryogenic system for the electrical characterization  (\ref{subsec:Cryostat}), the OLIMPO cryostat and optical system for the optical characterization (\ref{subsec:OLIMPOCryostat}), and the readout chain (\ref{subsec:Readout}); Section~\ref{sec:Results} summarizes the results of the electrical (\ref{subsec:Electrical_characterization}) and optical characterization (\ref{subsec:Optical_characterization}) performed on the four LEKID arrays: the measurements of resonator quality factors (\ref{subsec:Quality_Factors}), electrical responsivity and noise equivalent power (\ref{subsec:Electrical_Responsivity}), optical responsivity and noise equivalent temperature (\ref{subsec:optical_NET}), spectral response (\ref{subsec:Spectra}), and optical efficiency (\ref{subsec:Optical_Efficiency}). Section~\ref{sec:Conclusion} contains the concluding remarks.

\section{Design}
\phantomsection\label{sec:Design}

The design of the OLIMPO detector arrays has optimized the main characteristics of LEKIDs (high--Q RLC resonator coupled to a feedline, lumped element circuit, efficient absorption for incoming photons) given the base temperature and the optical system of OLIMPO, which constrained the size of the focal planes, the horn apertures, and the spectral bands.

In order to perform properly, KIDs have to be cooled well below the critical temperature of the superconducting film, $T_{c}$, usually $T \lesssim T_{c}/6$. The base temperature of the OLIMPO refrigerator is \SI{300}{mK}, which means that $T_{c}\gtrsim \SI{1.80}{K}$. On the other hand, as we said, only radiation with energy, $h\nu$, greater than the Cooper pair binding energy, $2\Delta_{0}$, can break Cooper pairs and then can be detected: $h\nu>2\Delta_{0}$. Since the \SI{150}{GHz} spectral band starts from \SI{135}{GHz} (lower limit of the full width half maximum, see tab.~\ref{tab:OLIMPO_perform}), in order to be safe, is reasonable to consider \SI{130}{GHz} as the minimum radiation frequency detectable. Therefore, using 
\begin{equation}
\Delta_{0}=1.764\;k_{B}T_{c}\;,
\phantomsection\label{eq:delta_0}
\end{equation}
from the BCS theory \citep{Bardeen1957}, where $k_{B}$ is the Boltzmann constant, the constraint $h\nu>2\Delta_{0}$ translates into $T_{c}<\SI{1.77}{K}$. The two constraints on the critical temperature are formally inconsistent: we prefer to relax the one given by the base temperature, and therefore to satisfy the one given by the minimum detectable radiation frequency.

We choose aluminum for the superconducting film. In fact, Al has a bulk critical temperature of \SI{1.20}{K}, which increases to \SI{1.40}{K} for tens of nm thick films. In addition, we choose silicon for the dielectric substrate, for its mechanical, thermal, and electrical properties. This selection of materials is common for LEKIDs detecting mm--waves. For such materials, performance and fabrication technologies have already been demonstrated \citep{Monfardini2011, McCarrick2014, Goupy2016}.

\subsection{Detector Requirements}
\phantomsection\label{subsec:Detector_Requirements}
OLIMPO is designed to operate from the stratosphere for about 15 days. Its effectiveness relies on the ability to achieve high signal--to--noise ratio measurements in relatively short integrations. On the other hand, the spectral coverage, and the need to switch between photometric and spectroscopic optical configurations, determine changes in the radiative background which are larger than those due to the elevation changes of the telescope. The focal plane must therefore be populated with large--dynamic--range, photon--noise limited detectors in the expected in-flight radiative environment. Tab.~\ref{tab:OLIMPO_perform} shows the target background limited performance (BLIP) of OLIMPO in both photometric and spectrometric configurations.

\begin{table*}[!h]
	\centering
		\fontsize{10pt}{18pt}\selectfont{
		\begin{tabular}{c|c|c|c|c|c|c|c|c}
		\hline
		\hline
		\multicolumn{1}{c|}{\multirow{1}{*}{OLIMPO configuration}}&
		\multicolumn{4}{c|}{\multirow{1}{*}{Photometric}}&
		\multicolumn{4}{c}{\multirow{1}{*}{Spectrometric}}\\
		\hline
		\multicolumn{1}{c|}{Channel $\left[\SI{}{GHz}\right]$}&150&250&350&460&150&250&350&460\\
		\hline
		\hline
		Background Power $\left[\SI{}{pW}\right]$&2.7& 14 & 4.9 & 13 & 7.6 & 50 & 15 & 33 \\
		optical ${\rm NEP}$ $\left[\SI{}{aW/\sqrt{\rm Hz}}\right]$& 65 & 140 & 90 & 140 & 110 & 260 & 160 & 240\\
		optical ${\rm NET_{CMB}}$ $\left[\SI{}{\mu K.\sqrt{\rm s}}\right]$&110&110&850&3300&190&550&1400&5200\\
		%60&60&420&1700&100 &280 &730 &2700 \\
		optical ${\rm NET_{RJ}}$ $\left[\SI{}{\mu K.\sqrt{\rm s}}\right]$&60&30&70&60&110&60&120&110\\
		%&90&40&100&90&160 &80 &170 &150 \\
		\hline
		\hline
		\end{tabular}		
		}
		\caption{\small Forecast for the photon--noise--limited performance of the OLIMPO experiment, for both photometric and spectrometric configurations, in the four measurement bands (a conservative 50\% absorption efficiency for the detector has been assumed): $\left[135,160\right]\SI{}{GHz}$, $\left[190,305\right]\SI{}{GHz}$, $\left[330,365\right]\SI{}{GHz}$ and $\left[430,490\right]\SI{}{GHz}$, calculated as the full width half maximum (FWHM) of the spectral bands. The background power is the power absorbed by the detector and includes the emission of the CMB, the residual atmosphere, and the optical components of the instrument. The noise numbers are optical, referred to the photometer window, and take into account the inefficiencies due to the transmission and emission of the entire filters chain.}
		\phantomsection\label{tab:OLIMPO_perform}
\end{table*}

The detector arrays are coupled to the optical system by means of feed--horns. The size of the aberration--corrected focal plane is such that the \SI{150}{GHz} and \SI{250}{GHz} arrays must be fabricated on a \SI{3}{''} diameter wafer, while the \SI{350}{GHz} and \SI{460}{GHz} arrays can fit on \SI{2}{''} wafers. The apertures of the horns are such that the \SI{150}{GHz}, \SI{250}{GHz}, \SI{350}{GHz}, and \SI{460}{GHz} arrays can host a maximum of 19, 37, 23, and 41 illuminated pixels, respectively. To these we added 4 dark pixels on the \SI{150}{GHz} array and 2 dark pixels on each of the other arrays.

The last requirement concerns the resonant frequency of the detectors. We choose resonant frequencies in the hundreds of \SI{}{MHz} range in order to have reasonably large capacitors, so that we can have a uniform current distribution at the resonant frequency along the inductors, thus reducing TLS (two--level system) noise \citep{Noroozian2009}. 

For the OLIMPO experiment, we decided to use two readout signal chains and electronics for the four arrays corresponding to approximately 60--70 detectors per readout chain. Each readout chain consists of two coaxial lines, one for carrying the readout bias tones into the detectors and one for carrying the signals transmitted through the feedline at each KID frequency after amplification by a cryogenic low noise amplifier. We discarded the hypothesis to read out the four arrays independently because the thermal load of four independent readout chains would become significant for the OLIMPO cryostat: the thermal load introduced by four cryogenic amplifiers would negatively affect the hold time of the cryostat and the temperature of the amplifier, degrading its performance. On the other hand, we discarded the hypothesis to use only one readout system as well, in order to avoid large signal losses transmitted through the arrays, and to provide a minimum level of redundancy.  
 
In order to combine two arrays in the same readout line, we avoid overlaps in the resonant frequencies among the detectors in both arrays. In addition, in order to evenly divide the detectors between the two readout lines, we paired together the \SI{150}{GHz} and \SI{460}{GHz} arrays and the \SI{250}{GHz} and \SI{350}{GHz} arrays.

\subsection{Optical Simulations}
\phantomsection\label{subsec:Optical_Simulations}

Optical simulations were performed using ANSYS HFSS\footnote{\url{https://www.ansys.com/products/electronics/ansys-hfss}} to optimize the absorber and radiation coupler geometry and size, the illumination configuration (front or back), and the thicknesses of the substrate and the superconducting film.

\begin{figure}[htb]
\centering
\includegraphics[scale=0.67]{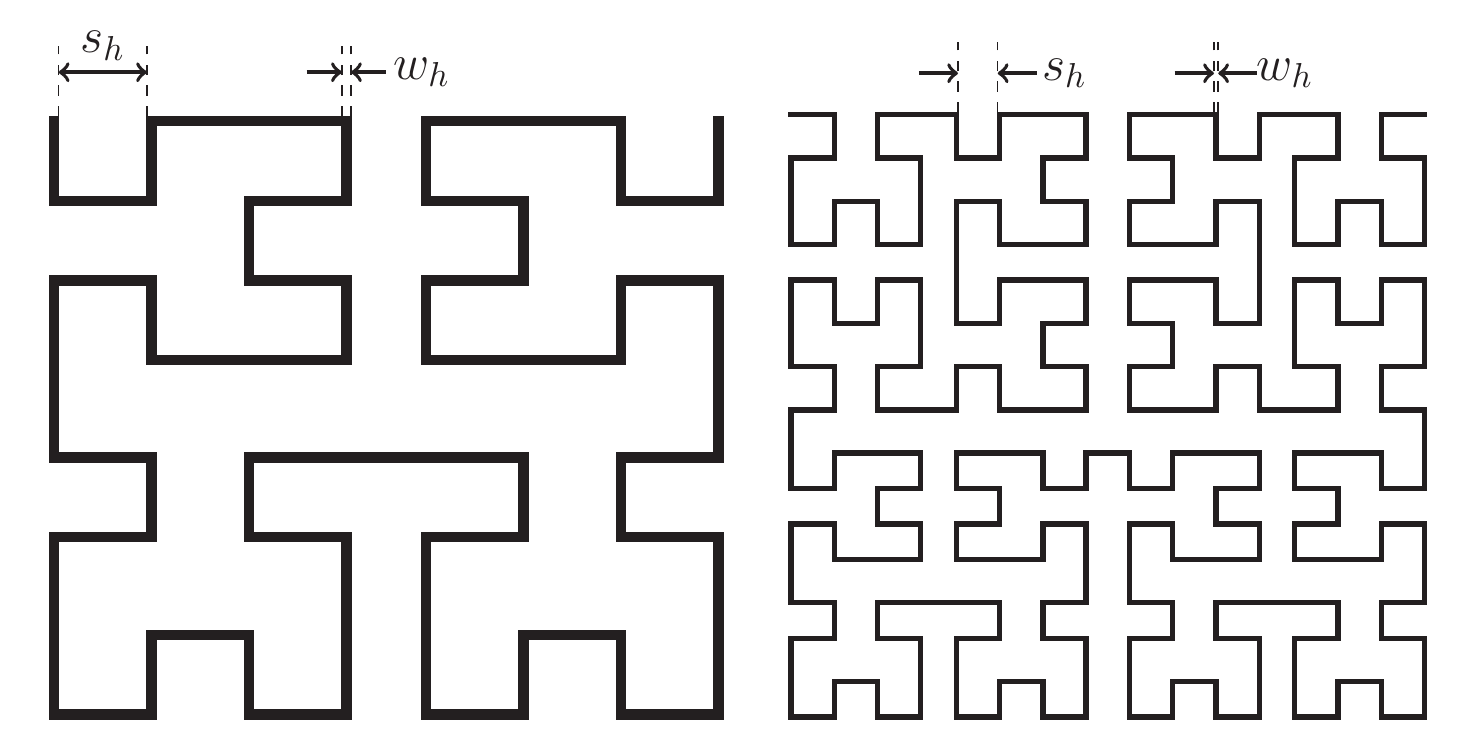}
\caption{\small Geometry of the inductor section optimized to absorb both polarizations of the incoming radiation: Hilbert fractal curve of the III (\emph{left panel}) and IV (\emph{right panel}) order.}
\phantomsection\label{fig:hilbert} 
\end{figure}

In non--polarimetric experiments as OLIMPO, the absorber geometry is optimized to absorb both polarizations of the incoming radiation: a good choice is the Hilbert fractal curve of order greater than or equal to the III; see fig.~\ref{fig:hilbert}. The Hilbert curve fills the absorbing area uniformly. This allows the detector to be sensitive to the two polarizations, with no preferential direction in absorption \citep{Monfardini2014}.

In general, radiation couplers are composed of two or three elements: a horn or planar antenna, a mode--filtering waveguide, and possibly a transition element, such as a flare, or a choke, or both of them, to ensure efficient power transfer to the detector. In our case, a horn antenna is used to couple the output of the cryogenic reimaging optics to the detector waveguide. The waveguide is the optical component selecting the modes and the lowest frequency of the radiation to be detected. The flare and the choke placed at the end of the waveguide have the main task of reducing the optical cross--talk between adjacent detectors.

For each frequency band of the OLIMPO experiment, we investigated 
\begin{itemize}
\item two different absorber geometries: the Hilbert fractal curve of the III and IV order, with different size (different values of $s_{h}$ and $w_{h}$, referring to fig.~\ref{fig:hilbert});
\item three different radiation couplers: waveguide, flared waveguide, or choked waveguide, with different size;
\item two illumination configurations: front--illuminated or back--illuminated;
\item different Si wafer and Al film thicknesses, $t_{Si}$ and $t_{Al}$ respectively (for the front illuminated configuration the Si wafer thickness coincides with the backshort distance);
\item different distances between the radiation coupler and the absorber for the front illuminated configuration, and different distance of the backshort for the back illuminated configuration, $d$.
\end{itemize}

The configurations are optimized and selected by maximizing the absorbance in the four spectral bands of OLIMPO, and minimizing the losses through the lateral surfaces of the Si wafer and the ``vacuum'' space between the radiation coupler and the absorber for the front illuminated configuration, or between the absorber and the backshort for the back illuminated configuration. Minimizing the losses corresponds to minimizing the optical cross--talk between adjacent detectors.

\begin{figure}[htb]
\centering
\includegraphics[scale=0.42]{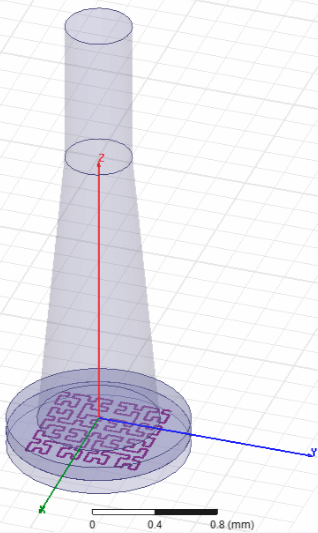}
\includegraphics[scale=0.5]{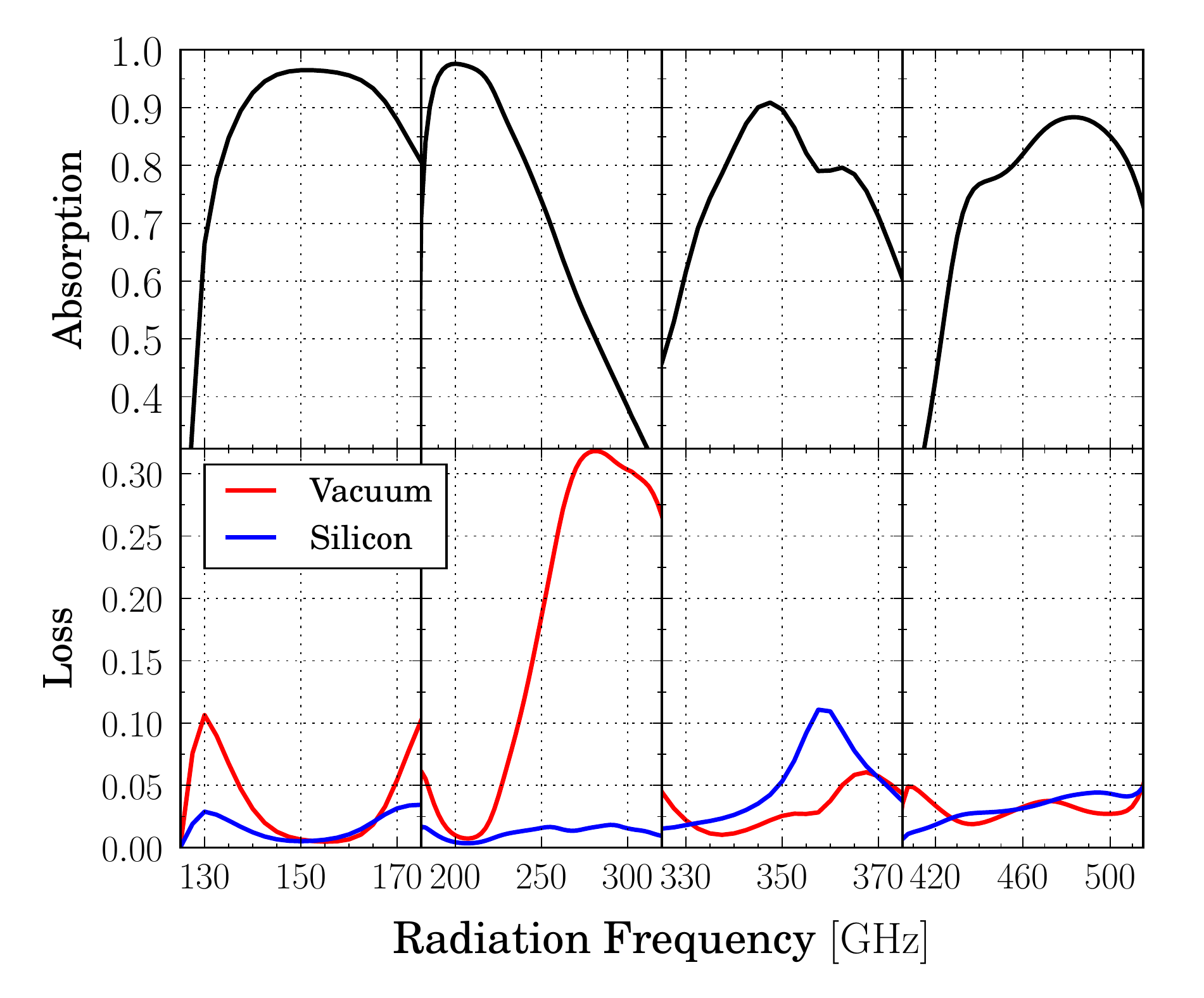}
\caption{\small \emph{Left panel}: HFSS design of a IV order Hilbert absorber, for the \SI{350}{GHz} channel of OLIMPO, coupled to the radiation through a circular flared waveguide. \emph{Right panel}: Frequency dependence of the absorption (\emph{top panels}) and the losses (\emph{bottom panels}) for the OLIMPO channels. The \emph{red lines} are for the ``vacuum'' and the \emph{blue lines} are for the silicon.}
\phantomsection\label{fig:design_results} 
\end{figure}

The \emph{left panel} of fig.~\ref{fig:design_results} shows the HFSS design of the \SI{350}{GHz} detector system in the front illuminated configuration (or in the back illuminated configuration): a circular flared waveguide, the ``vacuum'' space between the radiation coupler and the absorber (or the Si substrate), the IV order Hilbert absorber, and the Si substrate (or the ``vacuum'' space between the absorber and the backshort). The \emph{right panel} of fig.~\ref{fig:design_results} shows the results of the optical simulations of the four OLIMPO channels. The absorbance, integrated over the spectral band, is 94\% for the \SI{150}{GHz} channel, 71\% for the \SI{250}{GHz} channel, and 82\% for both the \SI{350}{GHz} channel and the \SI{460}{GHz} channel. The losses, integrated over the spectral band, are lower than the 2\% for the \SI{150}{GHz} channel, 19\% for the \SI{250}{GHz} channel, 5\% for the \SI{350}{GHz} channel, and 3\% for the \SI{460}{GHz} channel.

The results described above refer to the detector system configurations collected in tab.~\ref{tab:tot_result}. Simulations suggest that the best absorber is a front--illuminated IV order Hilbert, with the characteristic length $s_{h}$ scaling with the observed radiation wavelength. For the 150 and the \SI{250}{GHz} channels, it is possible to obtain high absorbance and low losses by constraining mainly the distance between the waveguide end and the absorber, without the need of a transition element. On the other hand, for the 350 and the \SI{460}{GHz} channels the transition element becomes important to reduce the losses, which cannot be reduced significantly by decreasing $d$.

\begin{table}[h]
	\centering
		\fontsize{8pt}{12pt}\selectfont{
		\begin{tabular}{c|c|c|c|c|c|c|c|c|c|c|c|c}
		\hline
		\hline
		\multicolumn{1}{c|}{\multirow{2}{*}{Channel}}&
		\multicolumn{1}{c|}{\multirow{2}{*}{Radiation}}&
		\multicolumn{1}{c|}{\multirow{3}{*}{Illumination}}&
		\multicolumn{1}{c|}{\multirow{2}{*}{$t_{Si}$}}&
		\multicolumn{1}{c|}{\multirow{2}{*}{$d$}}&
		\multicolumn{4}{c|}{\multirow{1}{*}{Absorber}}&
		\multicolumn{2}{c|}{\multirow{1}{*}{Waveguide}}&
		\multicolumn{2}{c}{\multirow{1}{*}{Flare}}\\
		\cline{6-13}
		\multicolumn{1}{c|}{\multirow{2}{*}{$\left[\SI{}{GHz}\right]$}}&
		\multicolumn{1}{c|}{\multirow{2}{*}{Coupler}}&
		\multicolumn{1}{c|}{}&
		\multicolumn{1}{c|}{\multirow{2}{*}{$\left[\SI{}{\mu m}\right]$}}&
		\multicolumn{1}{c|}{\multirow{2}{*}{$\left[\SI{}{\mu m}\right]$}}&
		\multicolumn{1}{c|}{\multirow{1}{*}{Hilbert}}&
		\multicolumn{1}{c|}{\multirow{1}{*}{$t_{Al}$}}&
		\multicolumn{1}{c|}{\multirow{1}{*}{$w_{h}$}}&
		\multicolumn{1}{c|}{\multirow{1}{*}{$s_{h}$}}&
		\multicolumn{1}{c|}{\multirow{1}{*}{$d_{wg}$}}&
		\multicolumn{1}{c|}{\multirow{1}{*}{$h_{wg}$}}&
		\multicolumn{1}{c|}{\multirow{1}{*}{$d_{f}$}}&
		\multicolumn{1}{c}{\multirow{1}{*}{$h_{f}$}}\\
		\multicolumn{1}{c|}{}&
		\multicolumn{1}{c|}{}&
		\multicolumn{1}{c|}{}&
		\multicolumn{1}{c|}{}&
		\multicolumn{1}{c|}{}&
		\multicolumn{1}{c|}{\multirow{1}{*}{order}}&
		\multicolumn{1}{c|}{\multirow{1}{*}{$\left[\SI{}{nm}\right]$}}&
		\multicolumn{1}{c|}{\multirow{1}{*}{$\left[\SI{}{\mu m}\right]$}}&
		\multicolumn{1}{c|}{\multirow{1}{*}{$\left[\SI{}{\mu m}\right]$}}&
		\multicolumn{1}{c|}{\multirow{1}{*}{$\left[\SI{}{mm}\right]$}}&
		\multicolumn{1}{c|}{\multirow{1}{*}{$\left[\SI{}{mm}\right]$}}&
		\multicolumn{1}{c|}{\multirow{1}{*}{$\left[\SI{}{mm}\right]$}}&
		\multicolumn{1}{c}{\multirow{1}{*}{$\left[\SI{}{mm}\right]$}}\\
		\hline
		\hline
		\multirow{1}{*}{150}&\multirow{1}{*}{Waveguide}&\multirow{1}{*}{Front}&\multirow{1}{*}{135}&\multirow{1}{*}{450}&\multirow{1}{*}{IV}&\multirow{1}{*}{30}&\multirow{1}{*}{2}&\multirow{1}{*}{162}&\multirow{1}{*}{1.4}&\multirow{1}{*}{6}&&\\
		\multirow{1}{*}{250}&\multirow{1}{*}{Waveguide}&\multirow{1}{*}{Front}&\multirow{1}{*}{100}&\multirow{1}{*}{350}&\multirow{1}{*}{IV}&\multirow{1}{*}{30}&\multirow{1}{*}{2}&\multirow{1}{*}{132}&\multirow{1}{*}{1.0}&\multirow{1}{*}{5}&&\\
		\multirow{2}{*}{350}&Flared &\multirow{2}{*}{Front}&\multirow{2}{*}{310}&\multirow{2}{*}{250}&\multirow{2}{*}{IV}&\multirow{2}{*}{30}&\multirow{2}{*}{2}&\multirow{2}{*}{72}&\multirow{2}{*}{0.60}&\multirow{2}{*}{2}&\multirow{2}{*}{1.0}&\multirow{2}{*}{7}\\
		&waveguide&&&&&&&&&&&\\
		\multirow{2}{*}{460}&Flared &\multirow{2}{*}{Front}&\multirow{2}{*}{135}&\multirow{2}{*}{150}&\multirow{2}{*}{IV}&\multirow{2}{*}{30}&\multirow{2}{*}{2}&\multirow{2}{*}{52}&\multirow{2}{*}{0.44}&\multirow{2}{*}{2}&\multirow{2}{*}{0.8}&\multirow{2}{*}{2}\\
		&waveguide&&&&&&&&&&&\\
				\hline
		\hline
		\end{tabular}
		}	
		\caption{\small Optimized parameters values for each OLIMPO channel. The table collects information about the radiation coupler geometry and size: the waveguide diameter $d_{wg}$, and height $h_{wg}$, the flare diameter $d_{f}$, and height $h_{f}$; the illumination configuration, the silicon wafer thickness $t_{Si}$, the distance between the radiation coupler and the absorber $d$, the absorber geometry and size: the aluminum film thickness $t_{Al}$, and $w_{h}$ and $s_{h}$ defined in fig.~\ref{fig:hilbert}. The results of the optical simulations over these detector systems are shown in the \emph{right panel} of fig.~\ref{fig:design_results}.}
		\phantomsection\label{tab:tot_result}	
\end{table}

\subsection{Electrical Simulations}
\phantomsection\label{subsec:Electrical_Simulations}

After fixing the geometry and the size of the absorber, we have to complete the design of the detectors by choosing the size of the capacitor, the bias coupling and the feedline. The resonator to feedline and resonator to ground couplings are two capacitors, as shown in the \emph{top panels} of fig.~\ref{fig:scheme_KIDs}. Electrical simulations are used to tune the feedline impedance and the resonant frequency, verify the lumped condition, constrain the coupling quality factor, and minimize the electrical cross--talk. 

The electrical cross--talk between adjacent resonators in the frequency domain can be minimized by suitably spacing the resonant frequencies. As we are going to see in Subsec.~\ref{subsec:Readout}, we use a readout electronics, based on a ROACH2 board, characterized by a total bandwidth of \SI{512}{MHz}. In order to populate a band of \SI{512}{MHz} with 66 (150/\SI{460}{GHz} line) or 64 (250/\SI{350}{GHz} line) detectors, we need to separate their resonant frequencies by a maximum of \SI{7.5}{MHz}. To be safe we prefer to reduce the spacing to \SI{4}{MHz} so that we can face inductive and capacitive parasitic effects, which could be present in the real device and are not included in the simulations. 

The coupling quality factor $Q_{c}$, which is a measure of the electrical losses external to the resonator, constrains the detector dynamics. Larger $Q_{c}$ values correspond to smaller detector dynamics and higher sensitivity. The OLIMPO requirements, described in Subsec.~\ref{subsec:Detector_Requirements}, translate into $Q_{c}$ of about \SI{1.5e4}{}. A first generation of detectors had been built to have $Q_{c}<\SI{5e3}{}$, with $t_{Al}=\SI{40}{nm}$. These detectors had indeed a huge dynamic range, but the responsivity was low \citep{Paiella2017}. Given the capacitance of each detector $C$, the coupling capacitance can be obtained from \citep{McCarrick2014}
\begin{equation}
\centering
C_{c}=\sqrt{\frac{8C}{2\pi\nu_{r}Q_{c}Z_{fl}}}\;,
\phantomsection\label{eq:coupling_cap_calc}
\end{equation}
where $\nu_{r}$ is the resonant frequency, $Q_{c}=1.5\times 10^{4}$, and $Z_{fl}$ is the feedline impedance.
 
Electrical simulations were performed with the SONNET\footnote{\url{http://www.sonnetsoftware.com/}} software, whose input is the layout of the detector, including the feedline, the coupling capacitors, and the ground planes. For each array we simulated the first, second, and last pixels in order to be sure that the lumped condition is verified for all the resonators (if the lumped condition is verified for the first and the last pixel, we can assume that it is verified for all the pixels), and control the spacing between the first two pixels and between the last and the first pixel of the arrays on the same readout line.

\begin{figure}[!h]
\centering
\includegraphics[scale=0.395]{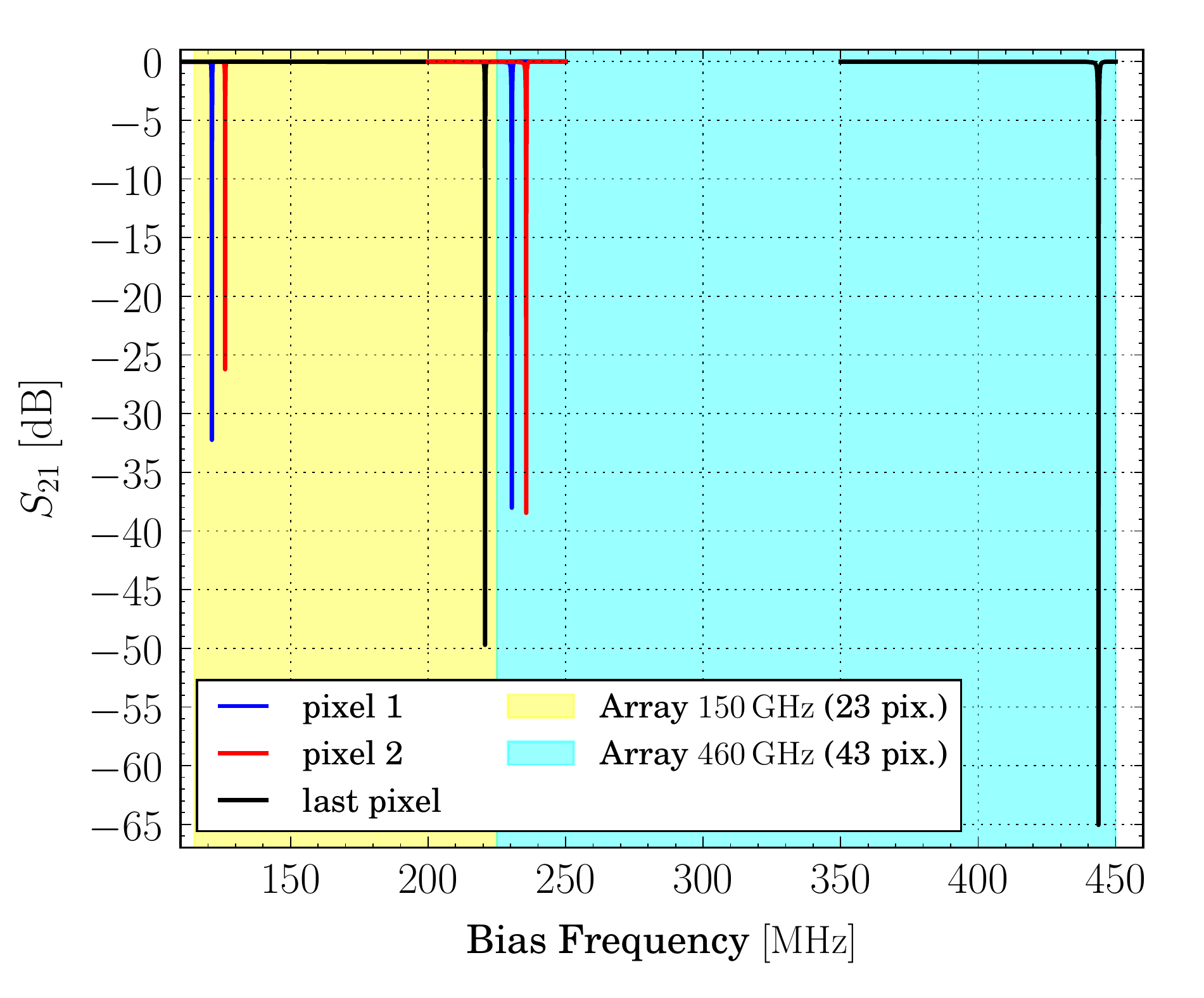}
\includegraphics[scale=0.395]{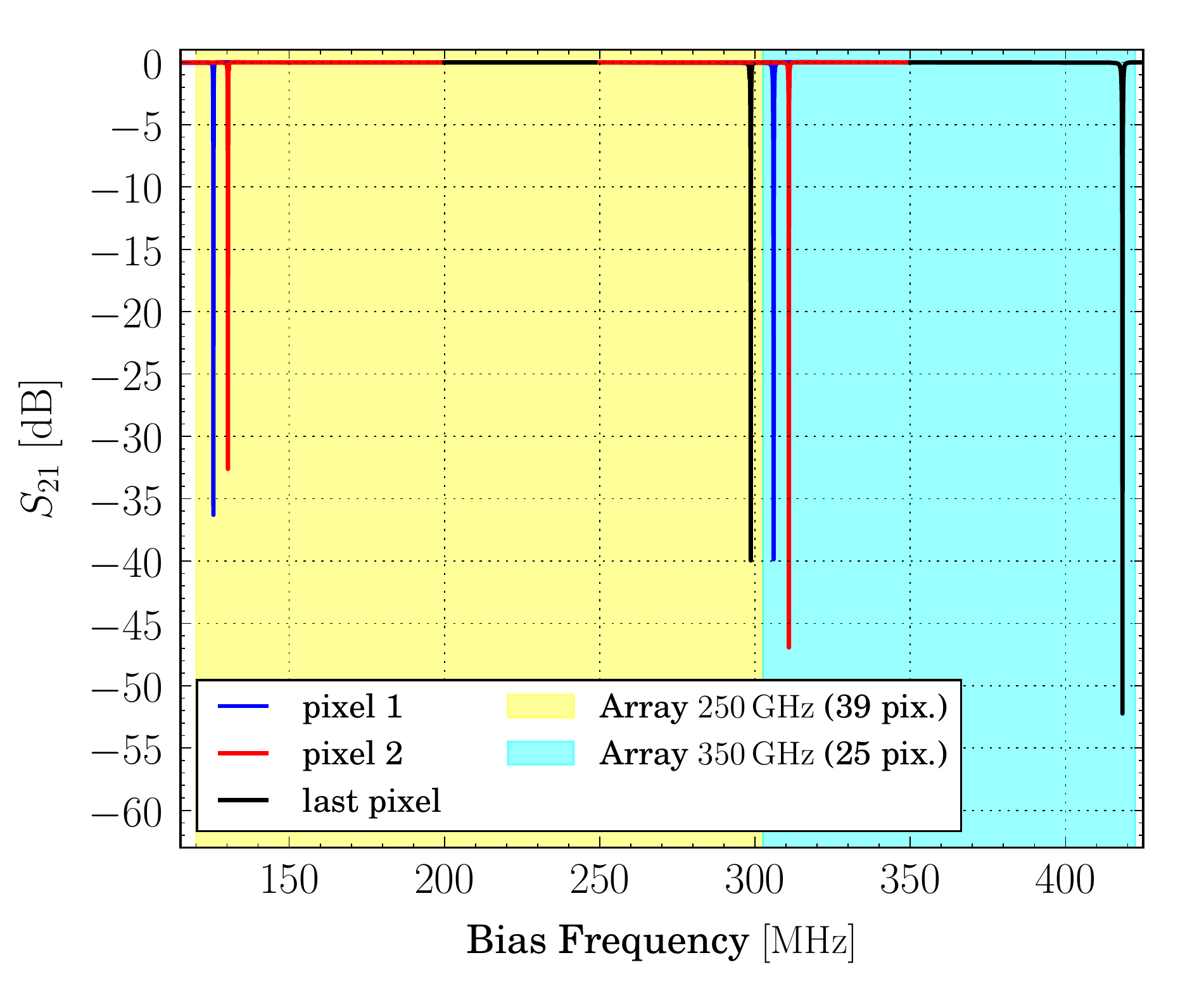}
\caption{\small Bias frequency dependence of the transmission scattering parameter $S_{21}$ for three simulated pixels of the 150 and the \SI{460}{GHz} arrays (\emph{left panel}), and of the 250 and \SI{350}{GHz} arrays (\emph{right panel}).}
\phantomsection\label{fig:electrical_simu} 
\end{figure}

Fig.~\ref{fig:electrical_simu} displays the SONNET results for the transmission scattering parameter $S_{21}$ of the three simulated pixels of each array: the 150/\SI{460}{GHz} line in the \emph{left panel}, and the 250/\SI{350}{GHz} line in the \emph{right panel}. Tab.~\ref{tab:ele_simul} collects the results of the simulations: frequency ranges, bandwidths, spacing of the resonators of the four arrays, and percentage non--uniformity in the current distribution along the inductor.
%For the \SI{150}{GHz} channel, the first and the second pixel frequencies are separated by \SI{5}{MHz}, and the total bandwidth is \SI{100}{MHz}, which leads to an average spacing of about \SI{4.3}{MHz}; for the \SI{460}{GHz} channel, the first and the second pixels are separated by \SI{5}{MHz}, the total bandwidth is \SI{215}{MHz}, and the average spacing is about \SI{5}{MHz}; for the \SI{250}{GHz} channel, the first and the second pixels are separated by \SI{5}{MHz}, the total bandwidth is \SI{175}{MHz}, and the average spacing is about \SI{4.5}{MHz}; for the \SI{350}{GHz} channel, the first and the second pixels are separated by \SI{5}{MHz}, the total bandwidth is \SI{115}{MHz}, and the average spacing is about \SI{4.6}{MHz}.

\begin{table}[htb]
	\centering
		\fontsize{10pt}{15pt}\selectfont{
		\begin{tabular}{c|c|c|c|c|c|c|c}
		\hline
		\hline
		\multicolumn{1}{c|}{\multirow{2}{*}{Channel}}&
        \multicolumn{1}{c|}{\multirow{2}{*}{\# of}}&
		\multicolumn{4}{c|}{\multirow{1}{*}{Resonant frequencies}}&
        \multicolumn{2}{c}{\multirow{1}{*}{Maximum non--uniformity in the}}\\
		%\multicolumn{1}{c|}{\multirow{1}{*}{Optimal}}&
		%\multicolumn{1}{c}{\multirow{1}{*}{$\#$3}}\\
		\cline{3-6}      
		\multicolumn{1}{c|}{\multirow{1}{*}{}}&
		\multicolumn{1}{c|}{\multirow{2}{*}{pixels}}&
        \multicolumn{1}{c|}{\multirow{1}{*}{Range}}&
		\multicolumn{1}{c|}{\multirow{1}{*}{Bandwidth}}&
		\multicolumn{2}{c|}{\multirow{1}{*}{Spacing $\left[\SI{}{MHz}\right]$}}&
        \multicolumn{2}{c}{\multirow{1}{*}{inductor current distribution}}\\
        \cline{5-8} 
        \multicolumn{1}{c|}{\multirow{1}{*}{$\left[\SI{}{GHz}\right]$}}&
        \multicolumn{1}{c|}{\multirow{1}{*}{}}&
		\multicolumn{1}{c|}{\multirow{1}{*}{$\left[\SI{}{MHz}\right]$}}&
		\multicolumn{1}{c|}{\multirow{1}{*}{$\left[\SI{}{MHz}\right]$}}&
		\multicolumn{1}{c|}{\multirow{1}{*}{1--2}}&
		\multicolumn{1}{c|}{\multirow{1}{*}{Average}}&
		\multicolumn{1}{c|}{\multirow{1}{*}{$\quad$First pixel$\quad$}}&
		\multicolumn{1}{c}{\multirow{1}{*}{Last pixel}}\\
		\hline
		\hline
		150&23&$\left[120;220\right]$&100&5&4.3&5\%&9\%\\
		250&39&$\left[125;300\right]$&175&5&4.5&4\%&10\%\\
		350&25&$\left[305;420\right]$&115&5&4.6&15\%&10\%\\
		460&43&$\left[230;445\right]$&215&5&5&13\%&18\%\\
		\hline
		\hline
		\end{tabular}
		}		
		\caption{\small Results of the SONNET simulations: frequency ranges, bandwidths, spacing of the resonators of the four arrays, and percentage non--uniformity in the current distribution along the inductor.}
		\phantomsection\label{tab:ele_simul}
\end{table}

\begin{figure}[htb]
\centering
\includegraphics[scale=0.275]{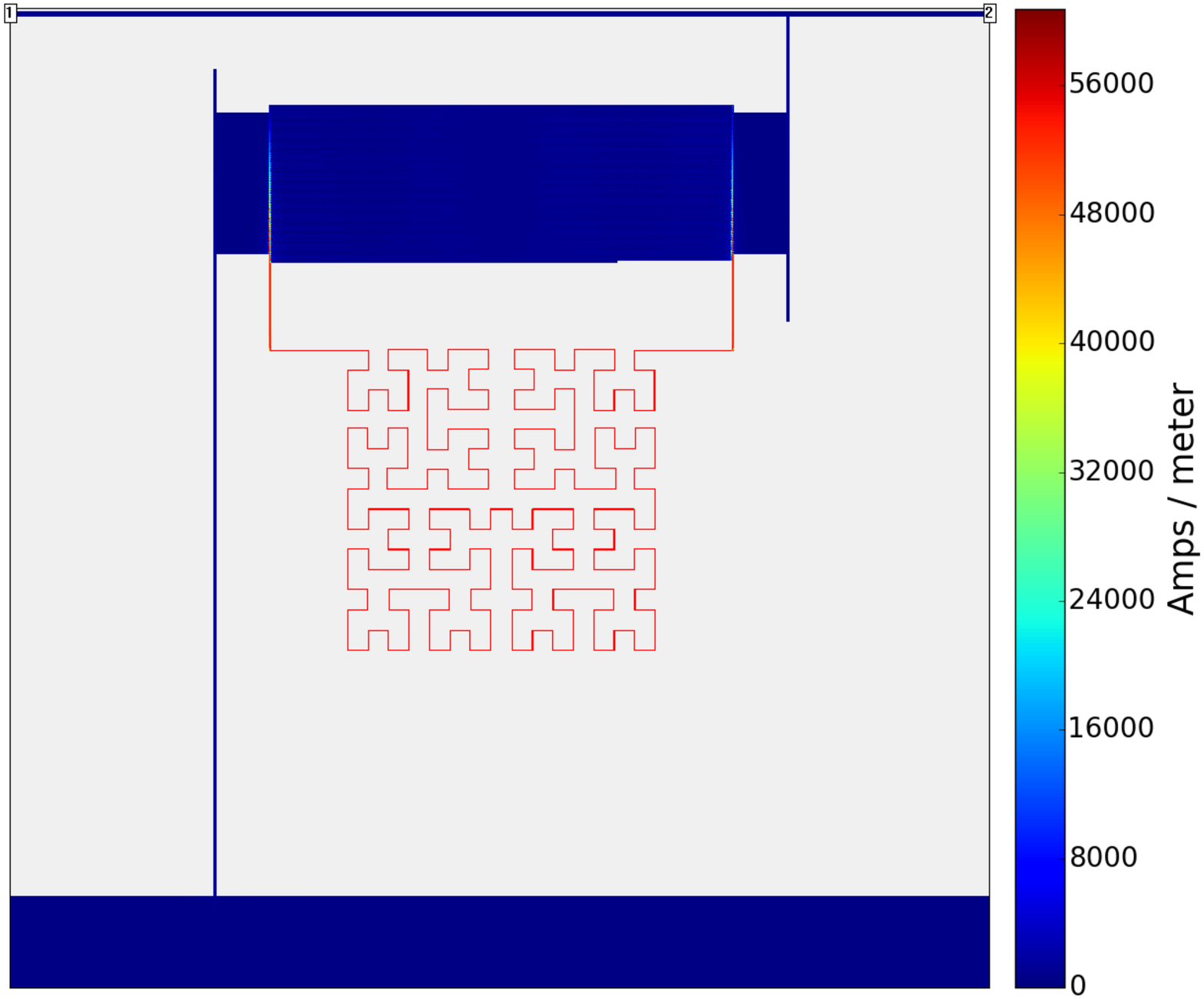}
\includegraphics[scale=0.275]{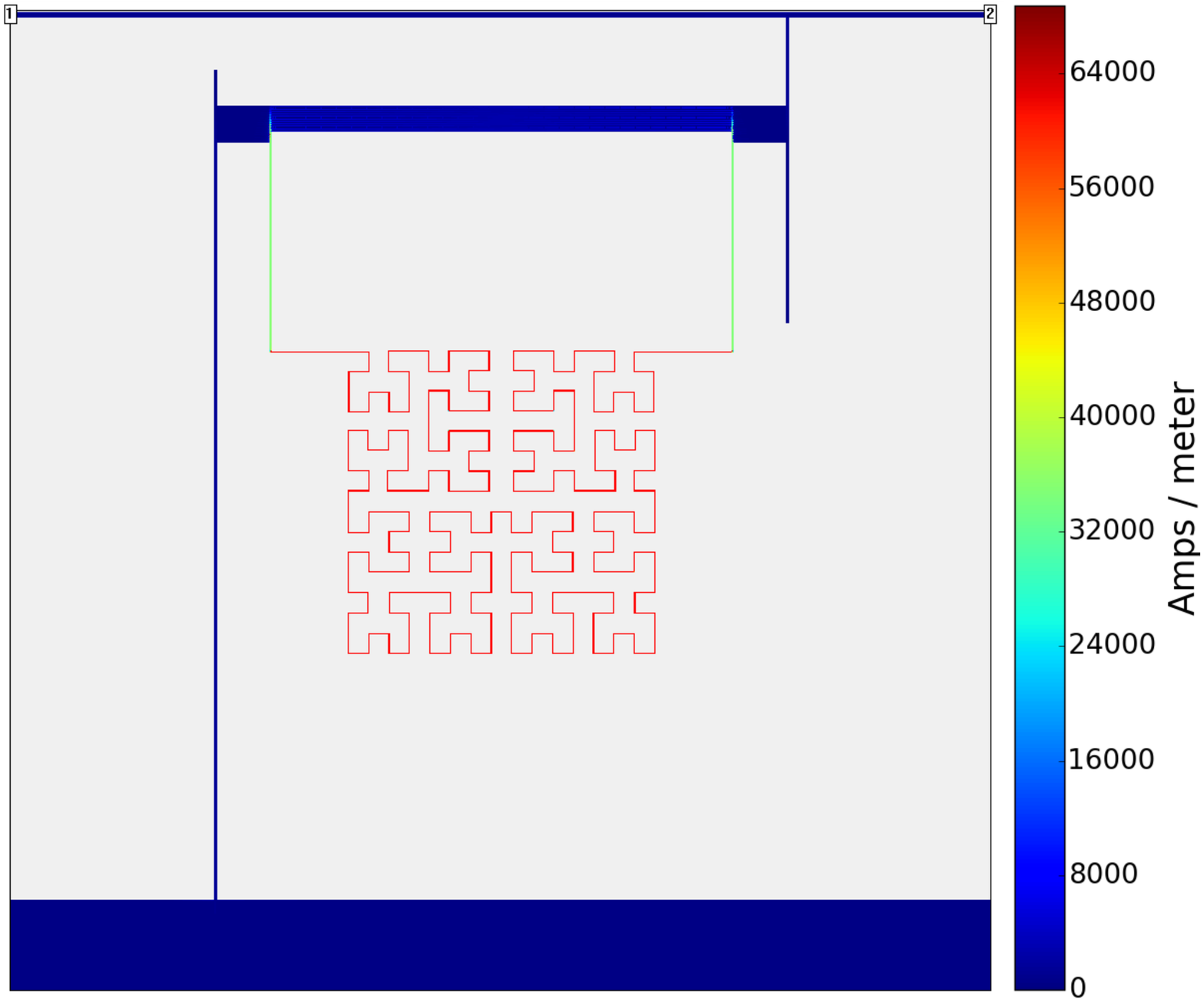}
\caption{\small Current distribution from SONNET for the first (\emph{left panel}) and the last (\emph{right panel}) pixels of the \SI{250}{GHz} array. The current is null in the capacitors and is uniform in the inductors. The maximum non--uniformity along the inductors is about 4\% for the first pixel, and about 10\% for the last pixel.}
\phantomsection\label{fig:current} 
\end{figure}

The lumped element condition is verified if the current distribution, at the resonant frequency, is uniform in the inductor and null in the capacitor. We consider this condition satisfied if the current variation between any two points along the inductor is always lower than 20\%. As an example, fig.~\ref{fig:current} shows the current distribution for the first and the last pixels of the \SI{250}{GHz} array.
%The lumped element condition is verified if the current distribution, at the resonant frequency, is uniform in the inductor and null in the capacitor. We consider this condition satisfied if the current variation between any two points along the inductor is always lower than 20\%. For the \SI{150}{GHz} first pixel the maximum non--uniformity in the current distribution along the inductor is 5\%, while for the last pixel is 9\%; for the \SI{460}{GHz} first pixel is 13\%, while for the last pixel is 18\%; for the \SI{250}{GHz} first pixel is 4\%, while for the last pixel is 10\%; and for the \SI{350}{GHz} first pixel is 15\%, while for the last pixel is 19\%. As an example, fig.~\ref{fig:current} shows the current distribution for the first and the last pixels of the \SI{250}{GHz} array.

\section{Fabrication process}
\phantomsection\label{sec:Fabrication_process} 

Our detectors are fabricated in the ISO5/ISO6 clean room of the Istituto di Fotonica e Nanotecnologie (IFN) of the Consiglio Nazionale delle Ricerche (CNR).

The layout of the KID array is first realized by electron beam lithography (EBL) on the polymethyl methacrylate (PMMA) film uniformly deposited on the Si wafer: the electron irradiation chemically modified the PMMA structure that is then developed in a solution (1:1) of methyl isobutyl ketone (MIBK) and isopropyl alcohol (IPA). A thin aluminium film is subsequently deposited by electron--gun evaporation on the substrate patterned with PMMA. The aluminium deposition rate ($\sim$\SI{10}{A/s}) and final thickness ($\sim$\SI{30}{nm}) have been controlled during the deposition by a quartz micro--balance and checked afterwards with a mechanical profilometer. Finally, the excess metal, deposited on the residual PMMA, is removed by a lift--off process \citep{Colantoni2016}.
 
The detectors are fabricated on high-quality (FZ method) intrinsic Si(100) wafers, with high resistivity ($\rho>\SI{10}{k\Omega.cm}$), double side polished. The face of the Si wafer opposite to that where the detectors have been realized is metalized with \SI{200}{nm} of Al. This film acts as a backshort for the incoming radiation. 

\begin{figure}[htb]
\centering 
\includegraphics[scale=0.5]{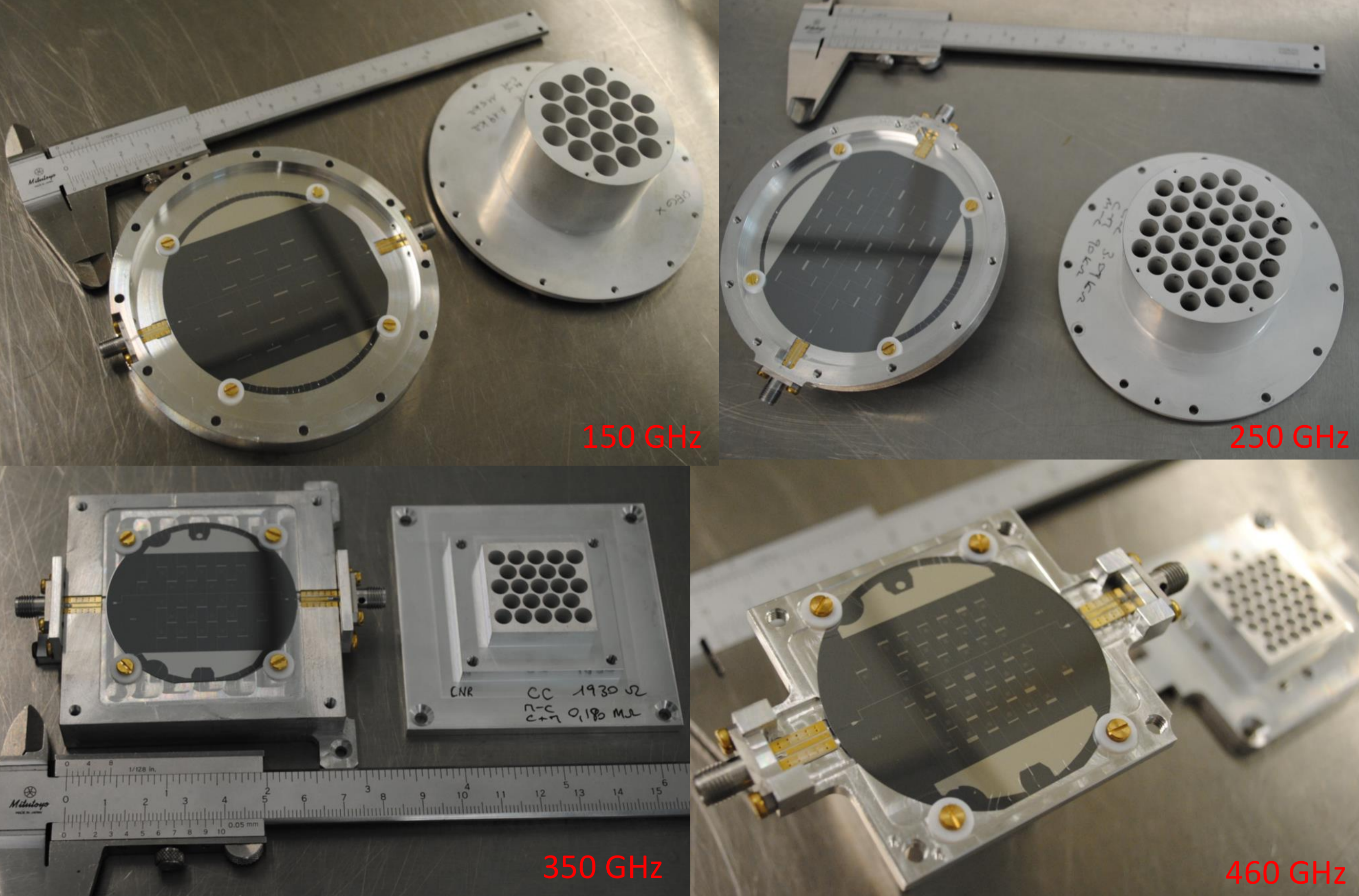}
\caption{\small Pictures of the four OLIMPO detector arrays mounted in their holders by means of four Teflon washers \SI{100}{\mu m} thick, and the horn arrays (including the waveguides and the flares where present). \emph{Clockwise from top left}: the \SI{150}{GHz}, \SI{250}{GHz}, \SI{460}{GHz}, and \SI{350}{GHz} arrays.}
\phantomsection\label{fig:photos}
\end{figure}

Fig.~\ref{fig:photos} shows the pictures of the four OLIMPO detector arrays, mounted in their holders by means of four Teflon washers \SI{100}{\mu m} thick, and their horn arrays. Both the sample holders and the horn arrays are made of ergal alloy (aluminum 7075), ensuring good thermalization, reducing the power losses through the horn arrays, and minimizing the interactions between the holders and the detectors.     

\section{Test Setup}
\phantomsection\label{sec:Experimental_Setup} 

The experimental setup for the electrical characterization is composed of a dark (no optical system, blanked detectors) laboratory cryogenic system and its readout electronics, while for the optical characterization is composed of the OLIMPO cryostat and its optical system and flight readout electronics.

\subsection{Dark laboratory cryostat}
\phantomsection\label{subsec:Cryostat} 
The cryogenic system is composed of a Sumitomo 062B pulse tube cryocooler\footnote{\url{http://www.shicryogenics.com/products/pulse-tube-cryocoolers}} (PTC), a $^{3}$He$/^{4}$He sorption fridge, and a custom dilution refrigerator. The PTC features two temperature stages: one at \SI{45}{K} and the other at \SI{3.5}{K}. The $^{3}$He$/^{4}$He fridge, mounted on the coldest stage of the pulse tube cryocooler, provides other two temperature stages: one at \SI{1}{K} and the other at \SI{350}{mK}. The last cooling stage, anchored on the coldest stage of the $^{3}$He$/^{4}$He fridge, is provided by a single--shot miniature $^{3}$He$/^{4}$He dilution refrigerator, whose mixing chamber can reach about \SI{136}{mK}, under an optical loading of about \SI{14}{\mu W} \citep{Melhuish2013}. 

The detector array in its holder is mounted on the coldest stage, with the horn entrance apertures closed by an aluminum foil to perform dark test. The sample holder is equipped with a heater and a thermometer to perform temperature sweeps and to reproduce the flight operating conditions (in the OLIMPO cryostat the detectors work at \SI{300}{mK}).

\subsection{OLIMPO cryostat and optical system}
\phantomsection\label{subsec:OLIMPOCryostat} 

\begin{figure}[!h]
\centering
\includegraphics[scale=0.32]{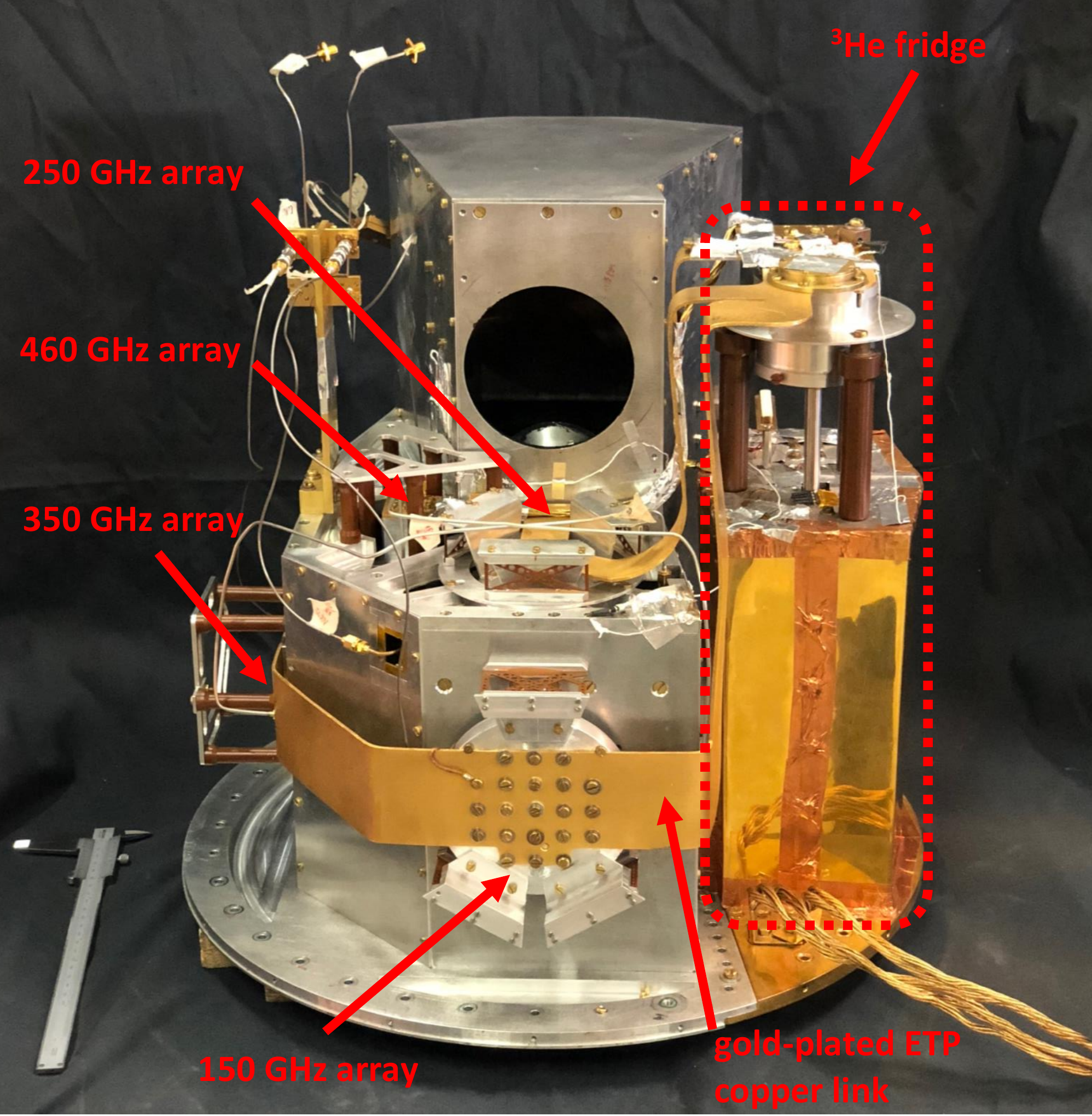}
\includegraphics[scale=0.42]{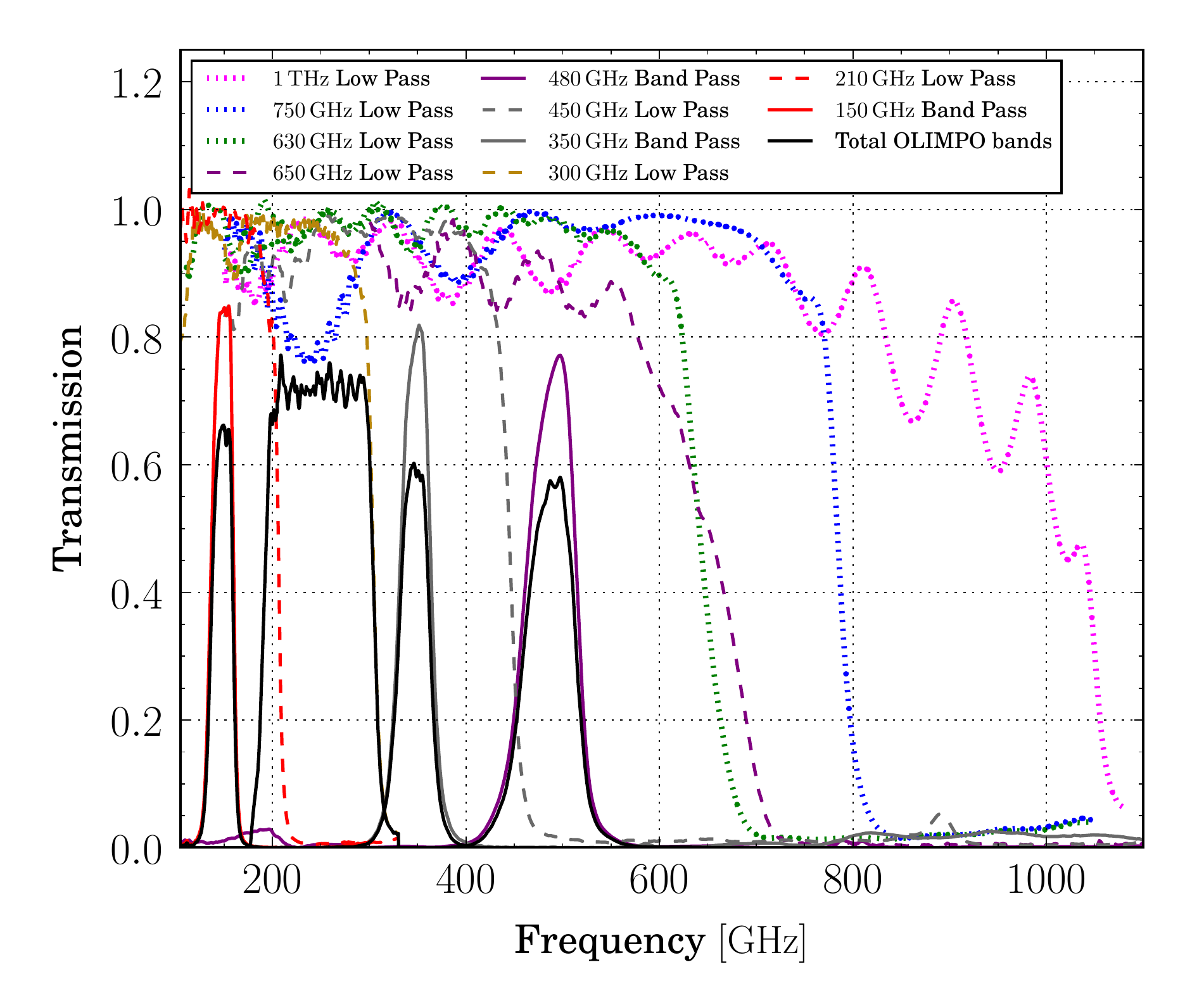}
\caption{\small \emph{Left panel}: Picture of the \emph{optics box} where the detector arrays, the $^{3}$He fridge, and the gold--plated ETP copper link are indicated. \emph{Right panel}: Transmission spectra of the filters mounted in the OLIMPO cryostat. The \emph{dotted lines} are the transmission spectra of the common filters to the four arrays: \SI{1}{THz} low pass filter, in \emph{fuchsia}, mounted on the N$_2$ shield at \SI{77}{K}; \SI{750}{GHz} low pass filter, in \emph{blue}, mounted on the vapor $^{4}$He shield at \SI{30}{K}; \SI{630}{GHz} low pass filter, in \emph{green}, mounted on the \emph{optics box} shield at \SI{1.8}{K}. The \emph{color dashed} and \emph{solid lines} are the transmission spectra of the filters mounted on the horn holders: different colors represent different detector arrays (in \emph{red} the \SI{150}{GHz} band, in \emph{goldenrod} the \SI{250}{GHz} band, in \emph{gray} the \SI{350}{GHz} band, and in \emph{purple} the \SI{460}{GHz} band), \emph{dashed lines} are for the low pass filters, and \emph{solid lines} are for band pass filters. The \emph{black solid line} represents the total transmission spectra of the OLIMPO bands obtained by the convolution of all the spectra on the same optical path (for the \SI{250}{GHz} band the lower cut--off is given by the \SI{190}{GHz} dichroic filter).}
\phantomsection\label{fig:op_and_filtri_OLIMPO}
\end{figure}

The cryogenic system of OLIMPO consists of a wet N$_2$ plus $^{4}$He cryostat, coupled to a $^{3}$He sub--K refrigerator, able to cool at about \SI{300}{mK}, for about \SI{16}{days}, the four detector arrays in their holders and their horn arrays, thanks to a gold--plated electrolytic tough pitch (ETP) copper link. The temperatures of liquid N$_2$ and $^{4}$He are \SI{77}{K} and \SI{4.2}{K} at standard pressure, respectively. The final temperature of the pumped $^{4}$He bath is \SI{1.8}{K}, useful to condensate efficiently the $^{3}$He and to reduce the emission of the \emph{optics box} (\emph{left panel} of fig.~\ref{fig:op_and_filtri_OLIMPO}). 
%The N$_2$ tank can host up to \SI{74}{l} of liquid N$_2$, while the $^{4}$He tank can host up to \SI{64}{l} of liquid $^{4}$He. The temperatures of liquid N$_2$ and $^{4}$He are \SI{77}{K} and \SI{4.2}{K} at standard pressure, respectively. The $^4$He bath is pumped during operations. About 40\% of the liquid $^4$He is used to cooldown the other 60\%.  The final temperature of the $^{4}$He bath is \SI{1.6}{K}, useful to condensate efficiently the $^{3}$He and to reduce the emission of the \emph{optics box}.
%In order to reduce the radiative heat load on the N$_2$ tank from the external shell, the \emph{superinsulation} approach has been performed by means of 30 layers of aluminized mylar interposed between the two stages \citep{masi1999}. 
A copper vapor $^{4}$He cooled intermediate shield further reduces the radiative thermal input on the $^{4}$He stage. In nominal thermal load conditions the regime temperature of this shield is about \SI{30}{K}. 

%When all is thermalized, the evaporation rate of N$_{2}$ is about \SI{0.22}{l/h} and that of $^{4}$He is about \SI{0.10}{l/h}, which guarantees an hold time of about \SI{15}{days} for both the liquids. During in-flight operations the Nitrogen bath is pumped, allowing a further thermal input reduction, and a corresponding increase in the cryogens hold time.

%The single shot $^{3}$He fridge is composed of a cryopump, an evaporator and a heat exchanger, which is in thermal contact with the $^{4}$He tank. The fridge is very similar to the one used in the BOOMERanG experiment \citep{Masi1998}, with the exception of the gas heat switch, used this time to connect when needed the cryopump to the main $^{4}$He bath. The fridge contains about \SI{40}{l} STP of $^{3}$He, and is able to cool at about \SI{300}{mK} the four detector arrays in their holders and their horn arrays for about \SI{16}{days}. The thermal contact between the evaporator and the array holders is obtained by means of a gold--plated electrolytic tough pitch (ETP) copper link, see the \emph{left panel} of fig.~\ref{fig:op_and_filtri_OLIMPO}.

The cryostat window is a disk of high density polyethylene (HDPE) \citep{DALESSANDRO201859}, with antireflection coating. The part of the OLIMPO optical system placed inside the cryostat is composed of a filter chain (see the \emph{right panel} of fig.~\ref{fig:op_and_filtri_OLIMPO}), three cold reimaging aluminum mirrors and the radiation couplers. %Incident radiation enters the cryostat through the window and encounters the first thermal shader, located behind the window. Then, radiation passes through the second thermal shader, the \SI{1}{THz} low pass filters, and the third thermal shader, all three anchored on the N$_{2}$ shield at \SI{77}{K}. After passing through the fourth thermal shader and the \SI{750}{GHz} low pass filter mounted on the vapor $^{4}$He shield at \SI{30}{K}, incoming radiation encounters the \SI{630}{GHz} low pass filter mounted on the \emph{optics box} shield at \SI{1.8}{K}. 

%Inside the \emph{optics box}, the low pass filtered radiation is reflected through three mirrors on the \SI{300}{GHz} dichroic filter, which splits the radiation into two beams: the low frequency beam component is transmitted, and the high frequency component is reflected. The low frequency beam is further split by the \SI{190}{GHz} dichroic, transmitting the lower frequencies towards the \SI{150}{GHz} array, and reflecting the higher frequencies towards the \SI{250}{GHz} array. The high frequency beam is split by the \SI{410}{GHz} dichroic filter, transmitting the lower frequencies towards the \SI{350}{GHz} array, and reflecting the higher frequencies towards the \SI{460}{GHz} array. All these optical components are in thermal contact with the \emph{optics box}, nominally at \SI{1.6}{K}. 

%For each array the beam passes through a low pass and a band pass filter, except for  the \SI{250}{GHz} array which uses only a low pass filter. These filters are anchored on the horn holders at $\sim$ \SI{300}{mK}. At this point, the filtered beams enter the horns, which include mode-selecting waveguides, and finally arrive to the detector arrays wafers.

\subsection{Readout}
\phantomsection\label{subsec:Readout} 

Our readout channels (a single one for the test cryostat, two for the OLIMPO cryostat) are composed of a bias line (input to the KID array) and a readout line (output of the KID array). The lines are made of stainless steel, Cu--Ni, and Nb--Ti coaxial cables with SMA connectors, and are run from the room--temperature connectors on the vacuum--jacket shell of the cryostat all the way down to the coldest stage. The signal on the input line is attenuated by three cryogenic attenuators (\SI{-10}{dB} each). The signal on the output line is amplified by a cryogenic low noise amplifier\footnote{\url{http://thz.asu.edu/products.html}} (LNA), developed by Arizona State University (ASU), mounted on the \SI{3.5}{K} plate, and is cleaned from DC through a DC block. In the OLIMPO cryostat, the LNAs of the two readout channels are mounted on the vapor $^{4}$He shield at \SI{30}{K}. All these components form the {\sl cold electronics}. 

\begin{figure}[!h]
\centering
\includegraphics[scale=0.45]{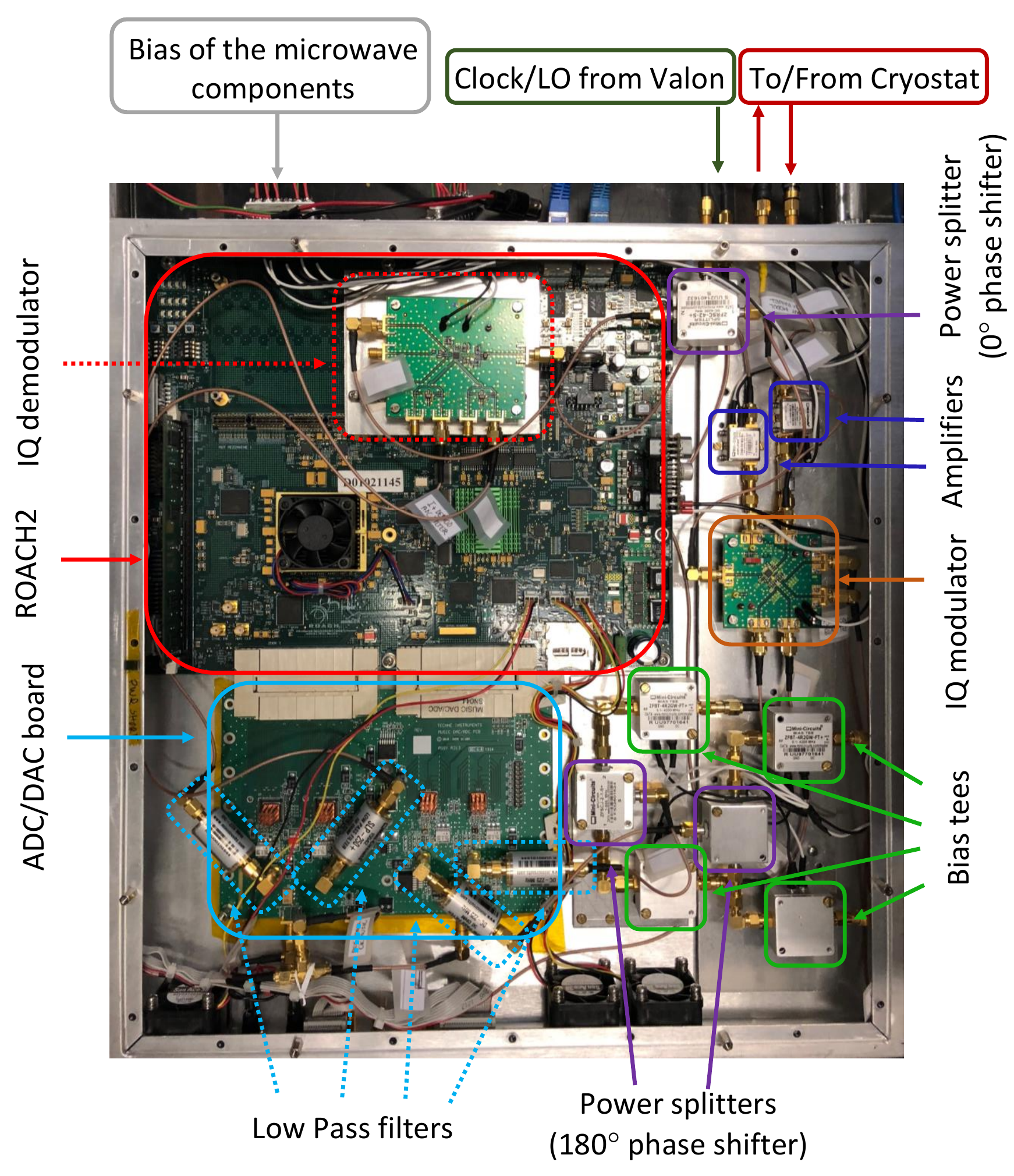}
\includegraphics[scale=0.555]{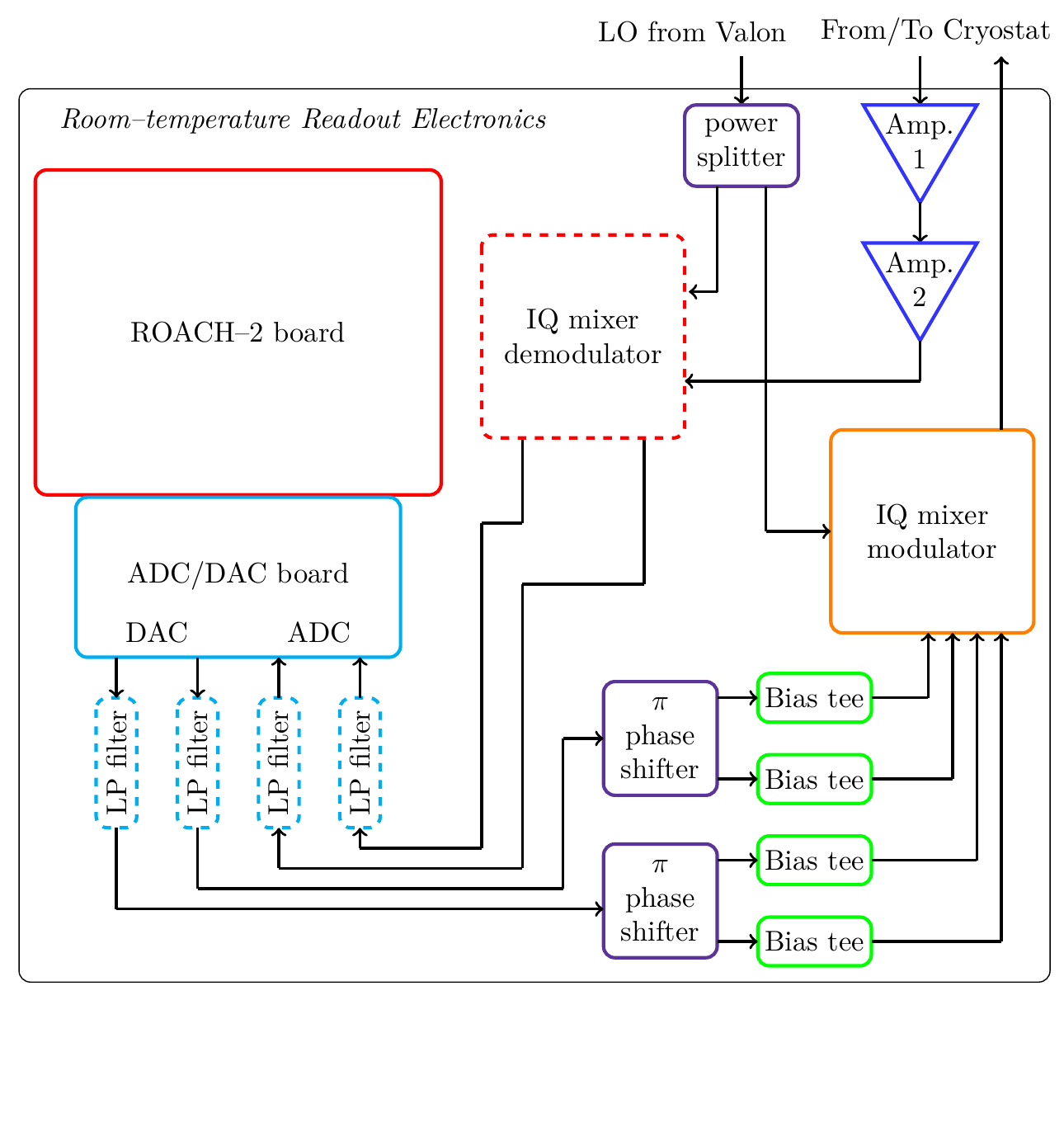}
\caption{\small Photo (\emph{left panel}) and block diagram (\emph{right panel}) of the laboratory OLIMPO readout \emph{room--temperature electronics}: this includes a ROACH2 (\emph{red solid box}), a MUSIC DAC/ADC board (\emph{cyan solid box}), three power splitters (\emph{violet solid boxes}), two room--temperature amplifiers (\emph{blue solid boxes}), an IQ modulator (\emph{red dotted box}), four bias tees (\emph{green solid boxes}), four low pass filters (\emph{cyan dotted boxes}), an IQ demodulator (\emph{orange solid box}), and a frequency synthesizer (not shown).}
\phantomsection\label{fig:photo_roach}
\end{figure}

The readout channel is completed by a {\sl room--temperature electronics} which generates a bias signal as the superposition of the tones matching the nominal resonant frequencies of all the KID pixels, and monitors the signal transmitted by each of the KID pixels when their resonances are shifted by the radiation flux. We use a ROACH2 board\footnote{\url{https://casper.berkeley.edu/wiki/ROACH2}}, including a MUSIC DAC/ADC board\footnote{\url{https://casper.berkeley.edu/wiki/MUSIC_Readout}}. Since the resonances of the OLIMPO KIDs extend to frequencies higher than \SI{256}{MHz}, our \emph{room--temperature electronics} (see fig.~\ref{fig:photo_roach}) includes up--conversion and down--conversion microwave components as described below. 

The baseband bias signals (in phase, $I$, and in quadrature, $Q$) output by the DACs are cleaned by low pass filters (\emph{Mini Circuits SLP$-$250$+$}), and up--converted by an IQ mixer modulator (\emph{Analog Devices ADL 5385}). The local oscillator (LO) is a frequency synthesizer model \emph{Valon technology 5009}. Its reference signal is split by a power splitter (\emph{Mini Circuits ZFRSC$-$42$-$S$+$}) and supplies the modulator cited above and the demodulator described below. The \emph{ADL 5385} IQ mixer requires as input the $I$ and $Q$ signals and their $\SI{180}{^{\circ}}$ phase shifts (obtained by means of a \emph{Mini Circuits ZFSCJ$-$2$-$1$-$S$+$}); all the input signals are offset positive by means of four bias tees (\emph{Mini Circuits ZFBT$-$4R2GW$-$1$-$FT$+$}). The up--converted signal output of the \emph{ADL 5385} is connected to the bias line input of the cryostat. 

The signal from the readout line output of the cryostat is amplified by two room--temperature amplifiers (\emph{Mini Circuits ZX60$-$P103LN$+$} and \emph{ZX60$-$3018G$+$}), and input to the IQ mixer demodulator (\emph{Analog Devices ADL 5387}) to be down--converted to the baseband. The baseband $I$ and $Q$ demodulator outputs are low--passed (via \emph{Mini Circuits SLP$-$250$+$} filters) and input to the ADCs of the MUSIC board. The converted signal is processed by the FPGA to measure the {\sl amplitudes} of the $I$ ad $Q$ signals transmitted by all the KID pixels. The FPGA firmware has been developed by ASU, and is able to generate up to 1000 tones over an up--converted \SI{512}{MHz} bandwidth, with a demodulated output sampling rate up to about \SI{1}{kHz} \citep{Gordon2016}.

\section{Results}
\phantomsection\label{sec:Results} 

\subsection{Electrical characterization}
\phantomsection\label{subsec:Electrical_characterization} 

Here we describe the electrical characterization of the four OLIMPO detector arrays. Each array has been individually cooled inside the dark cryostat, and characterized at both the base temperature and \SI{300}{mK}. Because of the different geometry, volume and mass of the sample holders, due to the different mounting needs of the holders in the OLIMPO cryostat, the base temperature reached is different for the four channels: \SI{185}{mK}, \SI{168}{mK}, \SI{155}{mK}, and \SI{255}{mK} for the 150, 250, 350, and \SI{460}{GHz} holders, respectively.  

\begin{figure}[!h]
\centering
\includegraphics[scale=0.265]{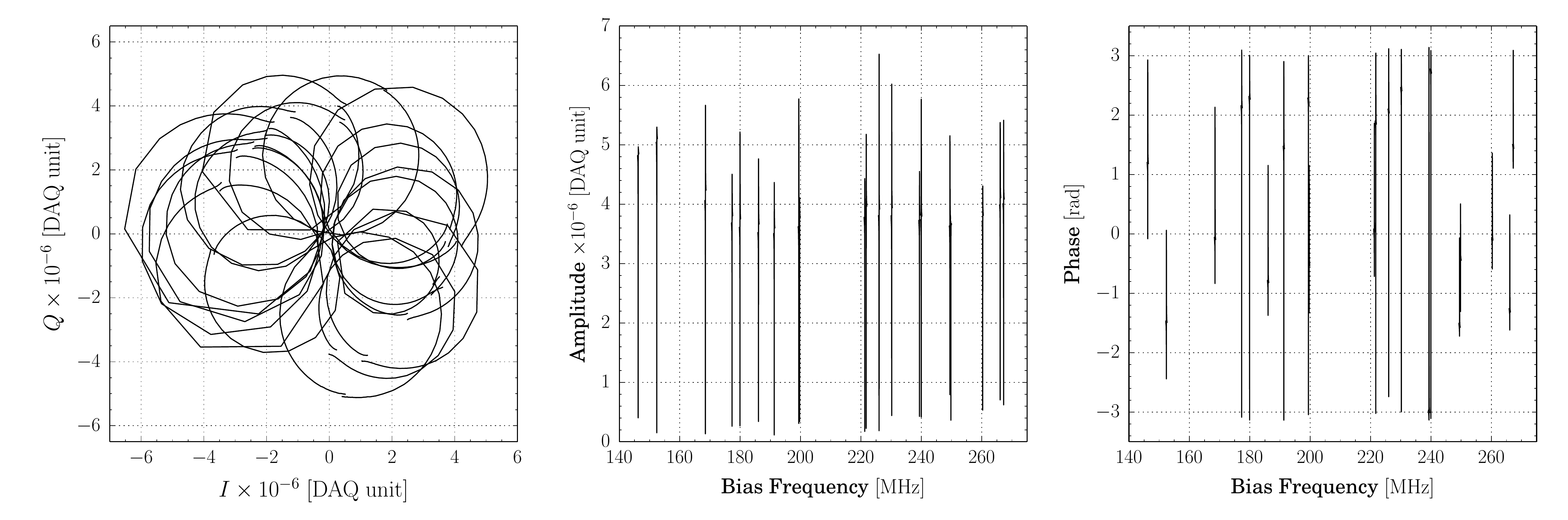}
\includegraphics[scale=0.265]{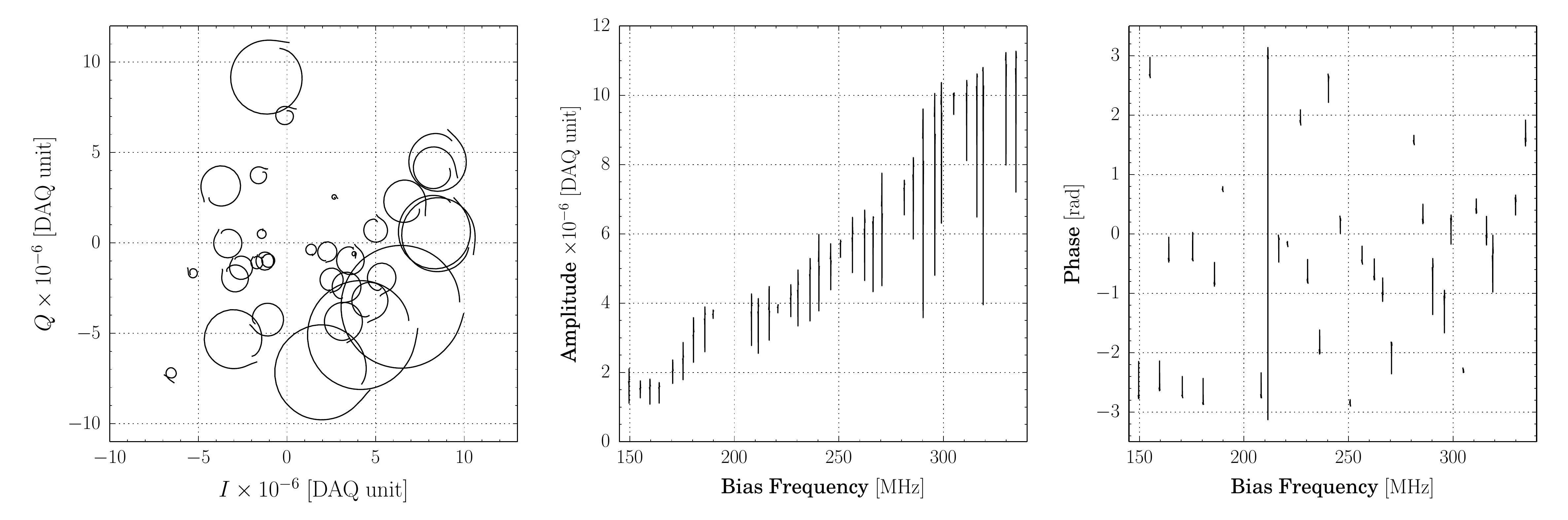}
\includegraphics[scale=0.265]{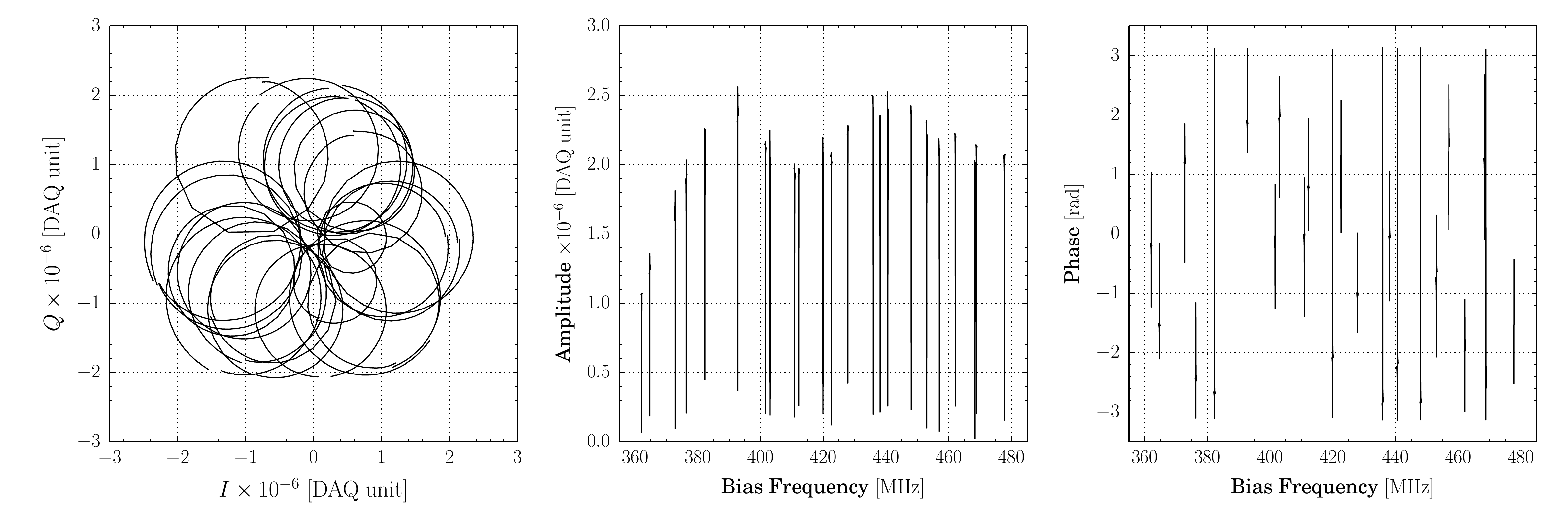}
\includegraphics[scale=0.265]{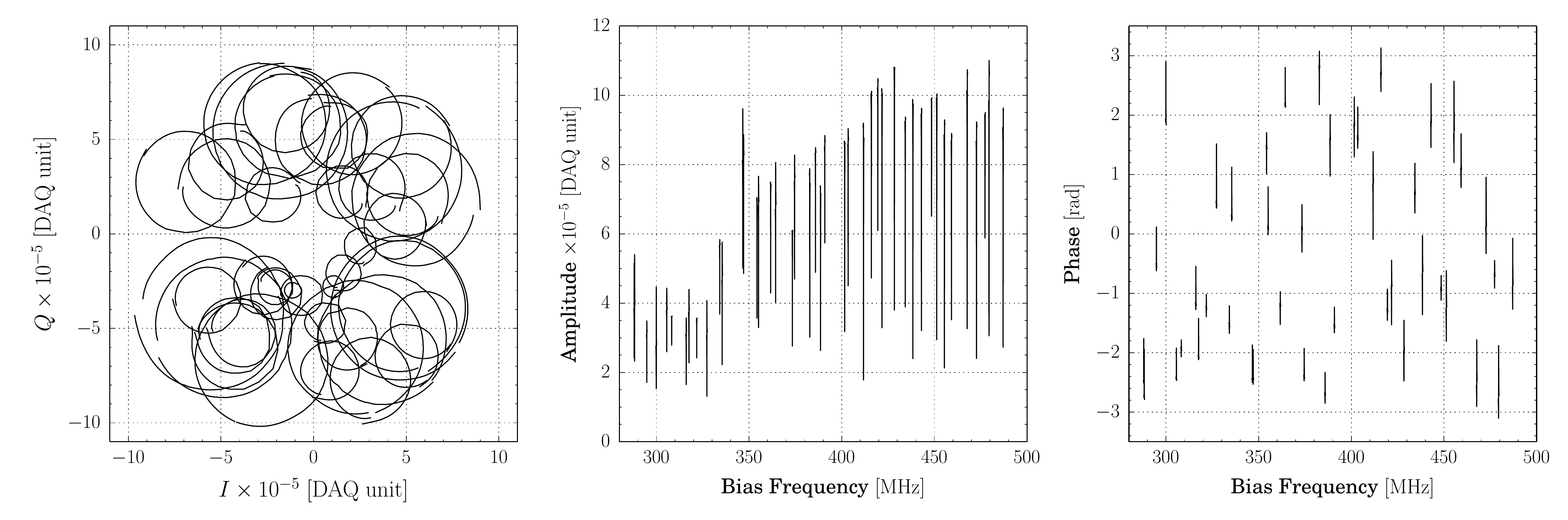}
\caption{\small $S_{21}$ parameter of the resonances of the four OLIMPO detector arrays. \emph{Left panels}: Complex $S_{21}$ parameter in the $IQ$ plane. \emph{Central panels}: Amplitude of the bias signal transmitted across the array. \emph{Right panels}: Phase of the bias signal transmitted across the array. \emph{First row}: \SI{150}{GHz} array at \SI{185}{mK}. The bias signal is composed of 20 tones with power $P_{bias}=-\SI{90}{dBm}$ each, sweeping in a frequency range of \SI{400}{kHz} around the resonant frequencies. \emph{Second row}: \SI{250}{GHz} array at \SI{168}{mK}. The bias signal is composed of 34 tones with power $P_{bias}=-\SI{79}{dBm}$ each, sweeping in a frequency range of \SI{400}{kHz} around the resonant frequencies. \emph{Third row}: \SI{350}{GHz} array at \SI{155}{mK}. The bias signal is composed of 23 tones with power $P_{bias}=-\SI{96}{dBm}$ each, sweeping in a frequency range of \SI{400}{kHz} around the resonant frequencies. \emph{Fourth row}: \SI{460}{GHz} array at \SI{255}{mK}. The bias signal is composed of 43 tones with power $P_{bias}=-\SI{99}{dBm}$ each, sweeping in a frequency range of \SI{400}{kHz} around the resonant frequencies.}
\phantomsection\label{fig:resonance_roach_all}
\end{figure}

For each array we measured the transmission $S_{21}$ scattering parameter, identifying the resonant frequencies and establishing, through a bias power sweep, the optimal bias power at the base temperature. From these measurements, shown in fig.~\ref{fig:resonance_roach_all}, we found also the frequency ranges, bandwidths and average spacing of the resonators of the four arrays. These values are collected in tab.~\ref{tab:ele_char} and they are in good agreement with the results of the simulations described in Subsec.~\ref{subsec:Electrical_Simulations}.

%From these measurements, we found that the \SI{150}{GHz} array has 20 operating pixels over 23 (87\%), with optimal bias power of \SI{-90}{dBm}; the \SI{250}{GHz} array has 34 operating pixels over 39 (87\%), with optimal bias power of \SI{-79}{dBm}; the \SI{350}{GHz} array has 23 operating pixels over 25 (92\%), with optimal bias power of \SI{-96}{dBm}; and the \SI{460}{GHz} array has all the 43 pixels alive (100\%), with optimal bias power of \SI{-99}{dBm}.

%As shown in fig.~\ref{fig:resonance_roach_all}, the resonators of the \SI{150}{GHz} array (\emph{first row}) populate the frequency range $\left[146;267\right]\SI{}{MHz}$, which results in a bandwidth of \SI{121}{MHz} and a mean spacing of about \SI{6}{MHz}; the resonators of the \SI{250}{GHz} array (\emph{second row}) populate the frequency range $\left[150;335\right]\SI{}{MHz}$, which results in a bandwidth of \SI{185}{MHz} and a mean spacing of about \SI{5.5}{MHz}; the resonators of the \SI{350}{GHz} array (\emph{third row}) populate the frequency range $\left[362;478\right]\SI{}{MHz}$, which results in a bandwidth of \SI{116}{MHz} and a mean spacing of about \SI{5}{MHz}; and the resonators of the \SI{460}{GHz} array (\emph{fourth row}) populate the frequency range $\left[288;487\right]\SI{}{MHz}$, which results in a bandwidth of \SI{199}{MHz} and a mean spacing of about \SI{4.5}{MHz}. These results are in good agreement with the results of the simulations described in~\ref{subsec:Electrical_Simulations}.

\begin{table}[htb]
	\centering
		\fontsize{10pt}{15pt}\selectfont{
		\begin{tabular}{c|c|c|c|c|c}
		\hline
		\hline
		\multicolumn{1}{c|}{\multirow{2}{*}{Channel}}&
        \multicolumn{1}{c|}{\multirow{2}{*}{Operating}}&
		\multicolumn{1}{c|}{\multirow{1}{*}{Optimal}}&
		\multicolumn{3}{c}{\multirow{1}{*}{Resonant frequencies}}\\
		%\multicolumn{1}{c|}{\multirow{1}{*}{Optimal}}&
		%\multicolumn{1}{c}{\multirow{1}{*}{$\#$3}}\\
		\cline{4-6}      
		\multicolumn{1}{c|}{\multirow{1}{*}{}}&
		\multicolumn{1}{c|}{\multirow{2}{*}{pixels}}&
		\multicolumn{1}{c|}{\multirow{1}{*}{Bias Power}}&
        \multicolumn{1}{c|}{\multirow{1}{*}{Range}}&
		\multicolumn{1}{c|}{\multirow{1}{*}{Bandwidth}}&
		\multicolumn{1}{c}{\multirow{1}{*}{Spacing}}\\
        \multicolumn{1}{c|}{\multirow{1}{*}{$\left[\SI{}{GHz}\right]$}}&
        \multicolumn{1}{c|}{\multirow{1}{*}{}}&
		\multicolumn{1}{c|}{\multirow{1}{*}{$\left[\SI{}{dBm}\right]$}}&
		\multicolumn{1}{c|}{\multirow{1}{*}{$\left[\SI{}{MHz}\right]$}}&
		\multicolumn{1}{c|}{\multirow{1}{*}{$\left[\SI{}{MHz}\right]$}}&
		\multicolumn{1}{c}{\multirow{1}{*}{$\left[\SI{}{MHz}\right]$}}\\
		\hline
		\hline
		150&20/23 (87\%)&$-$90&$\left[146;267\right]$&121&6\\
		250&34/39 (87\%)&$-$79&$\left[150;335\right]$&185&5.5\\
		350&23/25 (92\%)&$-$96&$\left[362;478\right]$&116&5\\
		460&43/43 (100\%)&$-$99&$\left[288;487\right]$&199&4.5\\
		\hline
		\hline
		\end{tabular}
		}		
		\caption{\small Number of operating pixels, optimal bias powers, and frequency ranges, bandwidths and average spacing of the resonators of the four arrays.}
		\phantomsection\label{tab:ele_char}
\end{table}

\subsubsection{Quality Factors}
\phantomsection\label{subsec:Quality_Factors} 

The quality factors have been estimated through the procedure described in appendix~\ref{sec:Fit}. This analysis was performed both at the base temperature, and at around \SI{300}{mK} (operating temperature of OLIMPO). The values of the quality factors averaged, for each array, over all the detectors are collected in tab.~\ref{tab:Q_factors}.

\begin{table}[htb]
	\centering
		\fontsize{10pt}{15pt}\selectfont{
		\begin{tabular}{c|c|c|c}
		\hline
		\hline
		\multicolumn{1}{c|}{\multirow{1}{*}{Channel}}&
        \multicolumn{1}{c|}{\multirow{1}{*}{Temperature}}&
		\multicolumn{2}{c}{\multirow{1}{*}{Average $Q$s}}\\
		\cline{3-4}
		\multicolumn{1}{c|}{\multirow{1}{*}{$\left[\SI{}{GHz}\right]$}}&
		\multicolumn{1}{c|}{\multirow{1}{*}{$\left[\SI{}{mK}\right]$}}&
		\multicolumn{1}{c|}{\multirow{1}{*}{$Q$}}&
        \multicolumn{1}{c}{\multirow{1}{*}{$Q_{i}$}}\\
		\hline
		\hline
		\multirow{2}{*}{150}&185&$2.8\times 10^{4}$&$5.8\times 10^{5}$\\
        &300&$1.7\times 10^{4}$&$1.3\times 10^{5}$\\
		\multirow{2}{*}{250}&168&$1.3\times 10^{4}$&$2.7\times 10^{4}$\\
		&295&$7.0\times 10^{3}$&$1.1\times 10^{4}$\\
        \multirow{2}{*}{350}&155&$3.2\times 10^{4}$&$5.1\times 10^{5}$\\
		&300&$1.0\times 10^{4}$&$1.5\times 10^{4}$\\
        \multirow{2}{*}{460}&255&$2.5\times 10^{4}$&$6.9\times 10^{4}$\\
		&310&$1.0\times 10^{4}$&$2.5\times 10^{4}$\\
        \hline
		\hline
		\end{tabular}
		}		
		\caption{\small Array--average of the quality factors for the base and the OLIMPO cryostat--like temperatures.}
		\phantomsection\label{tab:Q_factors}
\end{table}

In accordance with the design, the total quality factors are dominated by the coupling quality factors.

\subsubsection{Electrical Responsivity and NEP}
\phantomsection\label{subsec:Electrical_Responsivity} 

The electrical phase responsivity is given by
\begin{equation}
\mathcal{R}_{\vartheta}=-\frac{\eta_{pb}\tau_{qp}}{\Delta_{0}}4Q\frac{\delta x}{\delta N_{qp}}\;,
\phantomsection\label{eq:respon}
\end{equation}
where $\eta_{pb}$ is the pair--breaking efficiency, $\tau_{qp}$ is the quasiparticle lifetime, $\Delta_{0}$ is half Cooper pair binding energy, $x=\left(\nu_{r}-\nu_{r,0}\right)/\nu_{r,0}$ is the dimensionless resonant frequency shift, and $\delta x/\delta N_{qp}$ is the temperature responsivity. 

According to the BCS theory, at temperatures $T\ll T_{c}$ the Cooper pair binding energy is linked to the critical temperature of the superconductor by eq.~\eqref{eq:delta_0}, and the number of quasiparticles, $N_{qp}$, is linked to the superconductor temperature by equation
\begin{equation}
N_{qp}=2V\mathcal{N}_{0}\sqrt{2\pi k_{B}T\Delta_{0}}{\rm e}^{-\Delta_{0}/\left(k_{B}T\right)}\;,
\end{equation}
where $V$ is the volume of the absorber ($V_{\SI{150}{GHz}}=\SI{2478.6}{\mu m^{-3}}$, $V_{\SI{250}{GHz}}=\SI{2019.6}{\mu m^{-3}}$, $V_{\SI{350}{GHz}}=\SI{1101.6}{\mu m^{-3}}$, and $V_{\SI{460}{GHz}}=\SI{795.6}{\mu m^{-3}}$), and $\mathcal{N}_{0}$ is the density of states at the Fermi surface, which for aluminum is $\mathcal{N}_{0}=\SI{1.098e29}{J^{-1}.\mathsf{\mu}m^{-3}}$.

For the estimation of the electrical phase responsivity, we assume $\eta_{pb}=0.57$ \citep{PhysRevB.61.11807, 0953-2048-27-5-055012, doi:10.1063/1.4923097}. From the signal spikes due to the interaction between the detectors and cosmic rays we measured the quasiparticle lifetime $\tau_{qp}=\left(30.0\pm 0.89\right)\SI{}{\mu s}$ at \SI{300}{mK}. We measured the critical temperature $T_{c}=\left(1.310\pm 0.039\right)\SI{}{K}$ of the aluminum film \SI{30}{nm} thick in the same way described in \citep{Paiella2016}, and we measured the temperature responsivity by performing the detector temperature sweep. From this measurement, the temperature responsivity is determined using a linear fit.

\begin{figure}[!h]
\centering
\includegraphics[scale=0.395]{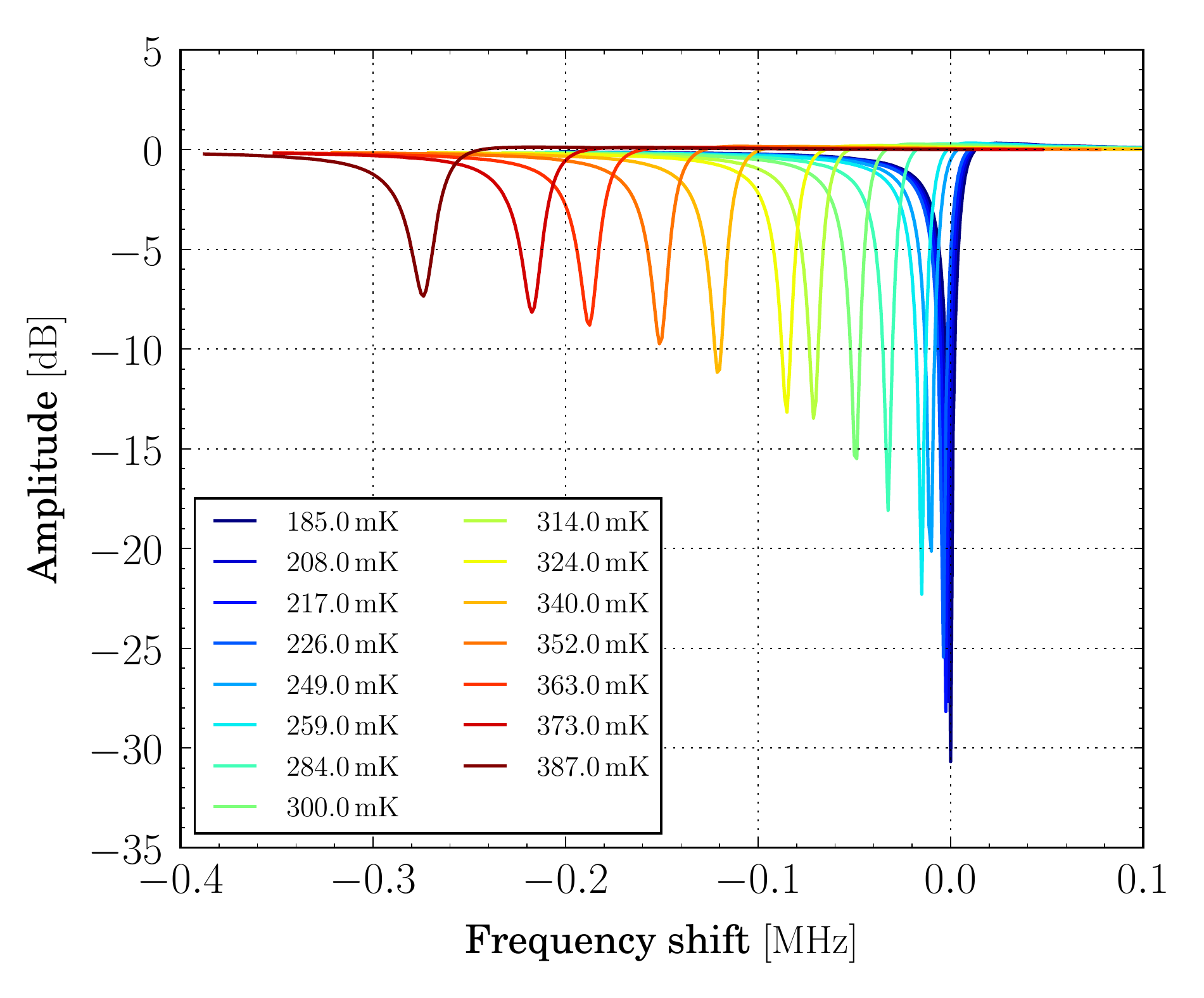}
\includegraphics[scale=0.395]{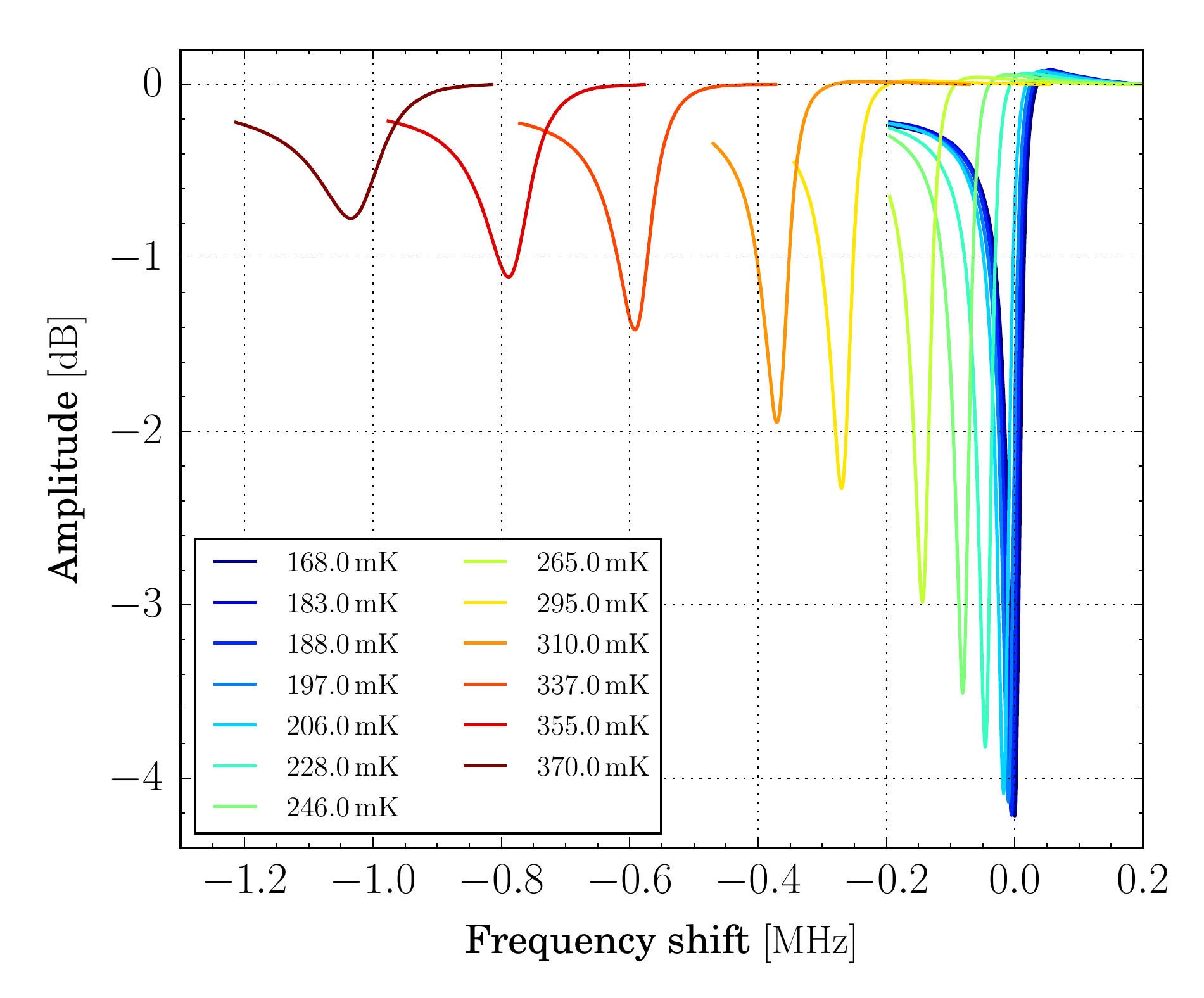}\\
\includegraphics[scale=0.395]{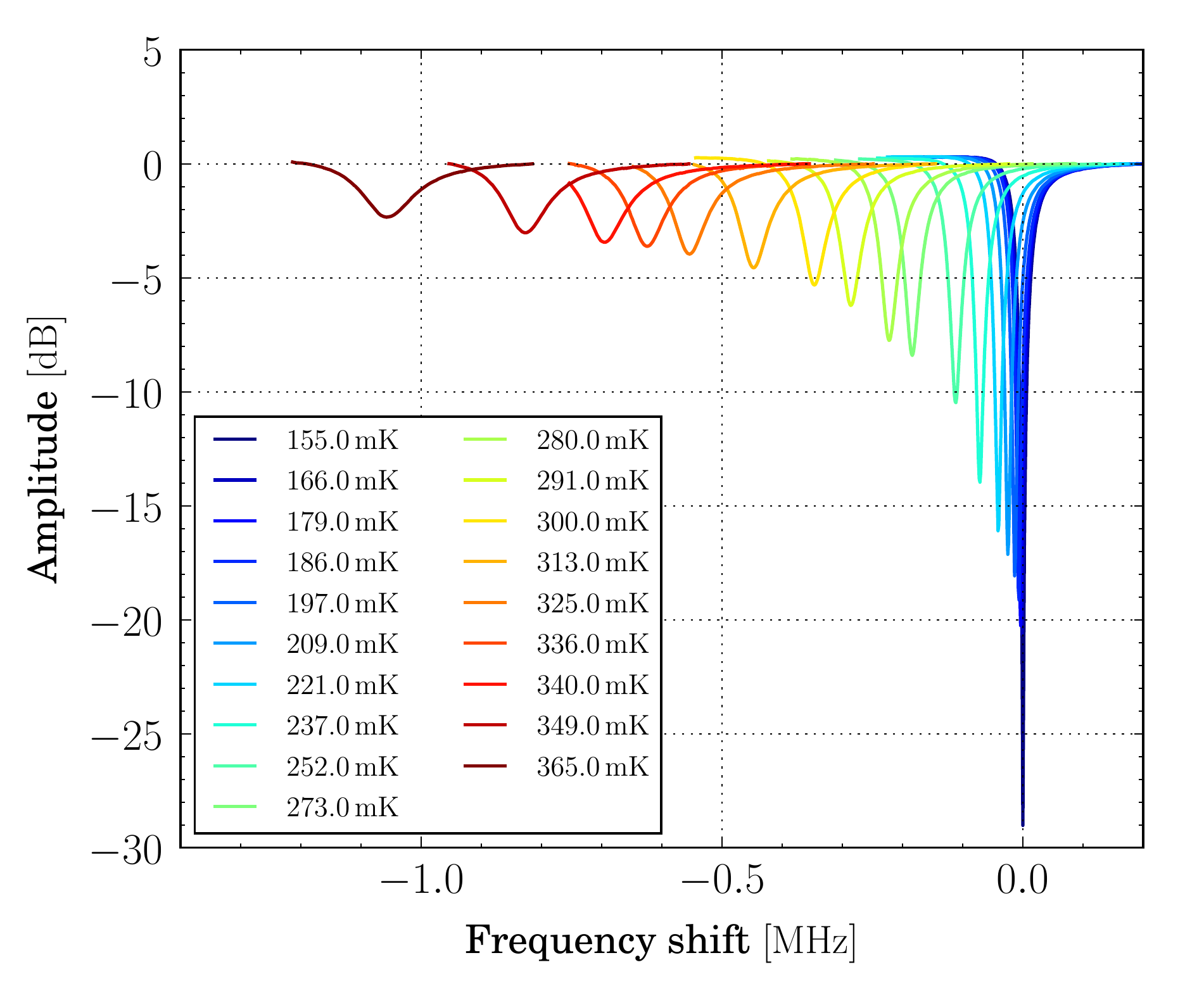}
\includegraphics[scale=0.395]{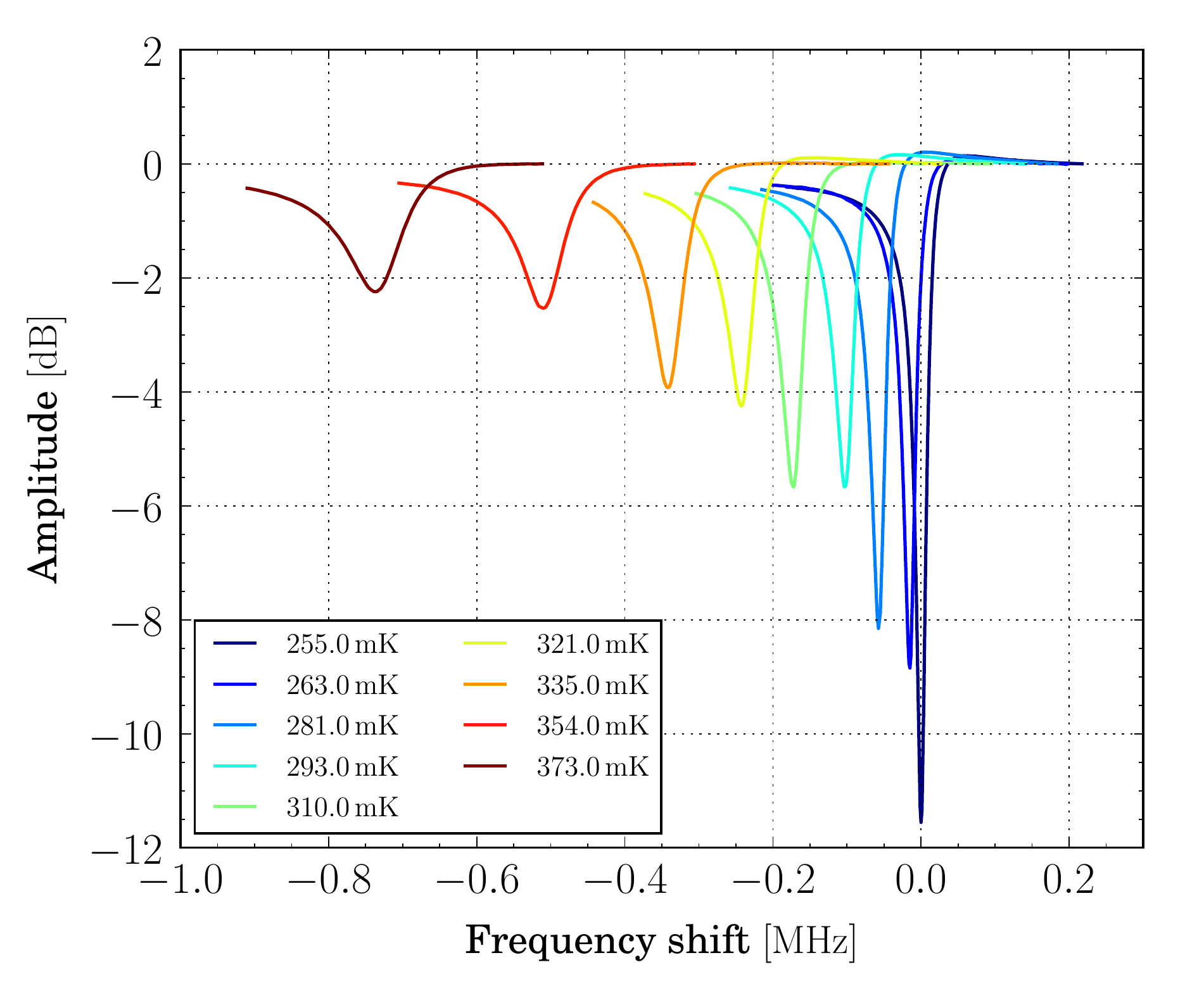}
\caption{\small \small Amplitude of the bias signal transmitted across the resonators for different temperatures of the arrays. \emph{Top--left panel}: $2^{\rm nd}$ pixel of the \SI{150}{GHz} array. \emph{Top--right panel}: $31^{\rm st}$ pixel of the \SI{250}{GHz} array. \emph{Bottom--left panel}: $19^{\rm th}$ pixel of the \SI{350}{GHz} array. \emph{Bottom--right panel}: $40^{\rm th}$ pixel of the \SI{460}{GHz} array.}
\phantomsection\label{fig:temp_sweep}
\end{figure}

Since we have cooled the arrays individually and because of the different thermal loads of their holders, the points of the temperature sweeps are different array by array. Fig.~\ref{fig:temp_sweep} collects the plots of the temperature sweeps in amplitude of one detector per array. In these plots, the temperatures of the sweeps for each array are indicated.

\begin{figure}[!h]
\centering
\includegraphics[scale=0.395]{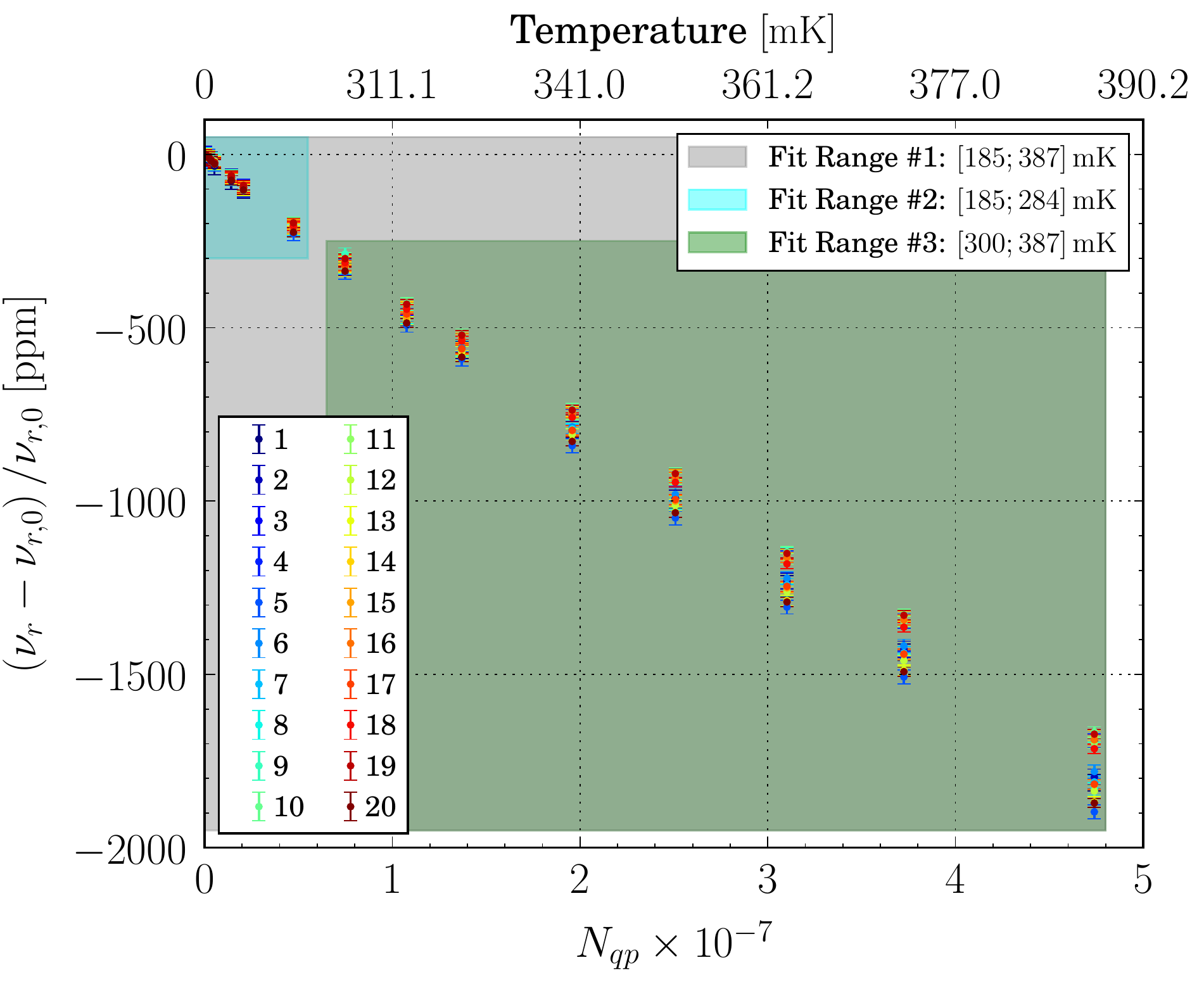}
\includegraphics[scale=0.395]{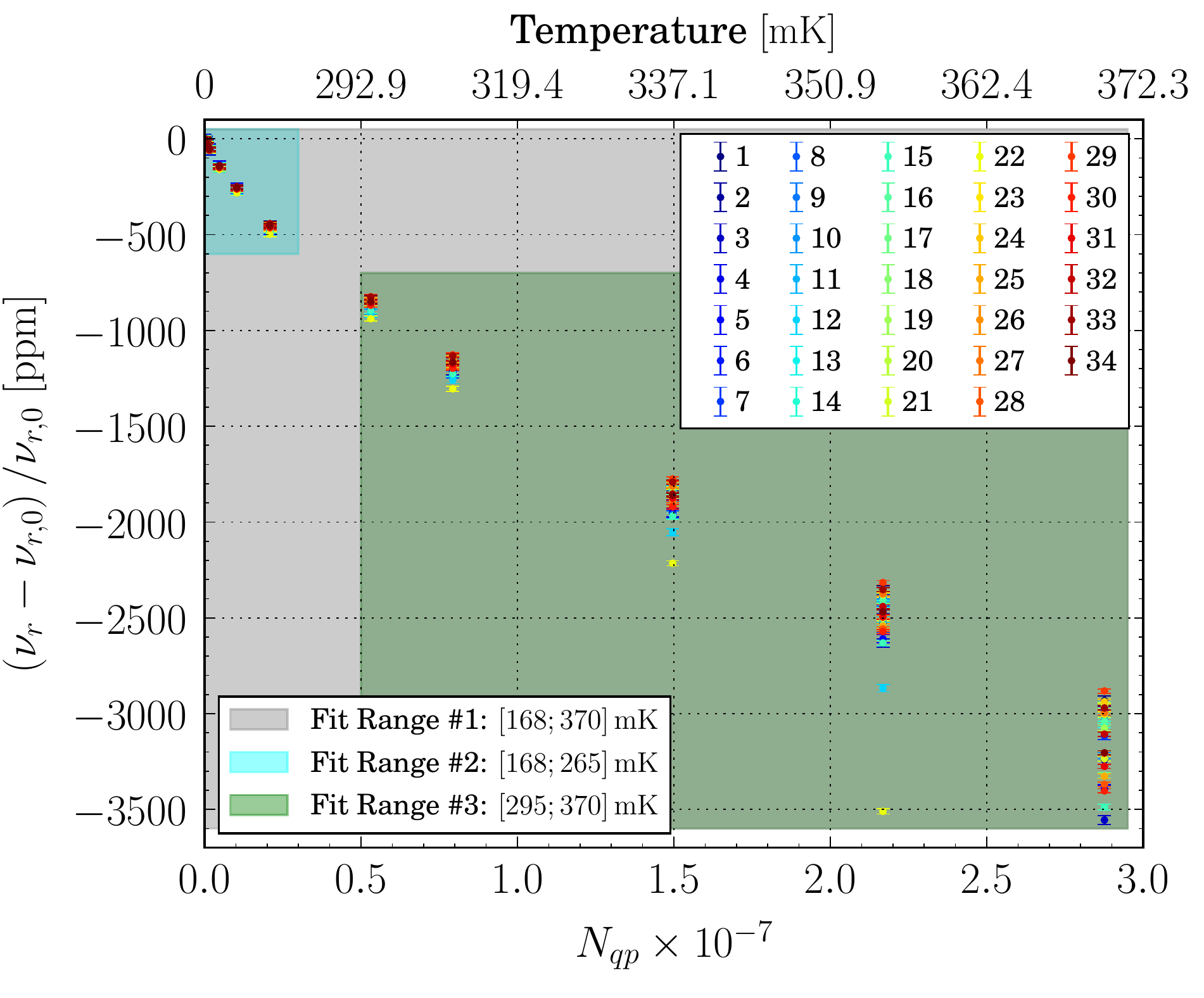}\\
\includegraphics[scale=0.395]{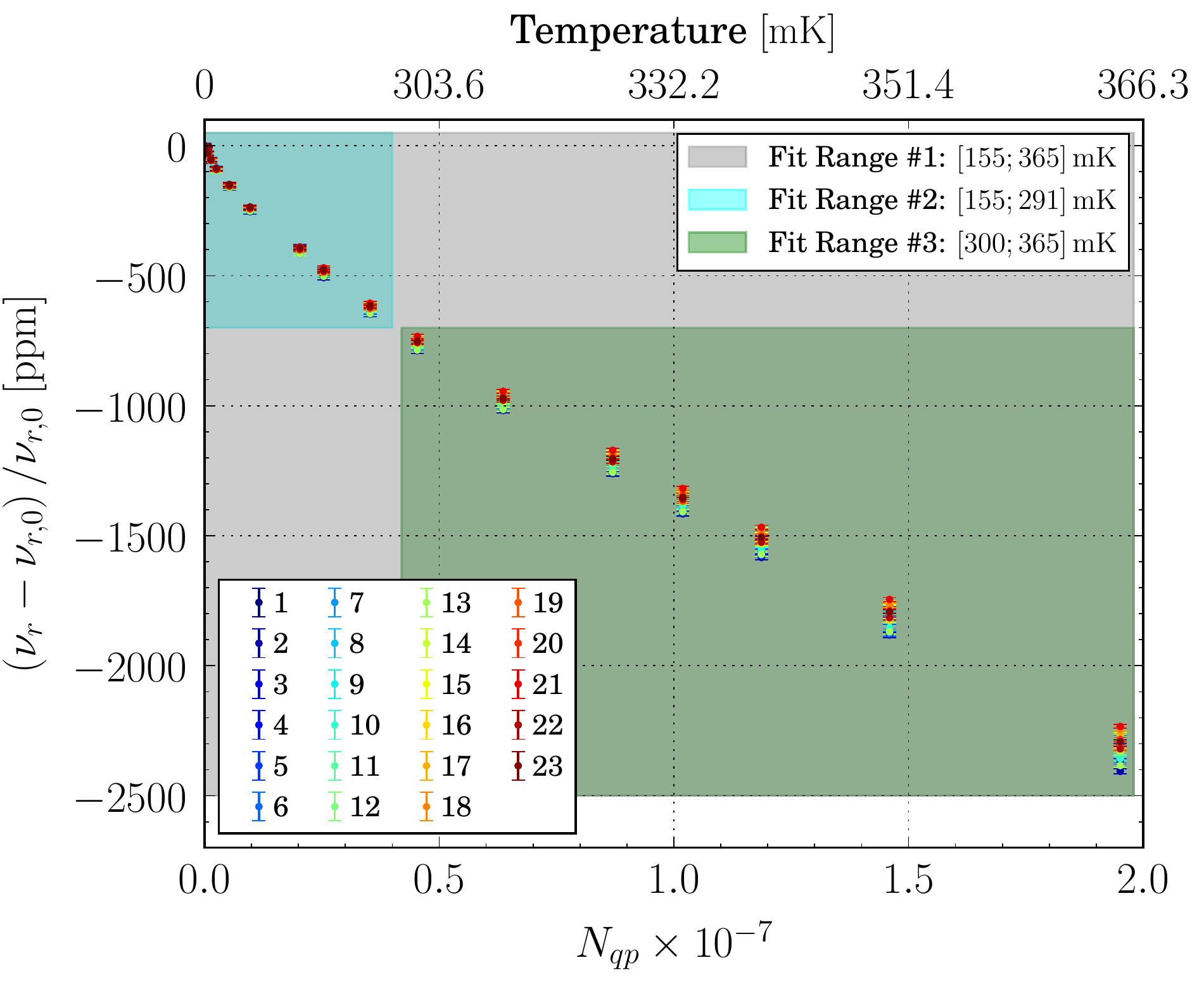}
\includegraphics[scale=0.395]{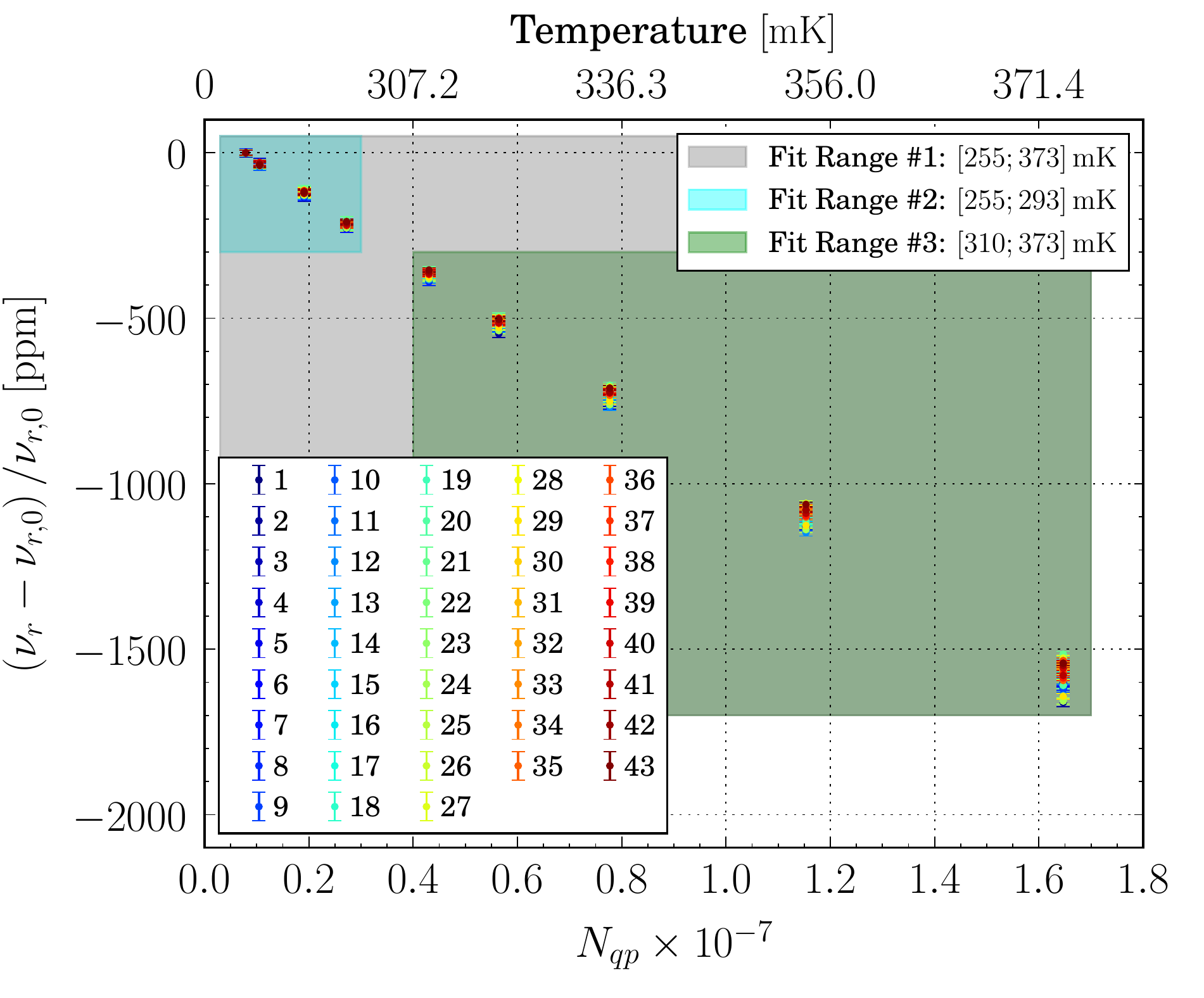}
\caption{\small Trends of $\left(\nu^{}_{r}-\nu^{}_{r,0}\right)/\nu^{}_{r,0}$ with the number of quasiparticles, $N_{qp}$. $\nu_{r,0}$ is the resonant frequency at the base temperature. The \emph{color dots} with the \emph{error bars} are the measured data, for which different colors indicate different resonators. The \emph{color areas} define the different ranges where the linear fits are performed in order to estimate the temperature responsivity $\delta x/\delta N_{qp}$. \emph{Top--left panel}: \SI{150}{GHz} array. \emph{Top--right panel}: \SI{250}{GHz} array. \emph{Bottom--left panel}: \SI{350}{GHz} array. \emph{Bottom--right panel}: \SI{460}{GHz} array.}
\phantomsection\label{fig:fit_range}
\end{figure}

Fig.~\ref{fig:fit_range} shows the measured trends for the dimensionless resonant frequency shift versus the number of quasiparticles (or equivalently versus temperature) for all the resonators of the four arrays. Due to low--level nonlinearity in the response over the whole temperature range we explored, a single linear fit is not adequate to reproduce the entire curve. Therefore, for each detector array, we considered three fit ranges. Specifically, \emph{Fit Range $\#$1} includes all the temperatures explored with the sweep; \emph{Fit Range $\#$2} includes the temperatures where the responsivity is maximum; and \emph{Fit Range $\#$3} starts from the OLIMPO base temperature. Tab.~\ref{tab:fit_range} specifies, in detail, the bounds in temperature of the fit ranges for each detector array.       

\begin{table}[htb]
	\centering
		\fontsize{10pt}{15pt}\selectfont{
		\begin{tabular}{c|c|c|c}
		\hline
		\hline
		\multicolumn{1}{c|}{\multirow{1}{*}{Channel}}&
		\multicolumn{3}{c}{\multirow{1}{*}{Fit range}}\\
		\cline{2-4}
		\multicolumn{1}{c|}{\multirow{1}{*}{$\left[\SI{}{GHz}\right]$}}&
		\multicolumn{1}{c|}{\multirow{1}{*}{$\#$1}}&
		\multicolumn{1}{c|}{\multirow{1}{*}{$\#$2}}&
		\multicolumn{1}{c}{\multirow{1}{*}{$\#$3}}\\
		\hline
		\hline
		150&$\left[185;387\right]\SI{}{mK}$&$\left[185;284\right]\SI{}{mK}$&$\left[300;387\right]\SI{}{mK}$\\
		250&$\left[168;370\right]\SI{}{mK}$&$\left[168;265\right]\SI{}{mK}$&$\left[295;370\right]\SI{}{mK}$\\
		350&$\left[155;365\right]\SI{}{mK}$&$\left[155;291\right]\SI{}{mK}$&$\left[300;365\right]\SI{}{mK}$\\
		460&$\left[255;373\right]\SI{}{mK}$&$\left[255;293\right]\SI{}{mK}$&$\left[310;373\right]\SI{}{mK}$\\
		\hline
		\hline
		\end{tabular}
		}		
		\caption{\small Definition of the fit ranges. \emph{Fit Range $\#$1}: all the temperatures. \emph{Fit Range $\#$2}: temperature where the responsivity is maximum. \emph{Fit Range $\#$3}: OLIMPO cryostat-like temperatures.}
		\phantomsection\label{tab:fit_range}
\end{table}

The results for the \emph{Fit Range $\#$1} and \emph{$\#$2} are described in appendix~\ref{sec:Electrical_responsivity_A}. Fig.~\ref{fig:fit_range3} shows the fit results in the \emph{Fit Range $\#$3}, for all the detectors of the OLIMPO detector arrays. For the \SI{250}{GHz} array, as shown in the \emph{top--right panel}, the resonances corresponding to the pixels 9 and 22 disappear at \SI{370}{mK}.

\begin{figure}[!h]
\centering
\includegraphics[scale=0.395]{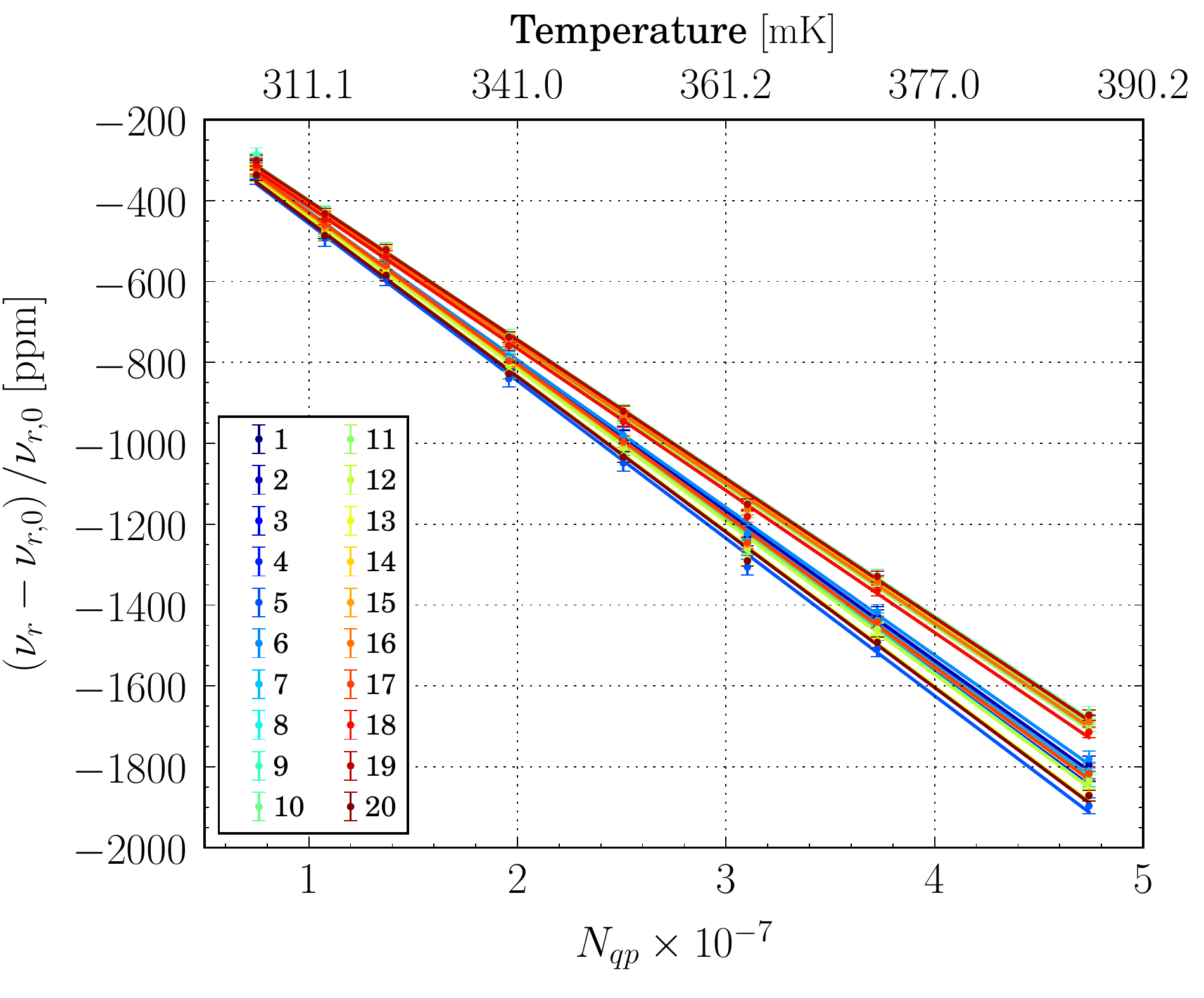}
\includegraphics[scale=0.395]{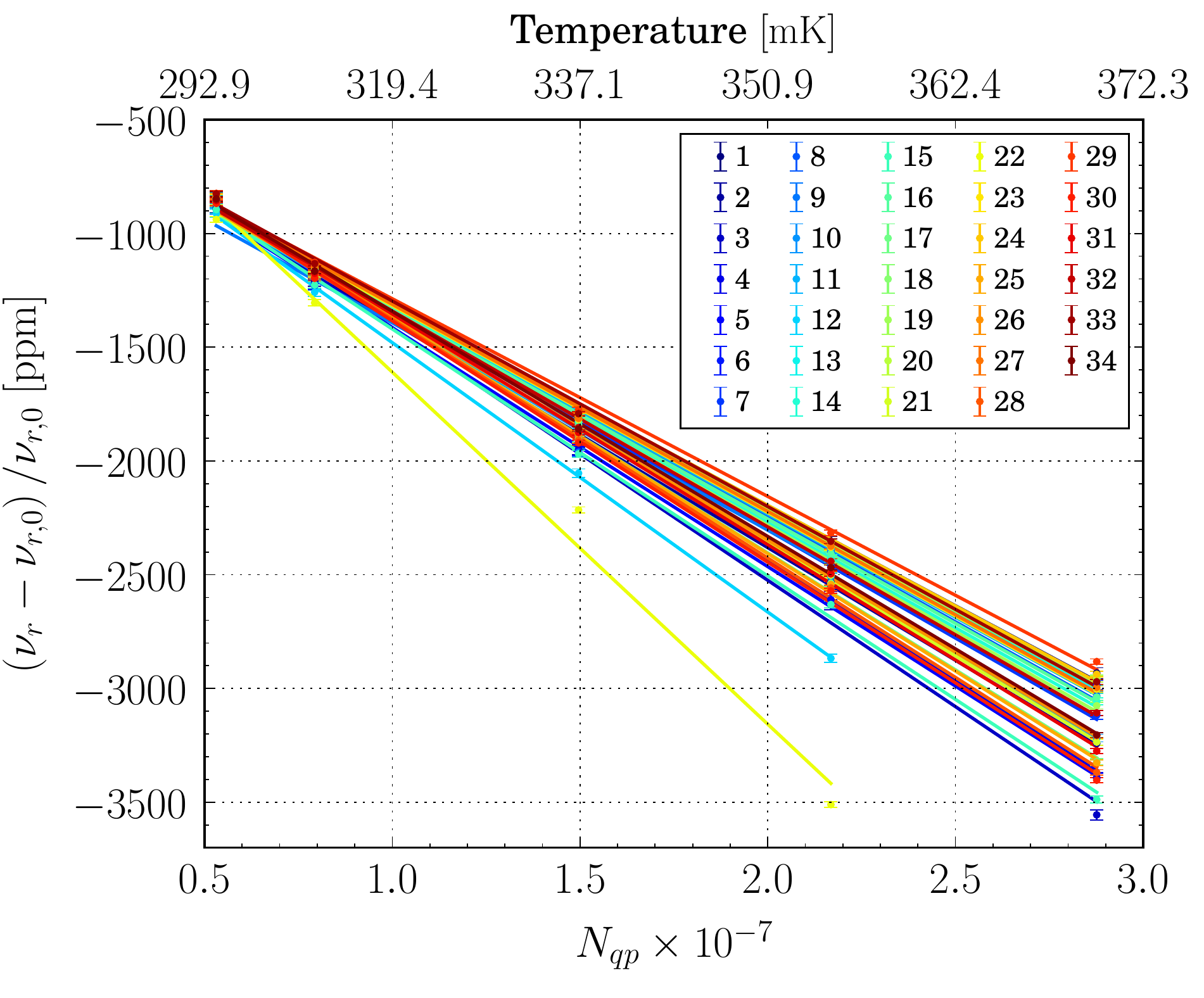}\\
\includegraphics[scale=0.395]{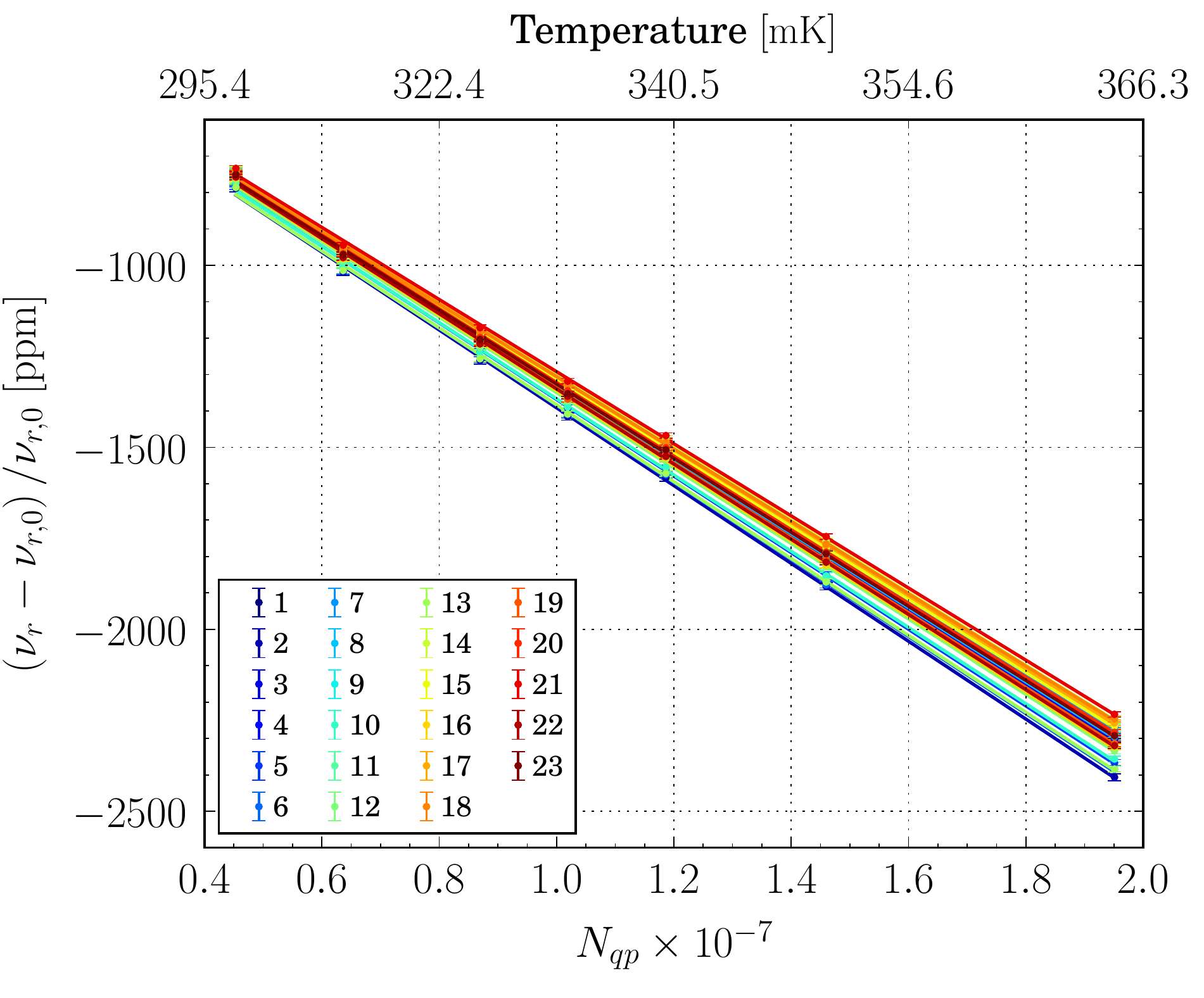}
\includegraphics[scale=0.395]{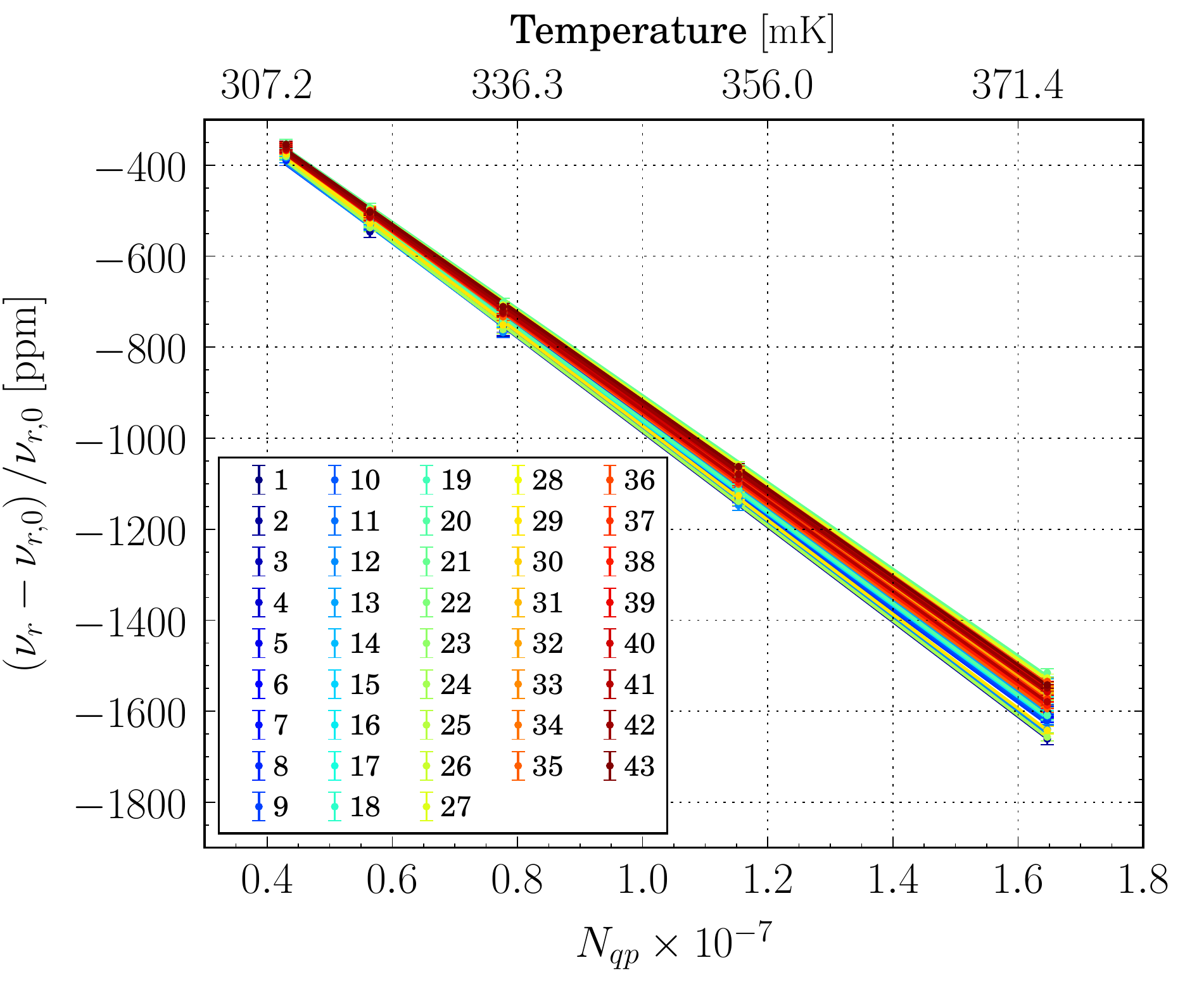}
\caption{\small Fractional frequency shift as a function of bath temperature for all the four arrays. Here, the temperatures varies from approximately 300 to \SI{400}{mK} (\emph{Fit Range $\#$3} in tab.~\ref{tab:fit_range}). The \emph{dots} with the \emph{error bars} are the measured data, and the \emph{solid lines} are the linear fit results. Different colors indicate different resonators. \emph{Top--left panel}: \SI{150}{GHz} array, fit range $\left[300;387\right]\SI{}{mK}$. \emph{Top--right panel}: \SI{250}{GHz} array, fit range $\left[295;370\right]\SI{}{mK}$. \emph{Bottom--left panel}: \SI{350}{GHz} array, fit range $\left[300;365\right]\SI{}{mK}$. \emph{Bottom--right panel}: \SI{460}{GHz} array, fit range $\left[310;373\right]\SI{}{mK}$.}
\phantomsection\label{fig:fit_range3}
\end{figure}

The electrical phase responsivity has been obtained through eq.~\eqref{eq:respon}, where we used the $Q$ values measured at around \SI{300}{mK}. The array--average electrical phase responsivity result to be \SI{1.35}{rad/pW}, \SI{1.49}{rad/pW}, \SI{2.09}{rad/pW}, and \SI{2.14}{rad/pW} for the 150, 250, 350, and \SI{460}{GHz} arrays respectively.

In order to estimate the electrical NEP, we measured the phase noise spectrum for the four arrays. We measured the noises at a modulation frequency of \SI{12}{Hz} and at a temperature of \SI{300}{mK}, which are the typical operating conditions in the OLIMPO receiver. The noises and the corresponding NEPs are collected in tab.~\ref{Tab:elec_NEP}, where the quoted uncertainties reflect the pixel variability over the arrays. These NEPs result lower than the BLIP NEPs of the OLIMPO experiment for both photometric and spectrometric configuration (see tab.~\ref{tab:OLIMPO_perform}), and are encouraging for the next step: the optical characterization in the OLIMPO cryostat.

\begin{table}[htb]
	\centering
		\fontsize{10pt}{18pt}\selectfont{
		\begin{tabular}{c|c|c}
		\hline
		\hline
		\multicolumn{1}{c|}{\multirow{1}{*}{Channel}}&
		\multicolumn{1}{c|}{\multirow{1}{*}{Average Noise}}&
		\multicolumn{1}{c}{\multirow{1}{*}{Average NEP}}\\
        \multicolumn{1}{c|}{\multirow{1}{*}{$\left[\SI{}{GHz}\right]$}}&
		\multicolumn{1}{c|}{\multirow{1}{*}{$\left[{\rm rad}/\sqrt{\rm Hz}\right]$}}&
		\multicolumn{1}{c}{\multirow{1}{*}{$\left[{\rm aW}/\sqrt{\rm Hz}\right]$}}\\
		\hline
		\hline
150& $6.07\times 10^{-5}$& $45.0\pm 3.3$\\
250& $2.35\times 10^{-4}$& $158\pm 23$\\
350& $1.71\times 10^{-4}$& $81.8\pm 3.6$\\
460& $3.08\times 10^{-4}$& $144\pm 10$\\
\hline
		\hline
		\end{tabular}		
		}
		\caption{\small Array--average of the phase noise and electrical noise equivalent power measure at \SI{12}{Hz}.}
	\phantomsection\label{Tab:elec_NEP}
\end{table}

\subsection{Optical Characterization}
\phantomsection\label{subsec:Optical_characterization}

Here we report the results of the optical measurements performed on the detector arrays in the OLIMPO cryostat. OLIMPO is equipped with two independent readout chains: the first one serves the \SI{150}{GHz} array and the \SI{460}{GHz} array, and the second serves the \SI{250}{GHz} array and the \SI{350}{GHz} array. 

In all the optical measurements, as well as in the flight configuration of the experiment, we choose to adopt the  convention to normalize the readout signals to unity after offsetting and rotating them in the $I$--$Q$ plane in order to maximize the relative contribution from $Q$ to the signal. This operation is performed at the client--level stage of the readout chain. It has the advantage of maximizing the sensitivity to smallest variations of the resonance circle, and to maximize the dynamics of the phase signal. For small variations of the resonance circle, the $Q$ fluctuations coincide with the phase fluctuations.  

\subsubsection{Optical Responsivity and NET}
\phantomsection\label{subsec:optical_NET}

In order to reproduce a radiative background comparable to the one expected on the detectors at about \SI{40}{km} of altitude, we inserted, in front of the cryostat window, a neutral density filter (NDF) with about 0.01 transmittance ($t_{\rm NDF}$), and a \emph{drilled plate}, designed in such a way to have a transmittance ranging from 0.04 to 0.05 for incident radiation from 150 to \SI{460}{GHz}. 

The NDF is a thin layer of polypropylene \SI{4}{\mu m} thick with a \SI{14}{nm} gold deposition. The \emph{drilled plate} consists of an aluminum disk \SI{5}{mm} thick, \SI{100}{mm} in diameter, populated with \SI{2}{mm} diameter holes, placed on the vertices of equilateral triangles (\SI{8}{mm} side). The transmittance of the \emph{drilled plate}, $t_{dp}$, has been measured for the four channels and the results are reported in tab.~\ref{Tab:drilled_plate_optical_sign}.

Under this background, the \SI{150}{GHz} and the \SI{250}{GHz} arrays reached a temperature of about \SI{290}{mK}, while the \SI{350}{GHz} and the \SI{460}{GHz} arrays reached a temperature of about \SI{320}{mK}.

The optical signal has been obtained by alternating at the cryostat window two blackbodies at different temperatures (room--temperature and \SI{77}{K}). This produces Rayleigh--Jeans signals ranging from 78 to \SI{92}{mK} (including the effect of the NDF and the \emph{drilled plate}). The optical signals, $\Delta T_{\rm RJ}$, for each channel are reported in tab.~\ref{Tab:drilled_plate_optical_sign}.

\begin{table}[htb]
	\centering
		\fontsize{10pt}{15pt}\selectfont{
		\begin{tabular}{c|c|c}
		\hline
		\hline
		\multicolumn{1}{c|}{\multirow{1}{*}{Channel $\left[\SI{}{GHz}\right]$}}&
		\multicolumn{1}{c|}{\multirow{1}{*}{$t_{dp}$ $\left[\%\right]$}}&
		\multicolumn{1}{c}{\multirow{1}{*}{$\Delta T_{\rm RJ}$ $\left[\SI{}{mK}\right]$}}\\
		\hline
		\hline
150& $3.74\pm 0.03$& $77.7\pm 0.6$\\
250& $4.24\pm 0.05$& $88.2\pm 1.1$\\
350& $4.44\pm 0.03$& $92.4\pm 0.6$\\
460& $4.27\pm 0.03$& $90.3\pm 0.6$\\
\hline
		\hline
		\end{tabular}		
		}
		\caption{\small Measured transmittance of the \emph{drilled plate} and total optical signals at the cryostat window for the four detector arrays, produced alternating two blackbodies (room-temperature and \SI{77}{K}), attenuated by a NDF and a \emph{drilled plate}. The optical signals are evaluated as $\Delta T_{\rm RJ}=\left(\SI{285}{K}-\SI{77}{K}\right)\:t_{\rm NDF}\:t_{dp}$.}
	\phantomsection\label{Tab:drilled_plate_optical_sign}
\end{table}

\begin{figure}[!h]
\centering
\includegraphics[scale=0.30]{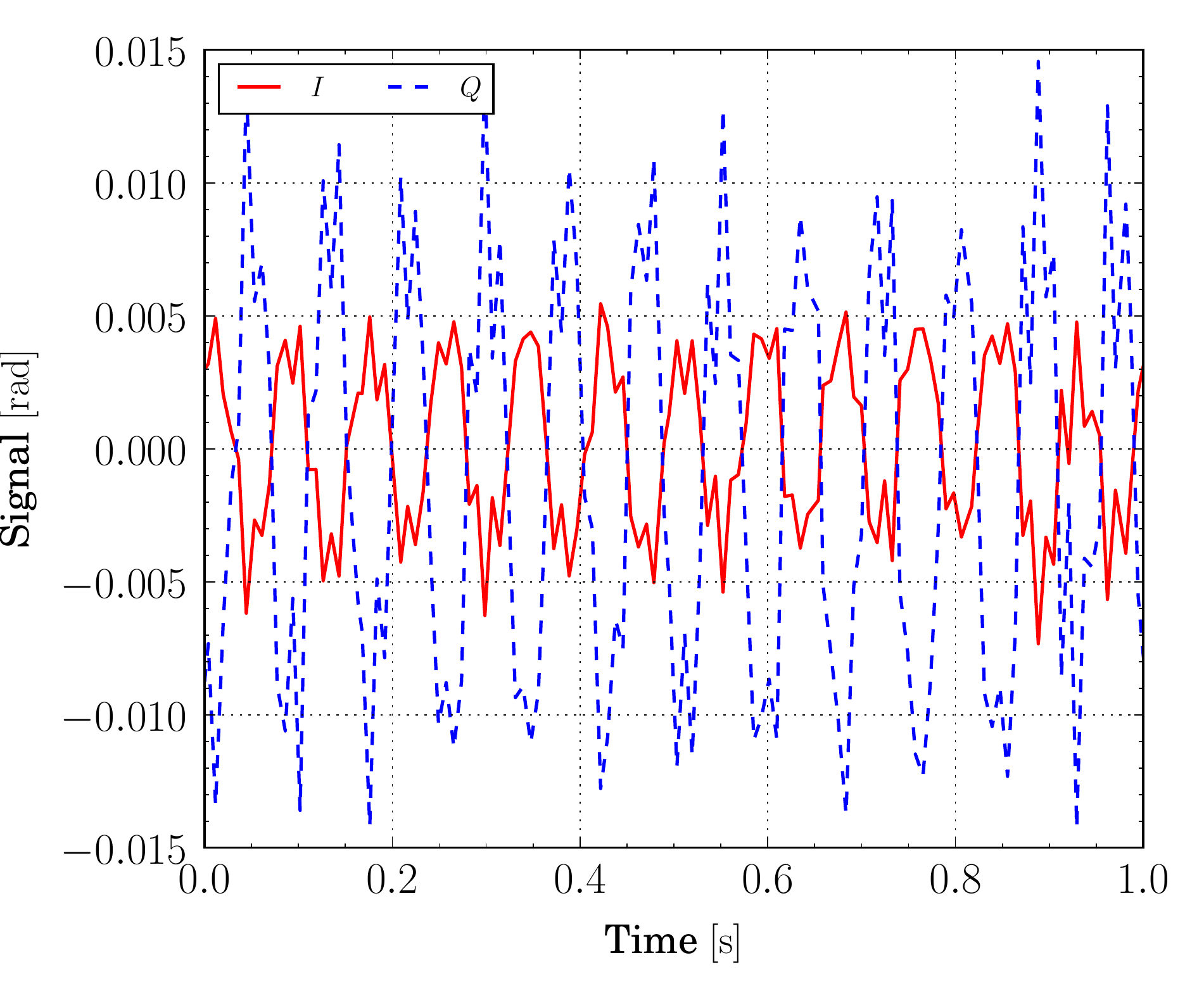}     
\includegraphics[scale=0.30]{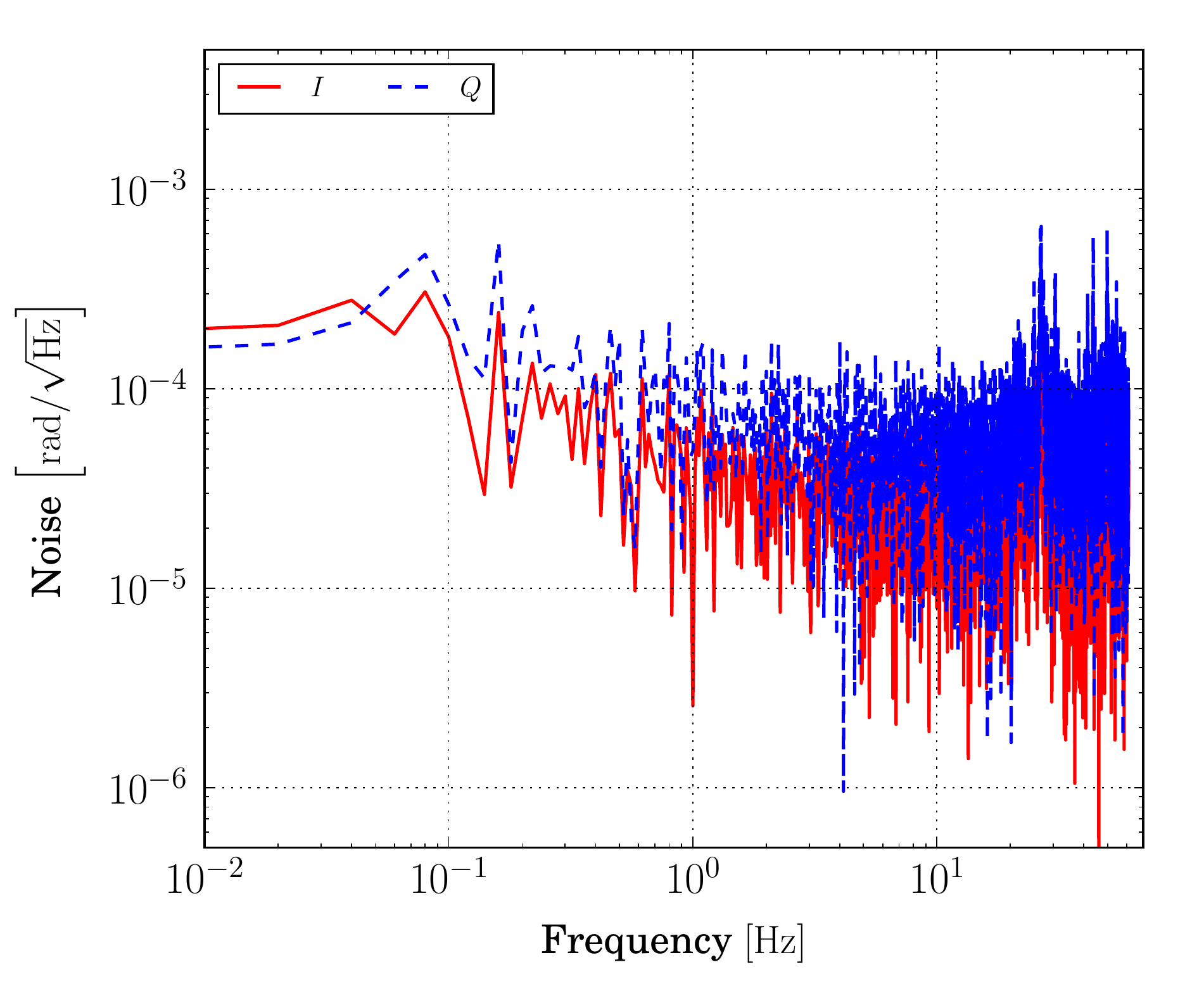}\\   
\includegraphics[scale=0.30]{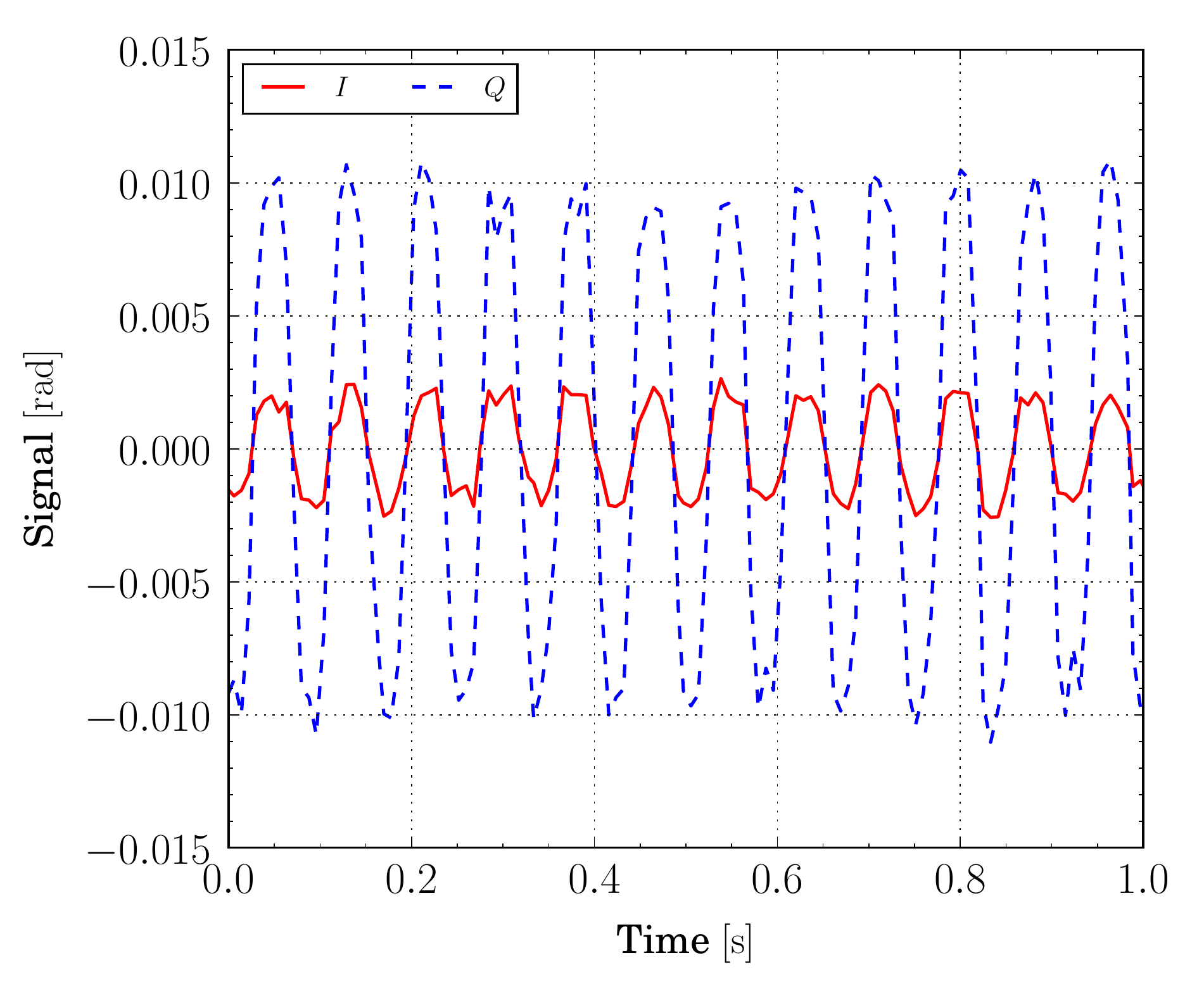}
\includegraphics[scale=0.30]{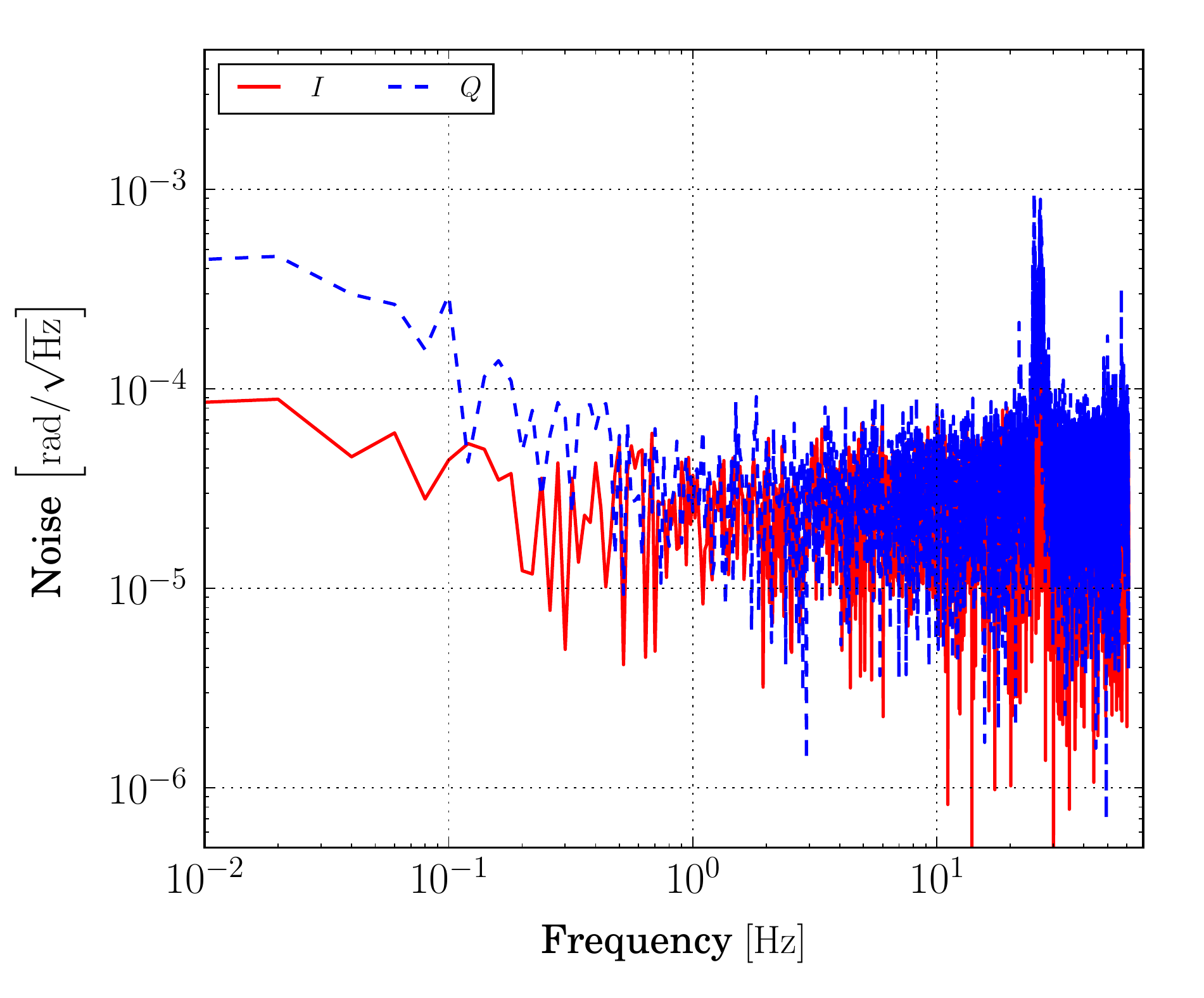}\\ 
\includegraphics[scale=0.30]{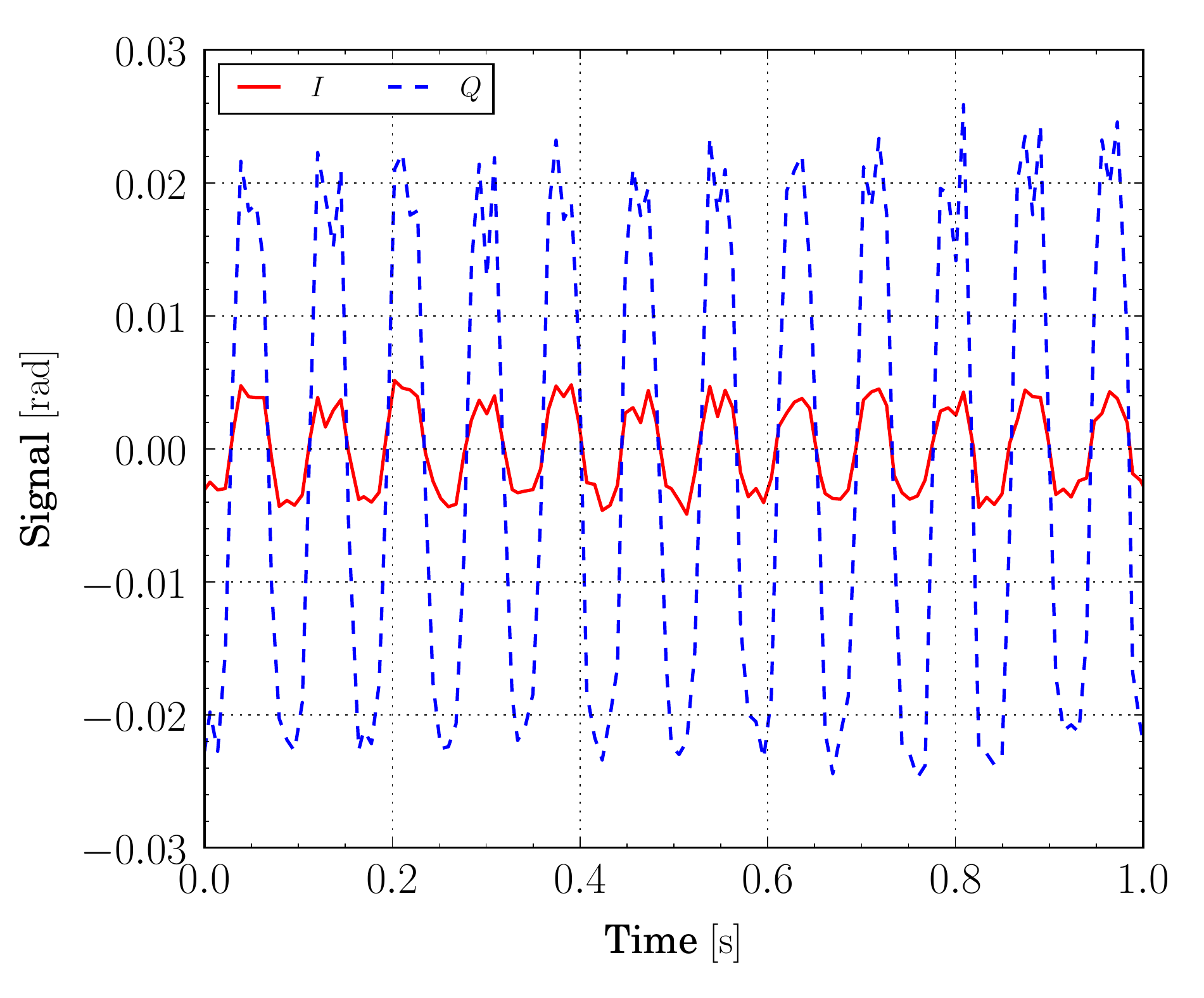}
\includegraphics[scale=0.30]{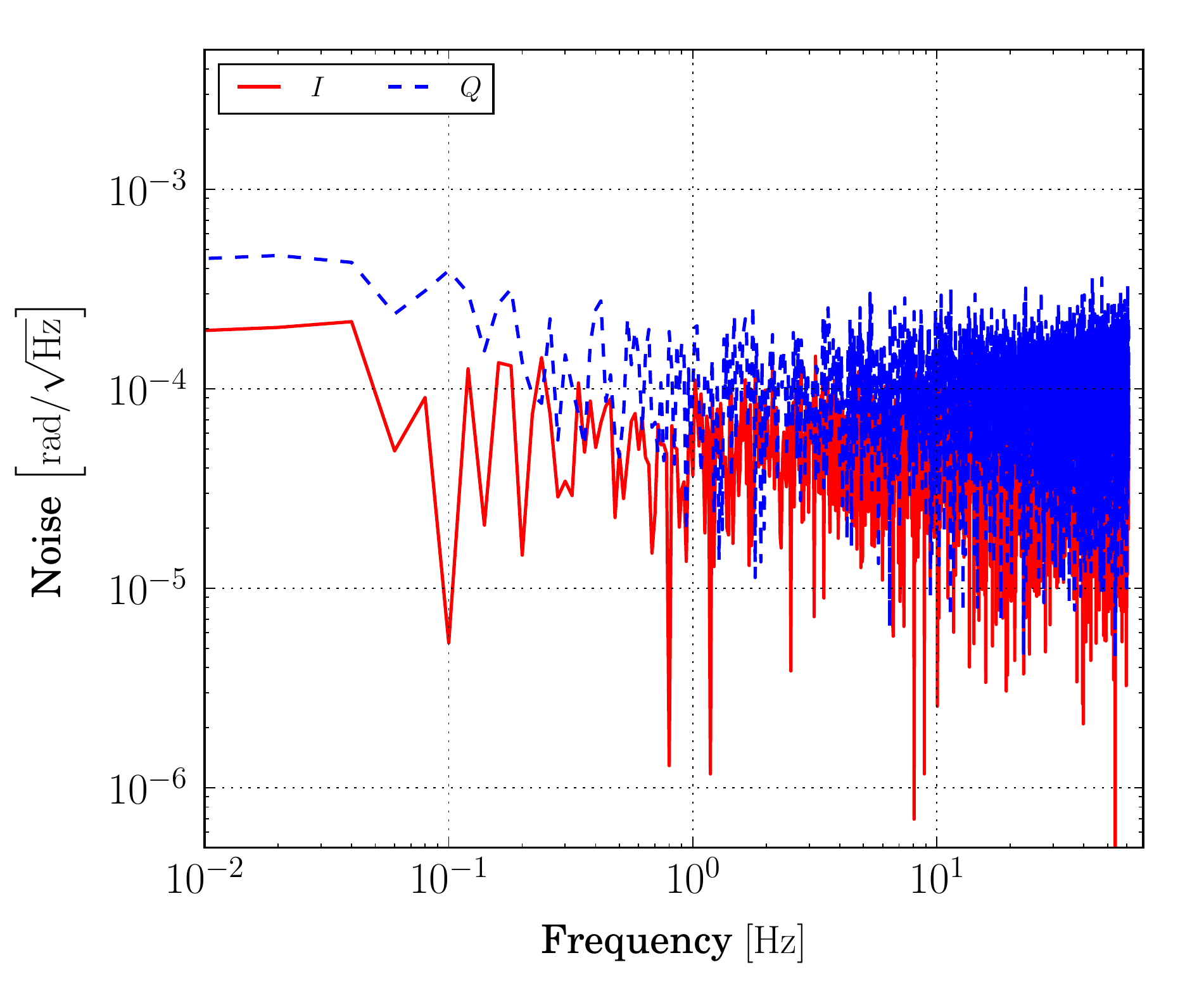}\\
\includegraphics[scale=0.30]{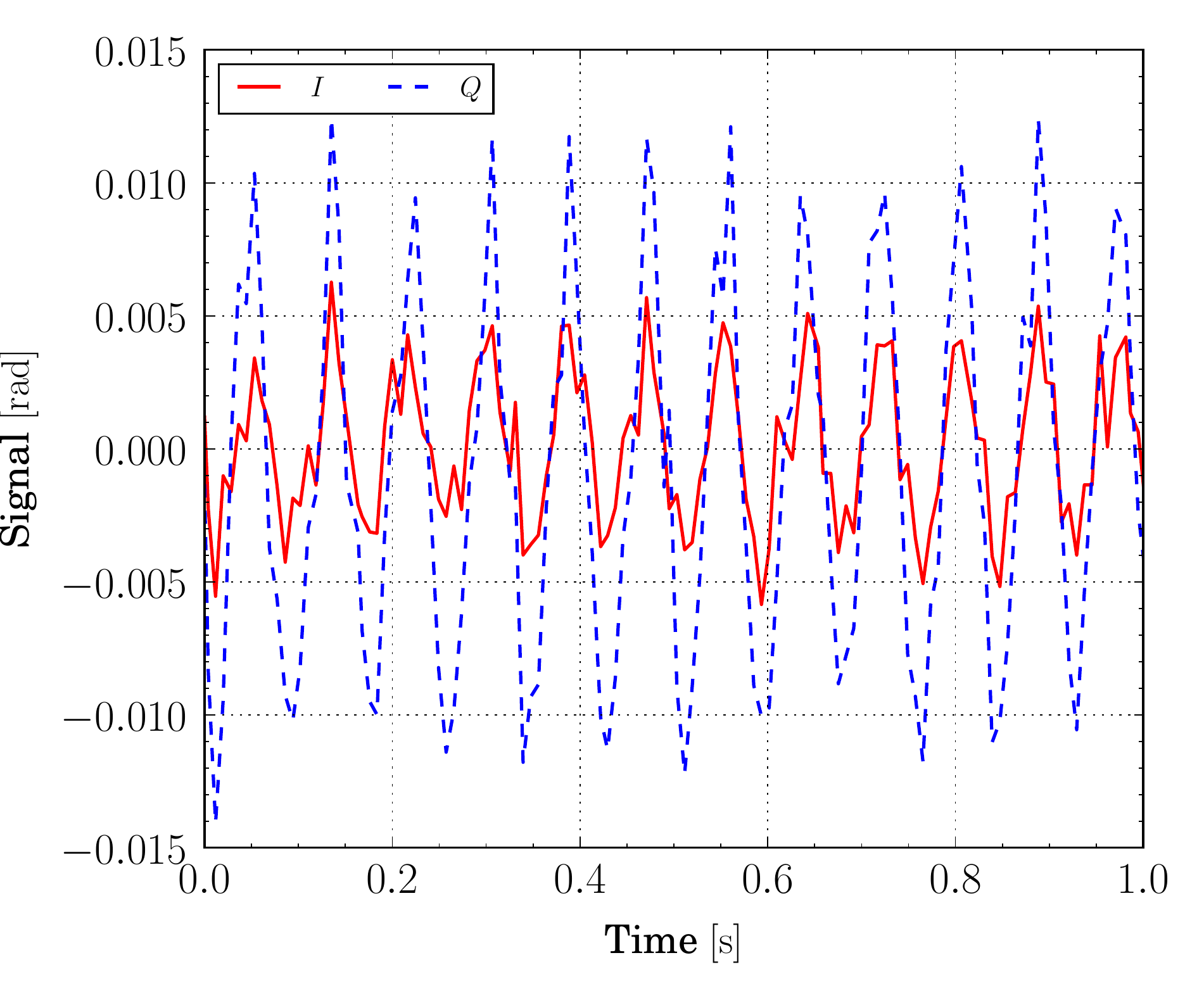}    
\includegraphics[scale=0.30]{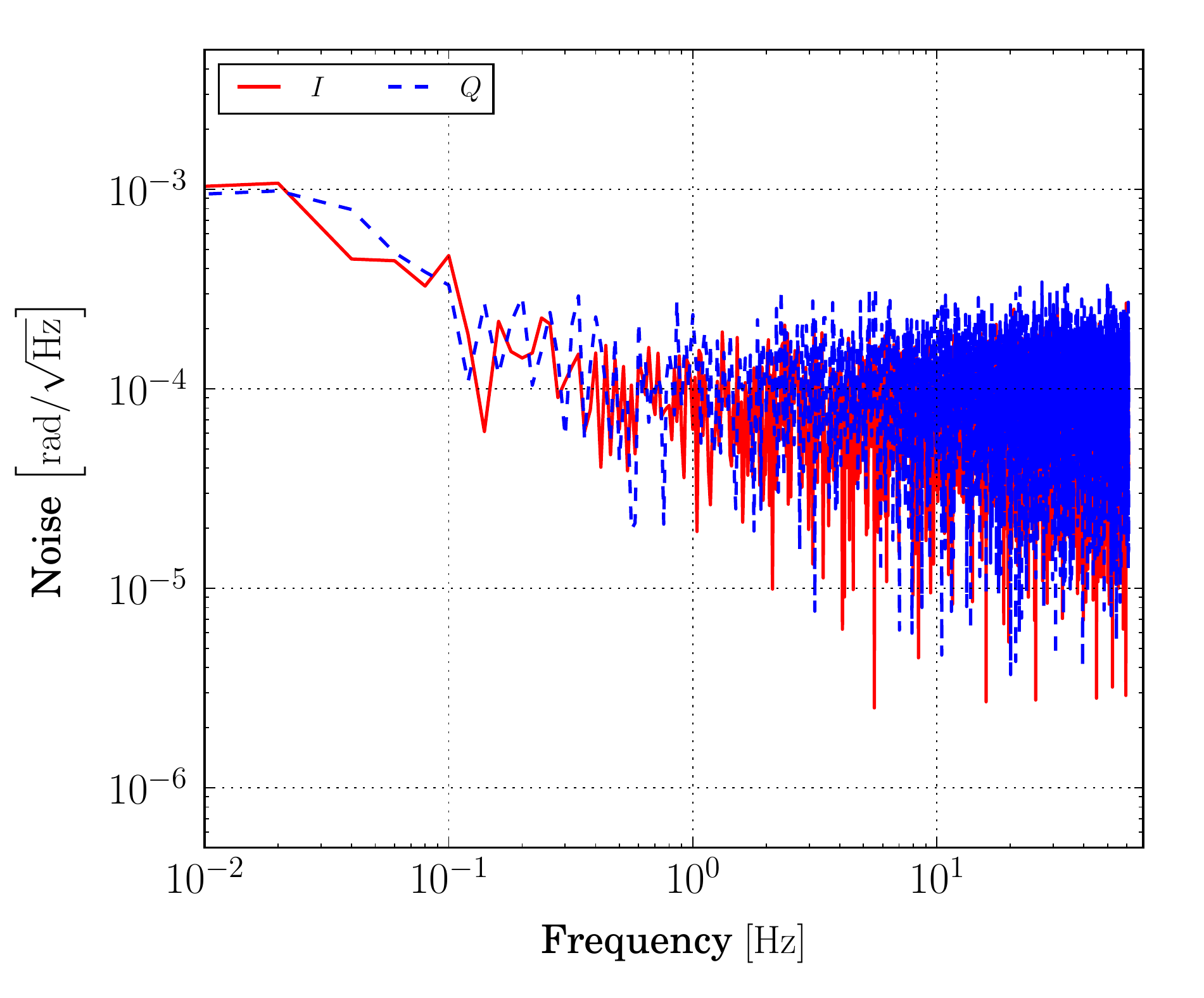}     
\caption{\small \emph{Left column}: \SI{1}{s} time streams of the $\Delta T_{\rm RJ}$ optical signal, modulated at \SI{12}{Hz}, read out in the $I$ (\emph{red solid line}) and $Q$ (\emph{blue dashed line}) channels for one pixel per array. \emph{Right column}: noise spectra of the $I$ (\emph{red solid line}) and $Q$ (\emph{blue dashed line}) channels (one pixel per array), obtained from \SI{50}{s} long signal time streams, with the cryostat window closed by a mirror. \emph{First row}: 8$^{th}$ pixel of the \SI{150}{GHz} array. \emph{Second row}: 7$^{th}$ pixel of the \SI{250}{GHz} array. \emph{Third row}: 17$^{th}$ pixel of the \SI{350}{GHz} array. \emph{Fourth row}: 7$^{th}$ pixel of the \SI{460}{GHz} array.}
\phantomsection\label{fig:optical_meas}
\end{figure}

Both blackbodies fill the beam of the optical system. The signal modulation has been performed by means of a large chopper, with an adjustable modulation frequency ranging from 5 to \SI{12}{Hz}. For the signal measurements, the modulation frequency has been set at \SI{12}{Hz}. The \emph{left panels} of fig.~\ref{fig:optical_meas} show the \SI{1}{s} time streams of the $\Delta T_{\rm RJ}$ signals, modulated at \SI{12}{Hz}, and read out in the $I$ and $Q$ channels, for one representative pixel of each array.

The noise measurements have been performed by closing the cryostat window with a metal mirror, and acquiring \SI{50}{s} long signal time streams for the $I$ and $Q$ channels. The noise spectra are obtained by performing the Fourier transforms of the these time streams. The resulting noise spectra are shown in the \emph{right panels} of fig.~\ref{fig:optical_meas}.

From these measurements, the noise equivalent temperatures for the $I$ and $Q$ channels in the Rayleigh--Jeans approximation (NET$_{\rm RJ}$), referred to the cryostat window, can be easily obtained from the ratio between the noises at \SI{12}{Hz}, $n_{I}$ and $n_{Q}$, and the responsivities, $\mathcal{R}_{I}$ and $\mathcal{R}_{Q}$. The responsivity is the ratio between the peak--to--peak of the measured modulated signal and the temperature difference of the two blackbody sources $\Delta T_{\rm RJ}$.

In order to maximize the signal, we defined a combined signal, $\Delta S_{comb}$, as
\begin{equation}
\Delta S_{comb}=\Delta I\cos\alpha+\Delta Q\sin\alpha\;,
\end{equation}
where $\alpha=\arctan\left(\Delta Q/\Delta I\right)$, and to which we can associate a combined responsivity, $\mathcal{R}_{comb}$, defined as
\begin{equation}
\mathcal{R}_{comb}=\mathcal{R}_{I}\cos\alpha+\mathcal{R}_{Q}\sin\alpha\;.
\end{equation}
The noise associated to $\Delta S_{comb}$ is obtained by performing the Fourier transforms of the \SI{50}{s} long combined signal time stream.

Tab.~\ref{tab:NET} collects the optical responsivity and NET$_{\rm RJ}$ averaged over all the the detectors of each array. 

\newcolumntype{C}{>{\centering\arraybackslash}p{5em}}
\begin{table}[htb]
	\centering
		\fontsize{10pt}{18pt}\selectfont{
		\begin{tabular}{c|c|c|C|C}
		\hline
		\hline
		\multicolumn{1}{c|}{\multirow{1}{*}{Channel}}&
		\multicolumn{2}{c|}{\multirow{1}{*}{average $\mathcal{R}$ $\left[\SI{}{rad/K}\right]$}}&
		\multicolumn{2}{c}{\multirow{1}{*}{average NET$_{\rm RJ}$ $\left[\SI{}{\mu K.\sqrt{\rm s}}\right]$}}\\ %$\left[\SI{}{\mu K/\sqrt{\rm Hz}}\right]$}}\\
		\cline{2-5}
		\multicolumn{1}{c|}{\multirow{1}{*}{$\left[\SI{}{GHz}\right]$}}&
		\multicolumn{1}{c|}{\multirow{1}{*}{$Q$}}&
		\multicolumn{1}{c|}{\multirow{1}{*}{combined}}&
		\multicolumn{1}{c|}{\multirow{1}{*}{$Q$}}&
		\multicolumn{1}{c}{\multirow{1}{*}{combined}}\\
		\hline
		\hline
150&$0.171\pm0.018$&$0.178\pm0.019$& $212\pm25$& $201\pm26$\\%$300\pm 35$& $285\pm 37$\\
250&$0.305\pm0.024$&$0.320\pm0.024$& $261\pm32$& $243\pm27$\\%$369\pm 45$& $344\pm 38$\\
350&$0.187\pm0.022$&$0.205\pm0.021$& $259\pm12$& $243\pm8$\\%$366\pm 17$& $343\pm 12$\\
460&$0.168\pm0.022$&$0.178\pm0.024$& $403\pm29$& $336\pm28$\\%$571\pm 41$& $475\pm 39$\\
\hline
		\hline
		\end{tabular}		
		}
		\caption{\small Array--averages of the optical responsivity and noise equivalent temperature in the Rayleigh--Jeans approximation, referred to the cryostat window.}
	\phantomsection\label{tab:NET}
\end{table}

\subsubsection{Spectral Response}
\phantomsection\label{subsec:Spectra}

The spectral response of the four detector arrays has been measured by using the OLIMPO differential Fourier Transform spectrometer \cite{schillaci2014}. A high--pressure mercury vapor lamp (\emph{Philips HPK--125}\footnote{\url{http://www.lamptech.co.uk/Spec\%20Sheets/D\%20MB\%20Philips\%20HPK125.htm}}) and an Eccosorb sheet have been used as the thermal Rayleigh--Jeans sources. The scan amplitude of the moving mirrors has been set in such a way that the DFTS resolution was $\Delta\nu=\SI{1.8}{GHz}$. 

The interferograms have been obtained scanning fast the delay optical lines around \SI{1}{cm/s} to avoid source drift systematic and a triangular apodization has been implemented to the interferograms to avoid spectral distortions due to the data analysis.

\begin{figure}[!h]
\centering
\includegraphics[scale=0.395]{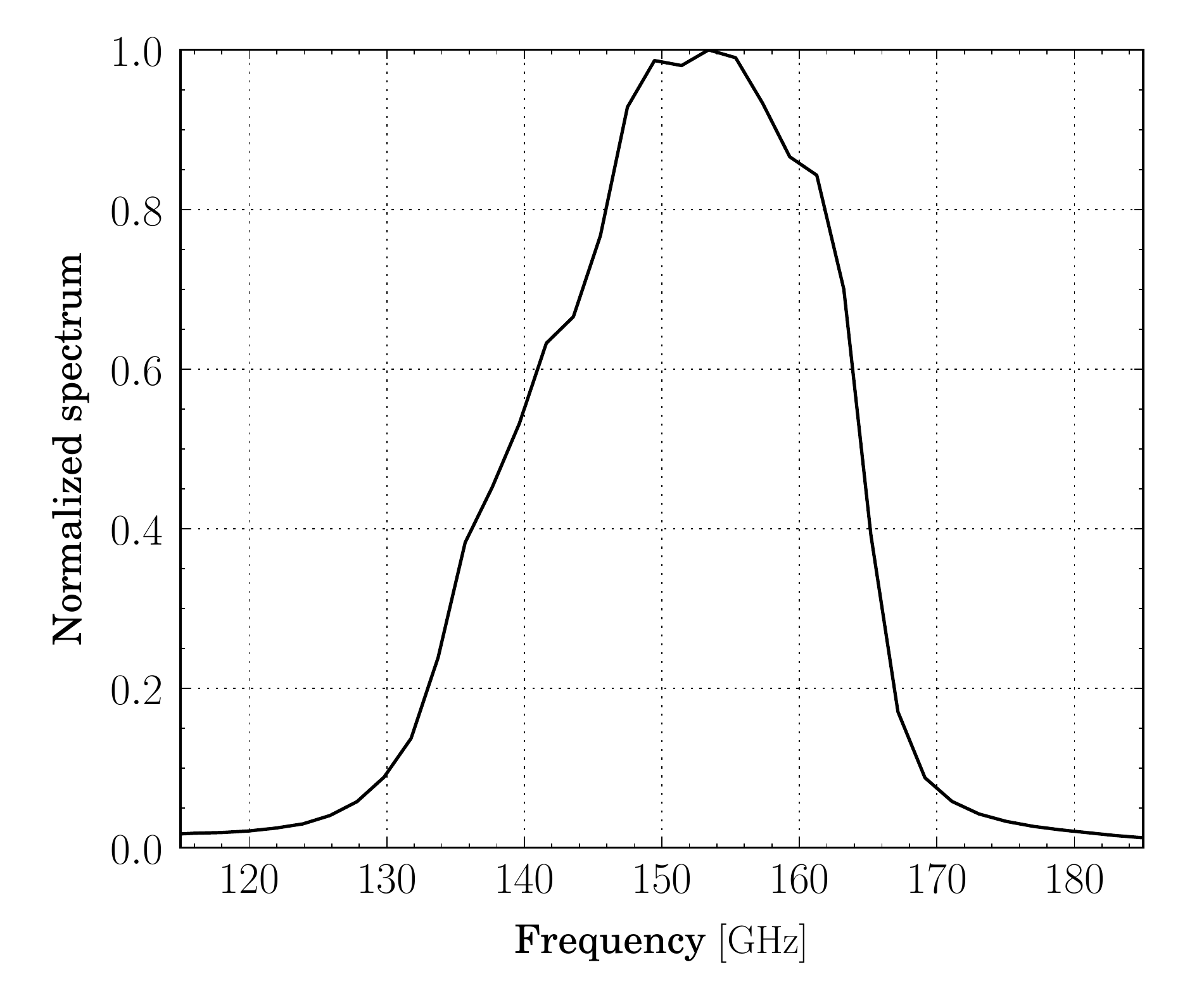}
\includegraphics[scale=0.395]{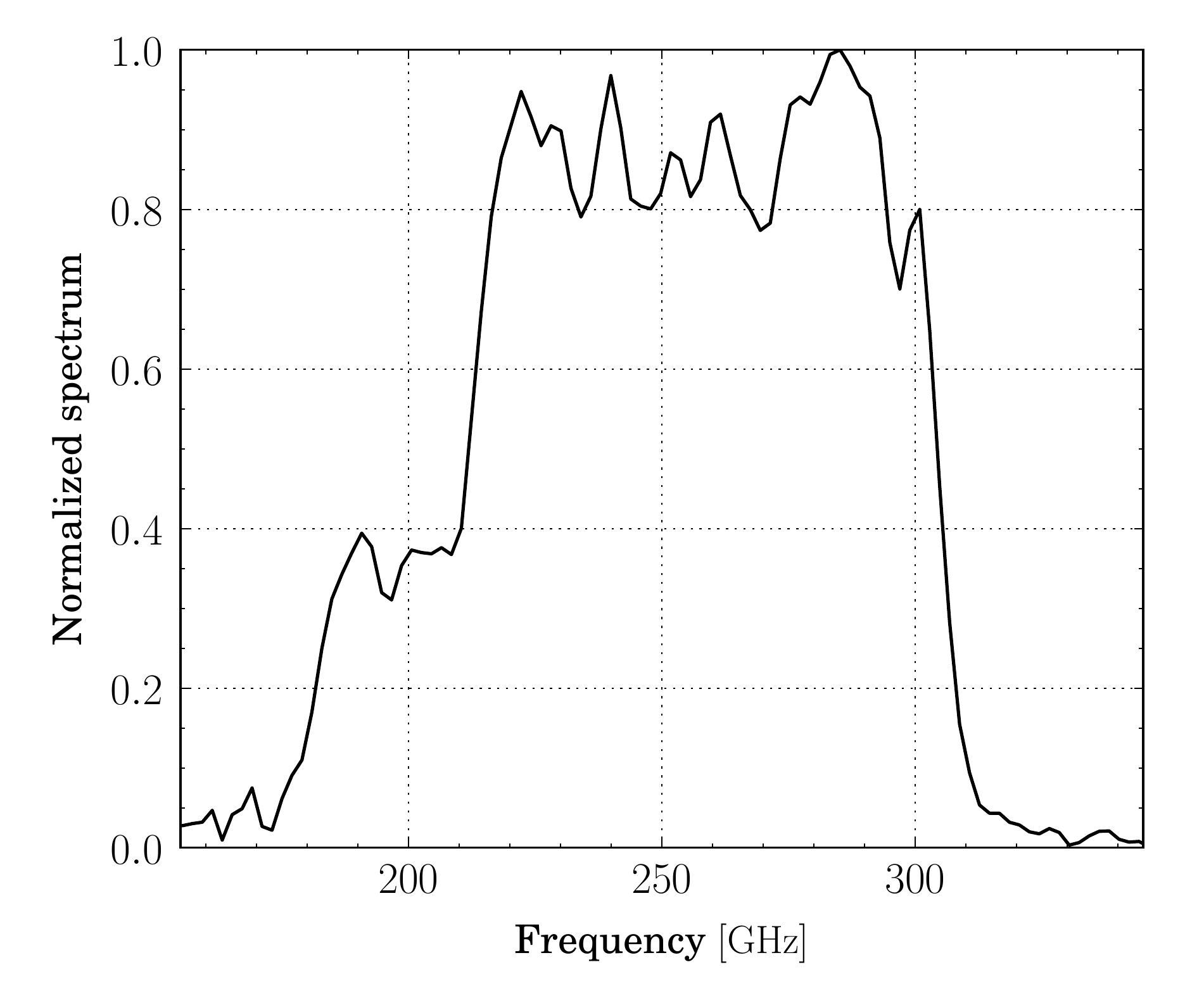}\\
\includegraphics[scale=0.395]{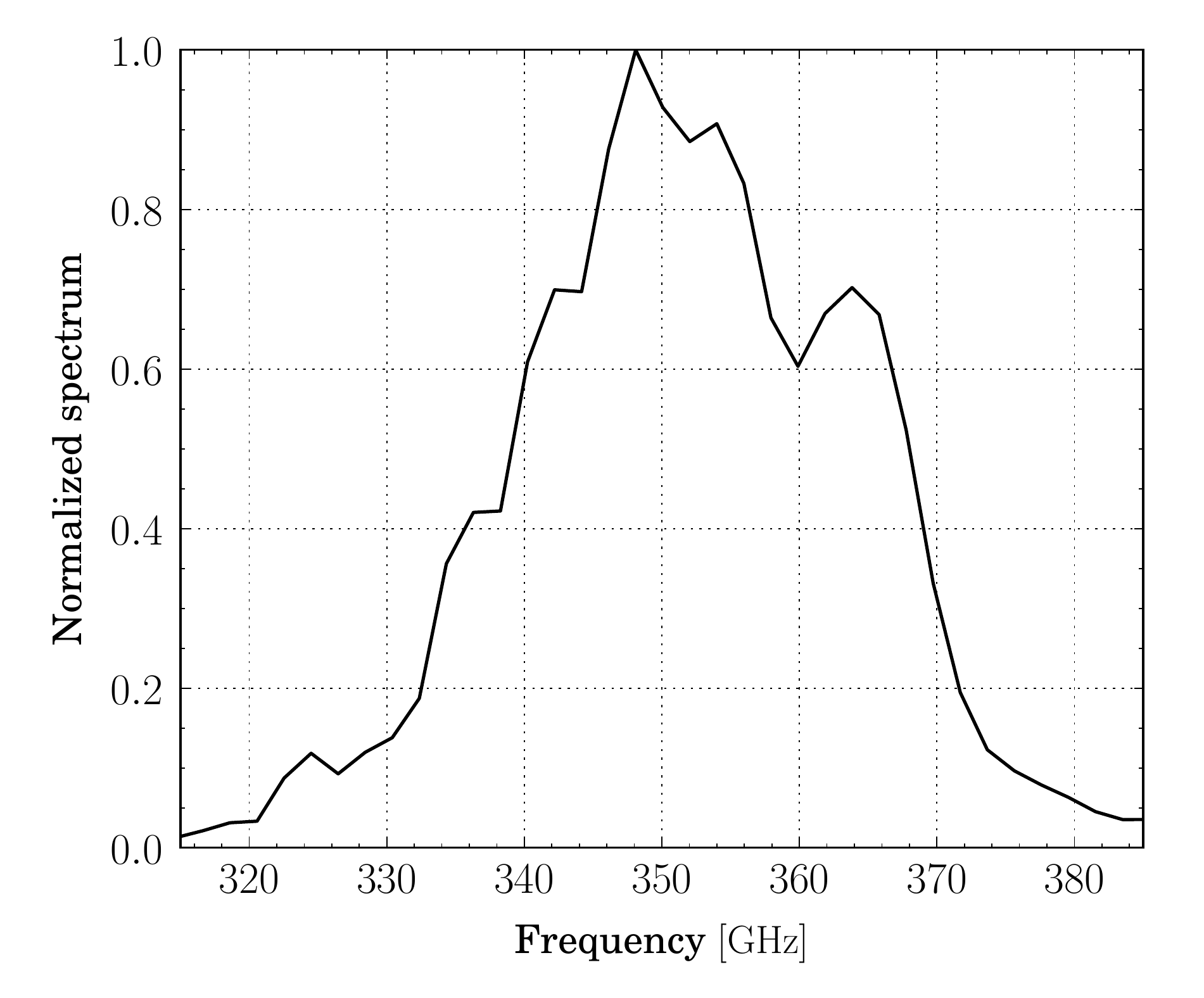}
\includegraphics[scale=0.395]{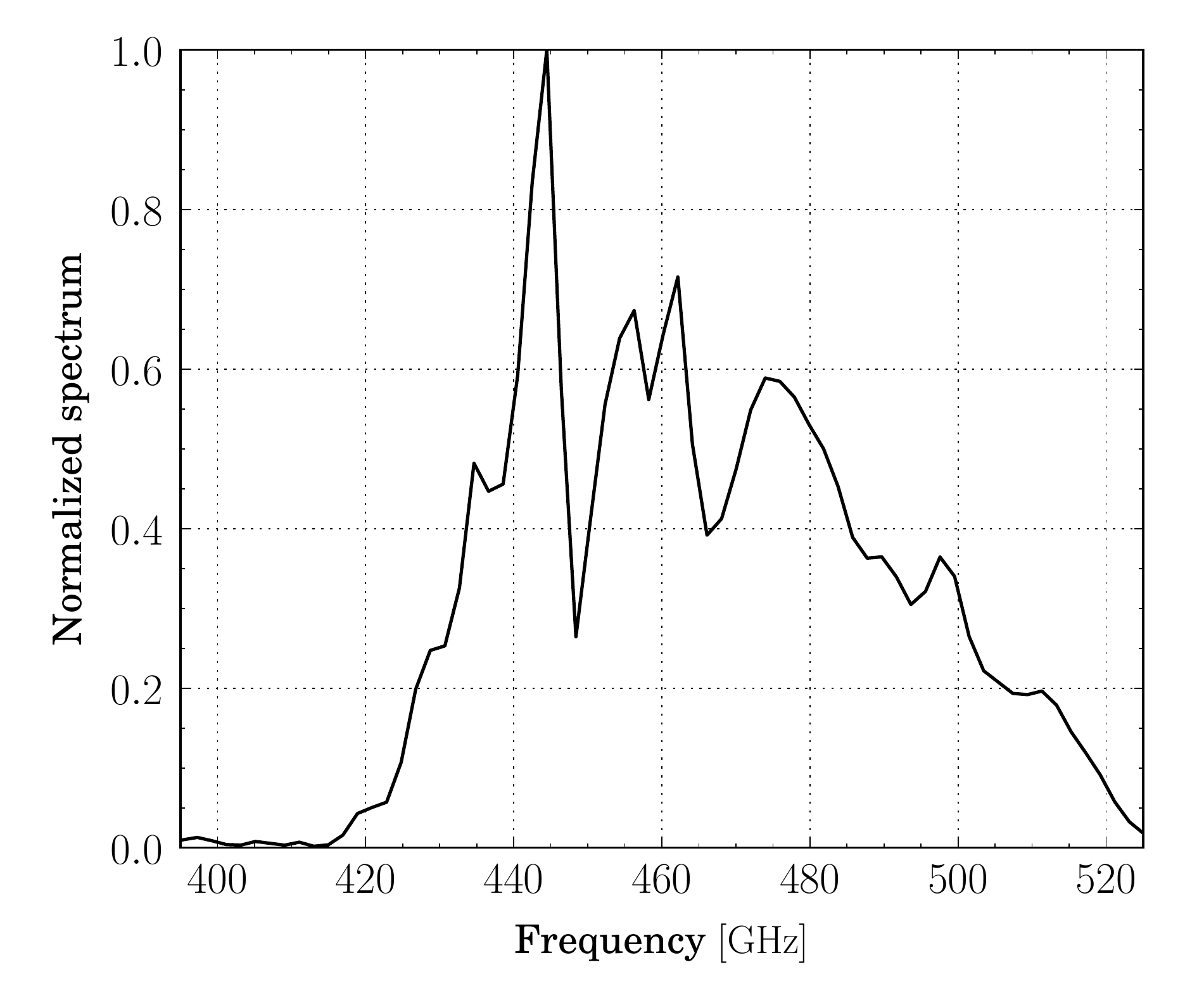}
\caption{\small Measured spectral response for the central pixels of each detector array. The spectra are normalized to the peak and corrected for the spectrum of the source (a high--pressure mercury vapor lamp \emph{Philips HPK--125}). \emph{Top--left panel}: \SI{150}{GHz} channel. \emph{Top--right panel}: \SI{250}{GHz} channel. \emph{Bottom--left panel}: \SI{350}{GHz} channel. \emph{Bottom--right panel}: \SI{460}{GHz} channel.}
\phantomsection\label{fig:spectra}
\end{figure}

Fig.~\ref{fig:spectra} reports the spectra measured for the central pixels of each array, corrected for the Rayleigh--Jeans spectrum of the source, and normalized to the peak signals. 

The FWHM of these spectra, and the effective frequencies for a CMB spectrum, an interstellar dust spectrum, and a SZ spectrum are collected in tab.~\ref{Tab:fwhm}. The effective frequencies are calculated as
\begin{equation}
\nu^{}_{eff,\;i}=\frac{\int\limits_{\Delta\nu} \nu B_{i}\left(T_{i},\nu\right) e\left(\nu\right)d\nu}{\int\limits_{\Delta\nu} B_{i}\left(T_{i},\nu\right) e\left(\nu\right)d\nu}
\end{equation}
where $i=$ CMB or dust or SZ, $B_{\rm CMB}=BB\left(T_{\rm CMB},\nu\right)$ with $T_{\rm CMB}=\SI{2.725}{K}$ \cite{0004-637X-473-2-576}, $B_{\rm dust}\sim \nu^{\:\alpha}BB\left(T_{\rm dust}, \nu\right)$ with $\alpha = 1.5$ and $T_d=\SI{20}{K}$ \cite{comp_sep},
\begin{equation}
B_{\rm SZ}=B_{\rm CMB}\:\frac{x\:{\rm e}^{x}}{{\rm e}^{x}-1}\left[x\:{\rm cotanh}\left(\frac{x}{2}\right)-4\right]
\end{equation}
with 
\begin{equation}
x=\frac{h\nu}{k_{B}T_{\rm CMB}}\;,
\end{equation}
$BB\left(T,\nu\right)$ is the blackbody spectrum, and $e\left(\nu\right)$ is the spectral response.

\begin{table}[htb]
	\centering
		\fontsize{10pt}{14pt}\selectfont{
		\begin{tabular}{c|c|c|c|c}
		\hline
		\hline
		\multicolumn{1}{c|}{\multirow{1}{*}{Channel}}&
		\multicolumn{1}{c|}{\multirow{1}{*}{FWHM}}&
		\multicolumn{1}{c|}{\multirow{1}{*}{$\nu^{}_{eff,\;{\rm CMB}}$}}&
		\multicolumn{1}{c|}{\multirow{1}{*}{$\nu^{}_{eff,\;{\rm dust}}$}}&
		\multicolumn{1}{c}{\multirow{1}{*}{$\nu^{}_{eff,\;{\rm SZ}}$}}\\
        $\left[\SI{}{GHz}\right]$&$\left[\SI{}{GHz}\right]$&$\left[\SI{}{GHz}\right]$&$\left[\SI{}{GHz}\right]$&$\left[\SI{}{GHz}\right]$\\
		\hline
		\hline
150& 25&151.0&153.4&150.1\\
250& 90&244.8&264.8&269.1\\
350& 30&350.8&353.4&352.3\\
460& 60&460.1&469.7&463.4\\
\hline
		\hline
		\end{tabular}		
		}
		\caption{\small FWHM of the measured spectra reported in fig.~\ref{fig:spectra} and effective frequencies for a CMB spectrum, an interstellar dust spectrum, and a SZ spectrum.}
	\phantomsection\label{Tab:fwhm}
\end{table}

\subsubsection{Optical Efficiency}
\phantomsection\label{subsec:Optical_Efficiency}

In order to estimate the optical efficiency, $\varepsilon$, of OLIMPO (excluding the primary and the secondary mirrors), we compare the electrical phase responsivity $\mathcal{R}_{\vartheta}^{\,elec}$ with the optical one $\mathcal{R}_{Q}^{\,opt}$. Since in the centered--rotated resonance circle readout scheme $Q$ coincides with the phase, converting $\mathcal{R}_{Q}^{\,opt}$ from ${\rm\left[rad/K\right]}$ to ${\rm\left[rad/W\right]}$, and dividing it by $\mathcal{R}_{\vartheta}^{\,elec}$, we obtain the optical efficiency:
\begin{equation}
\varepsilon =\frac{\mathcal{R}_{Q}^{\,opt}}{\mathcal{R}_{\vartheta}^{\,elec}}
\label{eq:effi}
\end{equation}

The electrical responsivity reported in Subsubsec.~\ref{subsec:Electrical_Responsivity} has been measured in dark conditions, and is thus overestimated with respect to the one in working conditions. For this reason the efficiency results reported here should be considered as lower limits for the real efficiency.

The conversion of $\mathcal{R}_{Q}^{\,opt}$ from ${\rm\left[rad/K\right]}$ to ${\rm\left[rad/W\right]}$ is given by
\begin{equation}
\mathcal{R}_{Q}^{\,opt}\;{\rm\left[\frac{rad}{W}\right]}=\mathcal{R}_{Q}^{\,opt}\;{\rm\left[\frac{rad}{K}\right]}\frac{dT}{dP_{\rm RJ}}\;,
\phantomsection\label{eq:NEP_opt}
\end{equation}
with 
\begin{equation}
\frac{dP_{\rm RJ}}{dT}=2k_{B}\Delta\nu\;,
\phantomsection\label{eq:conversion}
\end{equation}
where $\Delta \nu$ is the FWHM, which values are collected in tab.~\ref{Tab:fwhm}. Using eq.~\eqref{eq:NEP_opt} and \eqref{eq:conversion}, the values of the average $\mathcal{R}_{Q}^{\,opt}$ in ${\rm\left[rad/K\right]}$ collected in tab.~\ref{tab:NET} can be converted in ${\rm\left[rad/W\right]}$, shown in tab.~\ref{tab:resp_opt_W}, with together the optical NEPs, referred to the cryostat window.

\begin{table}[htb]
	\centering
		\fontsize{10pt}{18pt}\selectfont{
		\begin{tabular}{c|c|c|c|c}
		\hline
		\hline
		\multicolumn{1}{c|}{\multirow{1}{*}{Channel}}&
		\multicolumn{2}{c|}{\multirow{1}{*}{average $\mathcal{R}^{\,opt}$ $\left[\SI{}{rad/pW}\right]$}}&
		\multicolumn{2}{c}{\multirow{1}{*}{average NEP $\left[\SI{}{aW/\sqrt{\rm Hz}}\right]$}}\\
		\cline{2-5}
		\multicolumn{1}{c|}{\multirow{1}{*}{$\left[\SI{}{GHz}\right]$}}&
		\multicolumn{1}{c|}{\multirow{1}{*}{$Q$}}&
		\multicolumn{1}{c|}{\multirow{1}{*}{combined}}&
		\multicolumn{1}{c|}{\multirow{1}{*}{$Q$}}&
		\multicolumn{1}{c}{\multirow{1}{*}{combined}}\\
		\hline
		\hline
150& $0.248\pm 0.026$& $0.258\pm 0.028$& $207\pm25$& $196\pm26$\\
250& $0.130\pm 0.010$& $0.136\pm 0.010$& $866\pm105$& $808\pm90$\\
350& $0.226\pm 0.027$& $0.248\pm 0.026$& $303\pm14$& $284\pm10$\\
460& $0.102\pm 0.013$& $0.107\pm 0.014$& $857\pm68$& $787\pm65$\\
 \hline
 		\hline
 		\end{tabular}		
 		}
	\caption{\small Array--averages of the optical responsivity in $\left[\SI{}{rad/W}\right]$ and noise equivalent power, referred to the cryostat window.}
	\phantomsection\label{tab:resp_opt_W}
 \end{table}

Comparing $\mathcal{R}_{Q}^{\,opt}$ and $\mathcal{R}^{\,elec}_{\vartheta}$, we obtain the optical efficiencies for the four spectral bands. The values are shown in tab.~\ref{tab:opt_eff}.

 \begin{table}[htb]

 	\centering
		\fontsize{10pt}{18pt}\selectfont{
 		\begin{tabular}{c|c|c|c}
 		\hline
 		\hline
 		\multicolumn{1}{c|}{\multirow{1}{*}{Channel $\left[\SI{}{GHz}\right]$}}&
 		\multicolumn{1}{c|}{\multirow{1}{*}{$\mathcal{R}_{Q}^{\,opt}$ $\left[\SI{}{rad/pW}\right]$}}&
         \multicolumn{1}{c|}{\multirow{1}{*}{$\mathcal{R}^{\,elec}_{\vartheta}$ $\left[\SI{}{rad/pW}\right]$}}&
         \multicolumn{1}{c}{\multirow{1}{*}{$\varepsilon$ $\left[\%\right]$}}\\
 		\hline
 		\hline
 150& $0.248\pm0.026$ &$1.35\pm 0.10$&$ > 18\pm 2$\\
 250& $0.130\pm0.010$ &$1.49\pm 0.22$&$ > 9\pm 1$\\
 350& $0.226\pm0.027$ &$2.09\pm 0.10$&$ > 11\pm 1$ \\
 460& $0.102\pm0.013$ &$2.14\pm 0.11$&$ > 5\pm 1$ \\
 \hline
 		\hline
 		\end{tabular}		
 		}
		 \caption{\small Array--average of the optical efficiency, given by the ratio between the optical responsivity and the electrical responsivity measured in dark conditions.}
 	\phantomsection\label{tab:opt_eff}
 \end{table}

These measurements confirm the high sensitivity to mm-wave radiation of the four OLIMPO KID arrays and a reasonable efficiency of the entire receiver. 

\begin{figure*}[!h]
\centering
\includegraphics[scale=0.85]{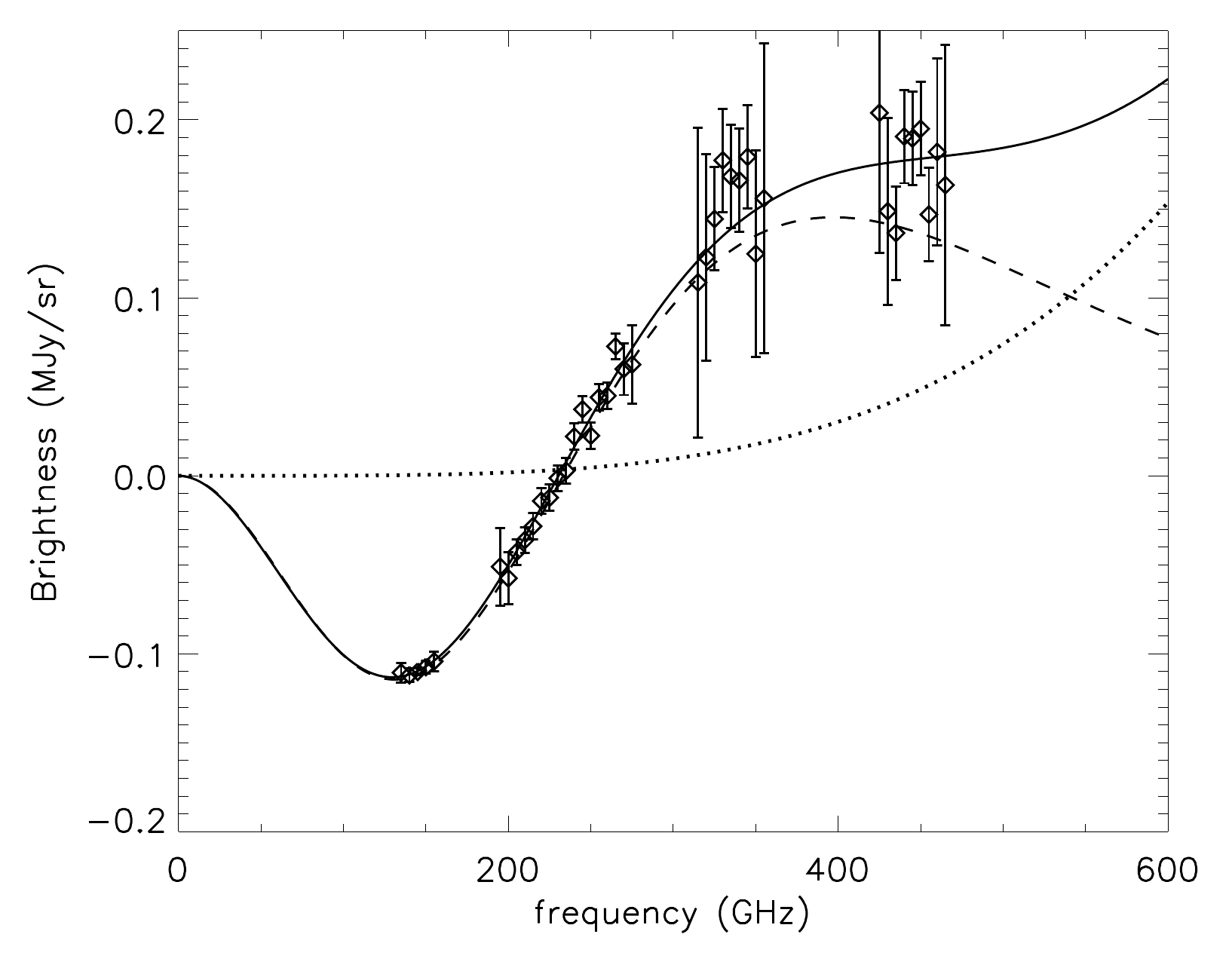}      
\caption{\small Expected sensitivity of the spectral observation of a rich cluster of galaxies with the OLIMPO DFTS. The \emph{solid line} represents the sum of thermal SZ brightness (\emph{dashed line}) and dust continuum (\emph{dotted line}). The filled regions represent the expected $\pm$1-$\sigma$ confidence intervals for the measured spectral data, based on the noise equivalent powers measured here and assuming an optical efficiency of the telescope + spectrometer + receiver system of 15\%. We assumed a DFTS resolution of \SI{5}{GHz}. The confidence intervals widen at both edges of each observation band, due to the reduced transmission of the band-selection filters at cut-on and cut-off. We have considered a 24--hours long integration on a cluster with  $\tau_{th}=0.005$, $T_e=\SI{8.5}{keV}$ and interstellar dust foreground contrast with respect to the reference sky region of \SI{0.0006}{MJy/sr} at \SI{150}{GHz}. We have considered only the center pixel of each array, pointing at the center of the cluster, since the other pixels will observe either the outskirts of the cluster, or the blank sky in the surrounding regions.}
\phantomsection\label{fig:simulation}
\end{figure*}

From the measured efficiencies and NETs we can forecast a high precision spectroscopic measurement of the SZ spectrum in rich clusters of galaxies. Assuming that the noise performance measured here is achieved also during the flight, and the typical optical efficiency is 15\% for all arrays, we have simulated a typical measurement of a rich cluster of galaxies ($\tau_{th}=0.005$ and $T_e=\SI{8.5}{keV}$) with the OLIMPO telescope and DFTS (set to $\Delta\nu=\SI{5}{GHz}$): the result is reported in fig.~\ref{fig:simulation}, and looks really promising.

\section{Conclusion}
\phantomsection\label{sec:Conclusion} 

In this paper we described the design, fabrication, electrical and optical characterization of the four detector arrays of the OLIMPO experiment. These detectors are front illuminated, horn--coupled, Al LEKIDs deposited on Si wafers. Optical and electrical simulations have brought to the definition of the entire detector systems, from the waveguide to the backshort, through the optical transition element, the absorber, the capacitor, the feedline, the coupling between the detector and the feedline, and the dielectric substrate.    

The electrical characterization has been performed in the laboratory with a dark setup based on a dilution cryostat, with a replica of the OLIMPO \emph{room--temperature electronics}. Measurements of the quasiparticle lifetime, of the critical temperature of the aluminum film, of the resonator quality factors, of the temperature responsivities, and of the phase noises were used to estimate the electrical NEPs. We obtained an average electrical NEP at \SI{12}{Hz} lower than the estimated BLIP NEP of the spectrometric configuration of OLIMPO, for all the detector arrays.  

The characterization of the detectors has been completed with the optical tests, performed after the integration of the arrays in the OLIMPO cryostat. We measured the NET$_{\rm RJ}$s under realistic loading conditions, by measuring signal time--streams and noise spectral densities. We characterized the spectral responses by using the OLIMPO DFTS, with a resolution a \SI{1.8}{GHz}. We obtained the optical efficiency of the OLIMPO receiver comparing the optical and the electrical responsivities.

Given the measured performance, we forecast a high precision spectroscopic measurement of the SZ spectrum in rich clusters of galaxies, which looks really promising.  

\acknowledgments{
\addcontentsline{toc}{section}{Acknowledgements}
We acknowledge the Italian Space Agency (ASI) for funding the OLIMPO experiment, and organizing the 2018 launch campaign. We thank Giorgio Amico for manufacturing the array sample holders. We are grateful to Angelo Cruciani for useful suggestions and discussions.}

\newpage{}
\appendix
\section{Fit of the complex $S_{21}$ parameter}
\phantomsection\label{sec:Fit}

In order to estimate the electrical properties of the resonators, each resonance circle (complex $S_{21}$ parameter) can be fit to the equation \citep{Gao2008}

\begin{equation}
S_{21}\left(\nu_{r}\right)=a\left[1-\frac{Q{\rm e}^{j\phi_{0}}/Q_{c}}{1+j2Q\displaystyle{\frac{\nu-\nu_{r}}{\nu_{r}}}}\right]\;,
\phantomsection\label{eq:fit}
\end{equation}  
where $a$ is a complex constant accounting for the gain and phase shift through the system, $Q$ is the total quality factor, and $\phi_{0}$ takes into account the rotation of the resonance circle due to the impedance mismatches in proximity of the resonance. The total and the coupling quality factors are linked to the internal quality factor, $Q_{i}$, through the equation

\begin{equation}
\frac{1}{Q}=\frac{1}{Q_{i}}+\frac{1}{Q_{c}}.
\phantomsection\label{eq:q_parameter}
\end{equation}

\begin{figure}[htb]
\centering
\includegraphics[scale=0.32]{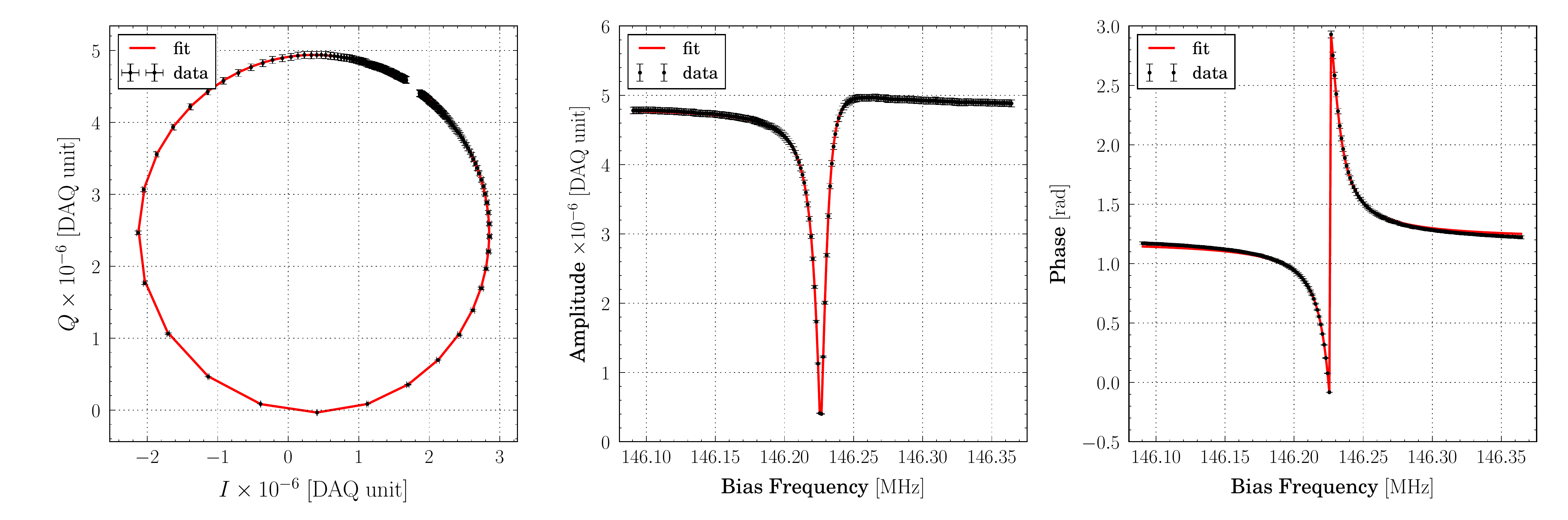}
\includegraphics[scale=0.32]{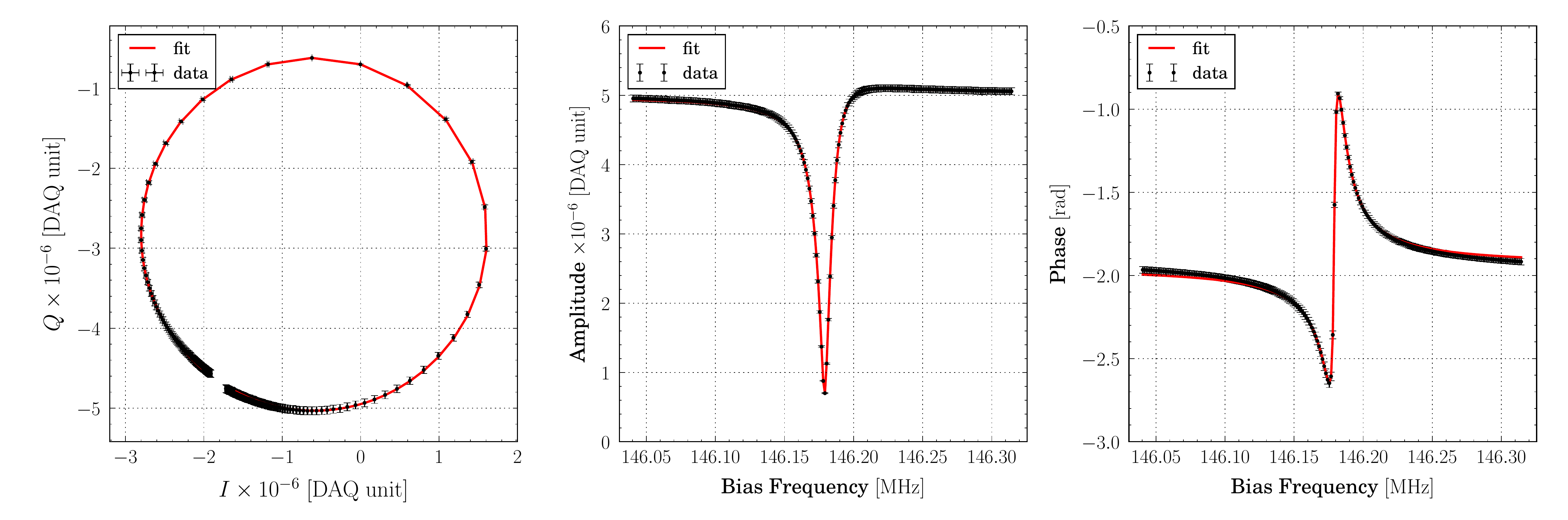}
\caption{\small Fit result for the first pixel of the \SI{150}{GHz} array at \SI{185}{mK}, in the \emph{top panels}, and at \SI{300}{mK}, in the \emph{bottom panels}. The fit function, eq.~\eqref{eq:fit}, is overplotted in \emph{red} to the measured data in \emph{black} (\emph{dots} with the \emph{error bars}) for the $IQ$ circle, in the \emph{left panel}, the amplitude, in the \emph{central panel}, and the phase, in the \emph{right panel}.}
\phantomsection\label{fig:fit_185_300mK}
\end{figure}

As an example, fig.~\ref{fig:fit_185_300mK} shows the results of the complex fit through eq.~\eqref{eq:fit} of the resonance circle of the first pixel of the \SI{150}{GHz} array at \SI{185}{mK} and \SI{300}{mK}.

\section{Electrical responsivity in \emph{Fit Range $\#$1} and \emph{$\#2$}}
\phantomsection\label{sec:Electrical_responsivity_A}

For completeness, in this appendix we report the estimate of the electrical phase responsivity in the \emph{Fit Range $\#$1} and \emph{$\#$2}.

\begin{figure}[!h]
\centering
\includegraphics[scale=0.395]{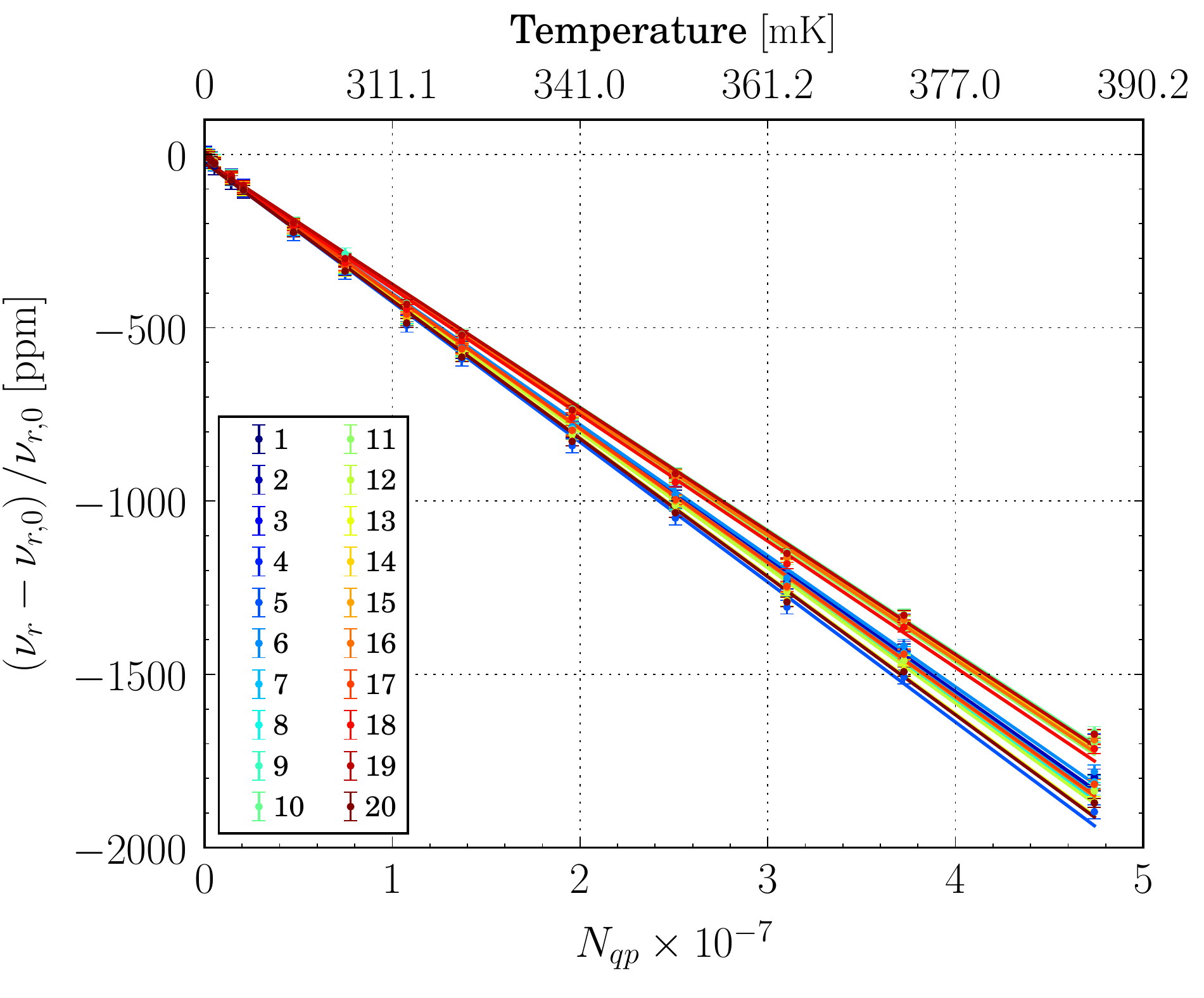}
\includegraphics[scale=0.395]{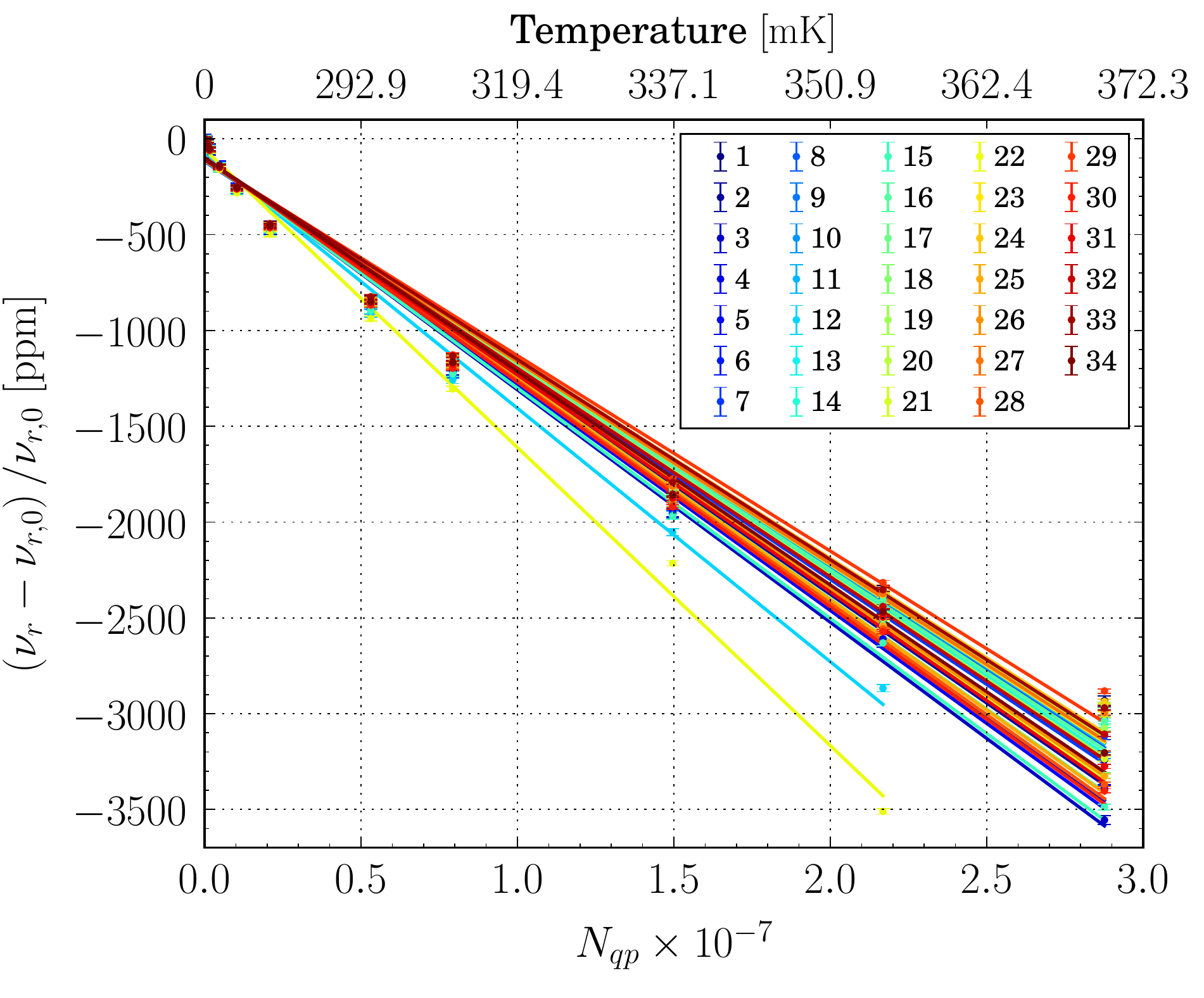}\\
\includegraphics[scale=0.395]{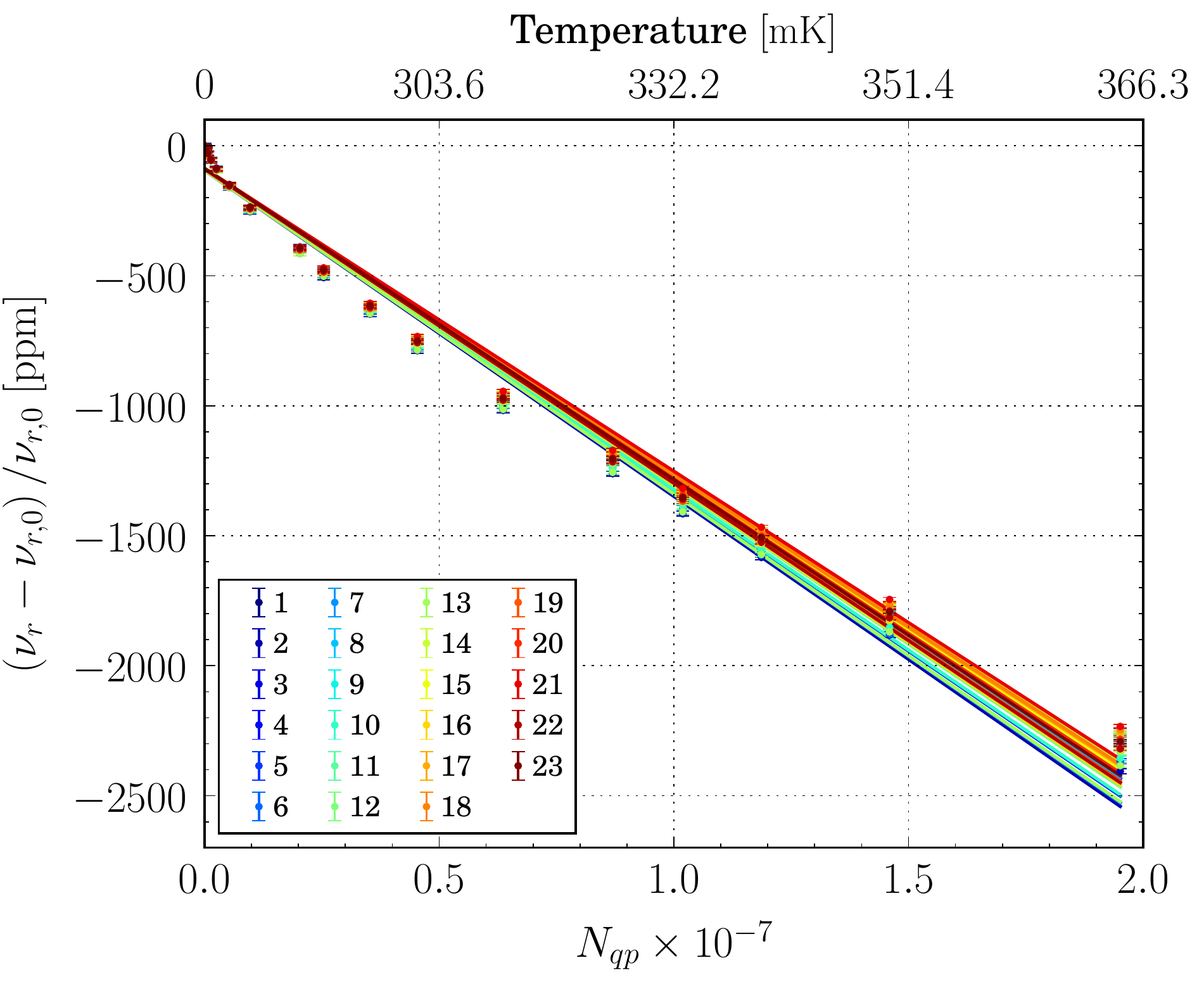}
\includegraphics[scale=0.395]{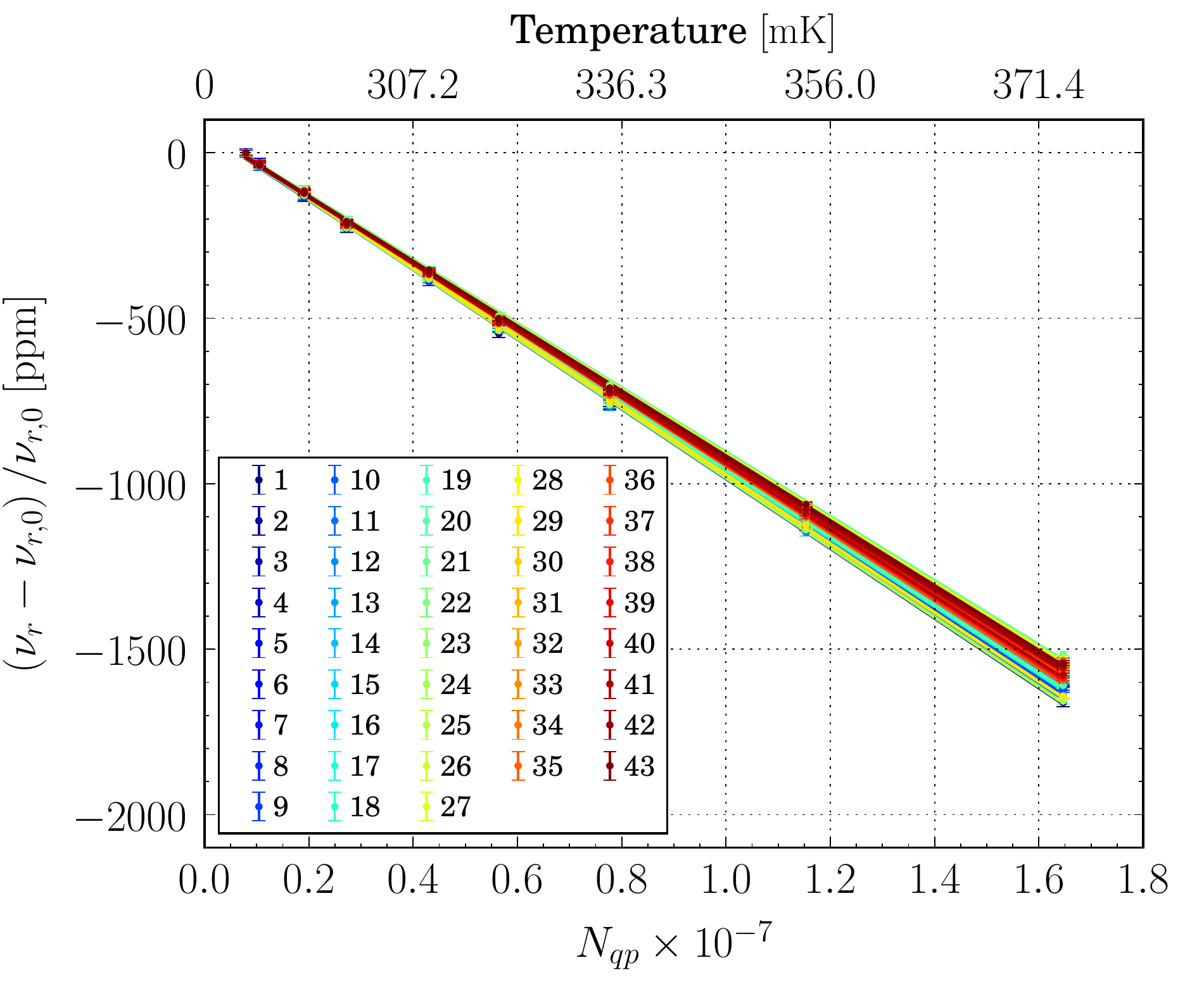}
\caption{\small Fractional frequency shift as a function of bath temperature for all the four arrays. Here, the temperatures varies from approximately 200 to \SI{400}{mK} (\emph{Fit Range $\#$1} in tab.~\ref{tab:fit_range}). The \emph{dots} with the \emph{error bars} are the measured data, and the \emph{solid lines} are the linear fit results. Different colors indicate different resonators. \emph{}\emph{Top--left panel}: \SI{150}{GHz} array, fit range $\left[185;387\right]\SI{}{mK}$. \emph{Top--right panel}: \SI{250}{GHz} array, fit range $\left[168;370\right]\SI{}{mK}$. \emph{Bottom--left panel}: \SI{350}{GHz} array, fit range $\left[155;365\right]\SI{}{mK}$. \emph{Bottom--right panel}: \SI{460}{GHz} array, fit range $\left[255;373\right]\SI{}{mK}$.}
\phantomsection\label{fig:fit_range1}
\end{figure}

\begin{figure}[!h]
\centering
\includegraphics[scale=0.395]{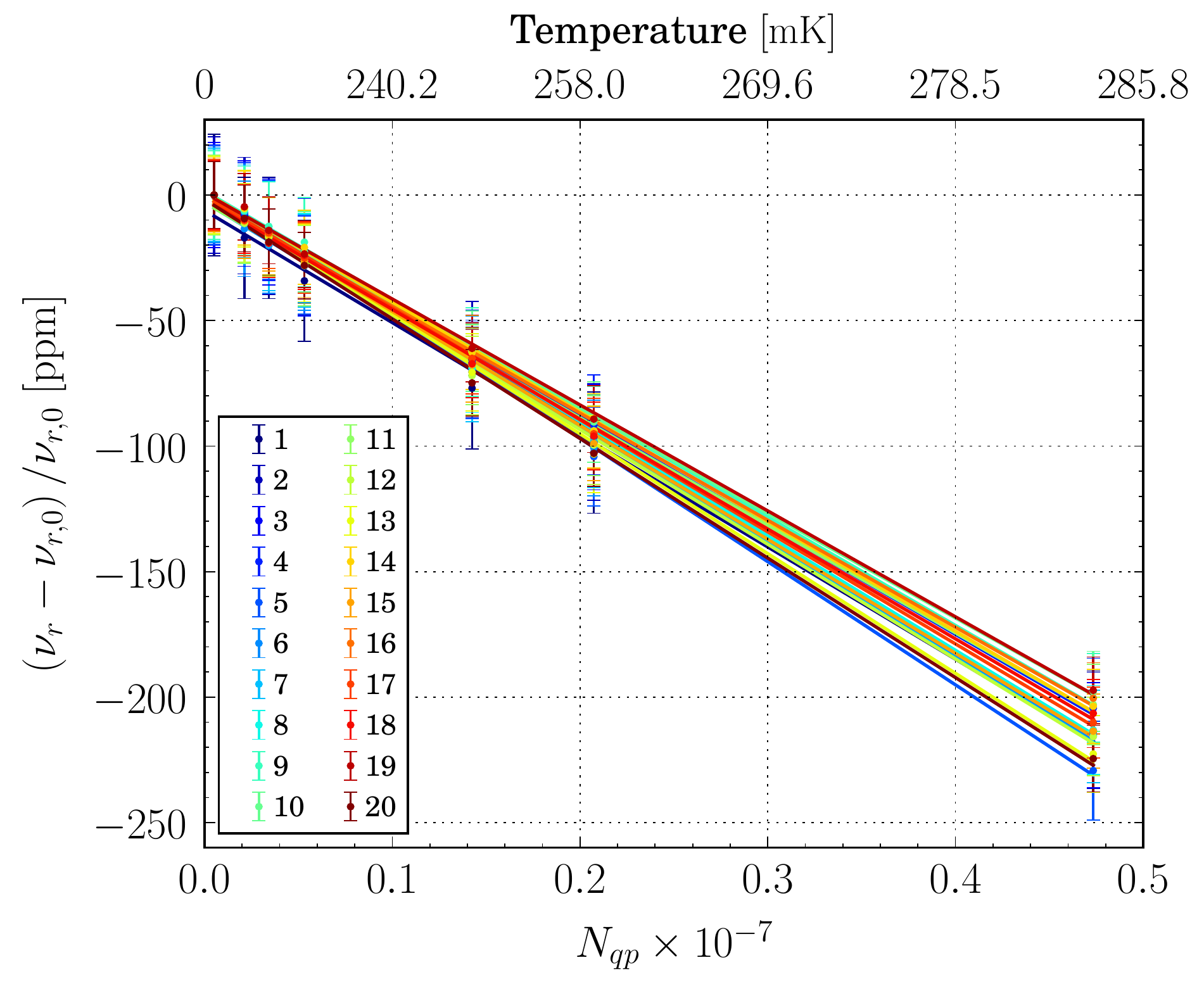}
\includegraphics[scale=0.395]{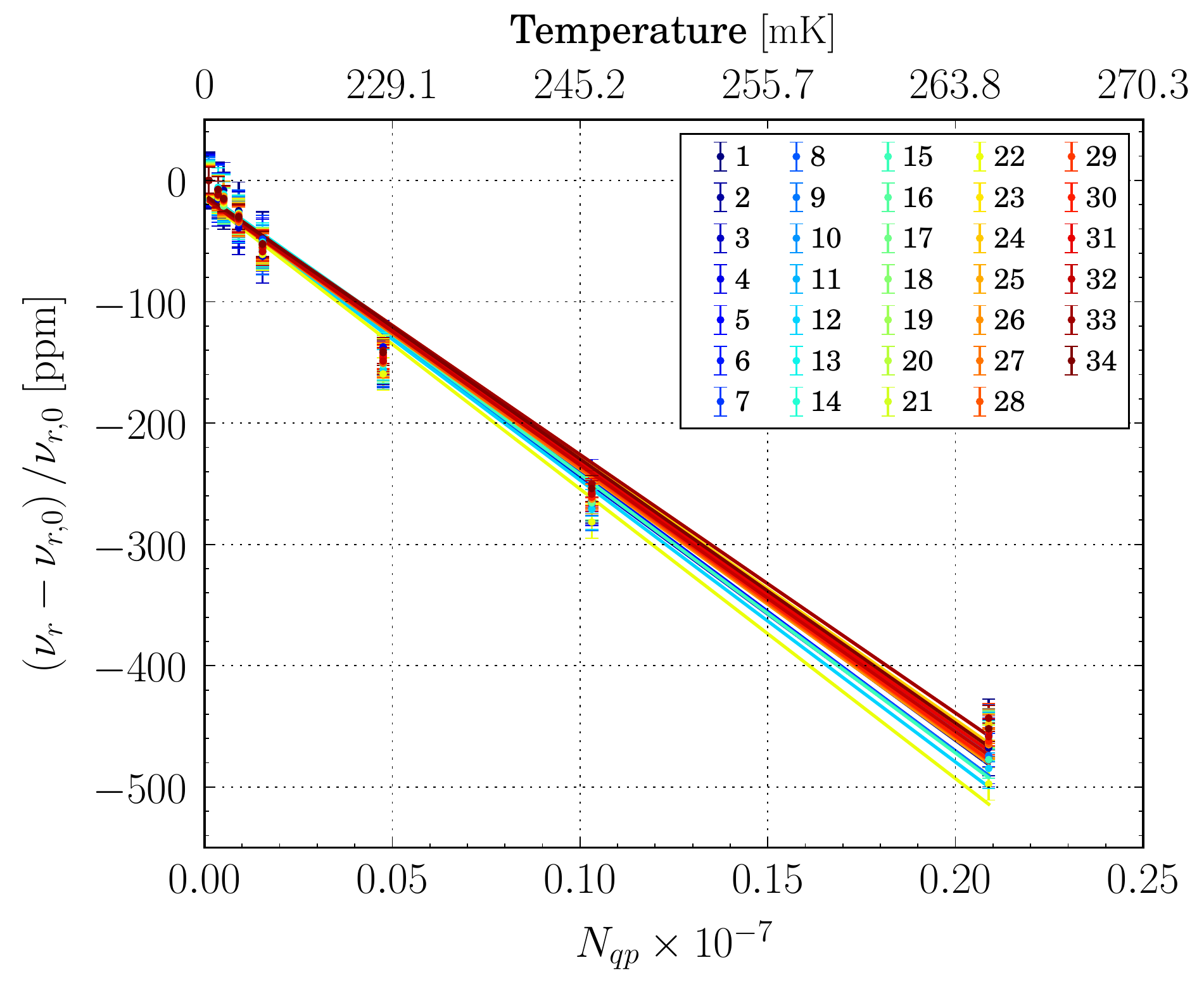}\\
\includegraphics[scale=0.395]{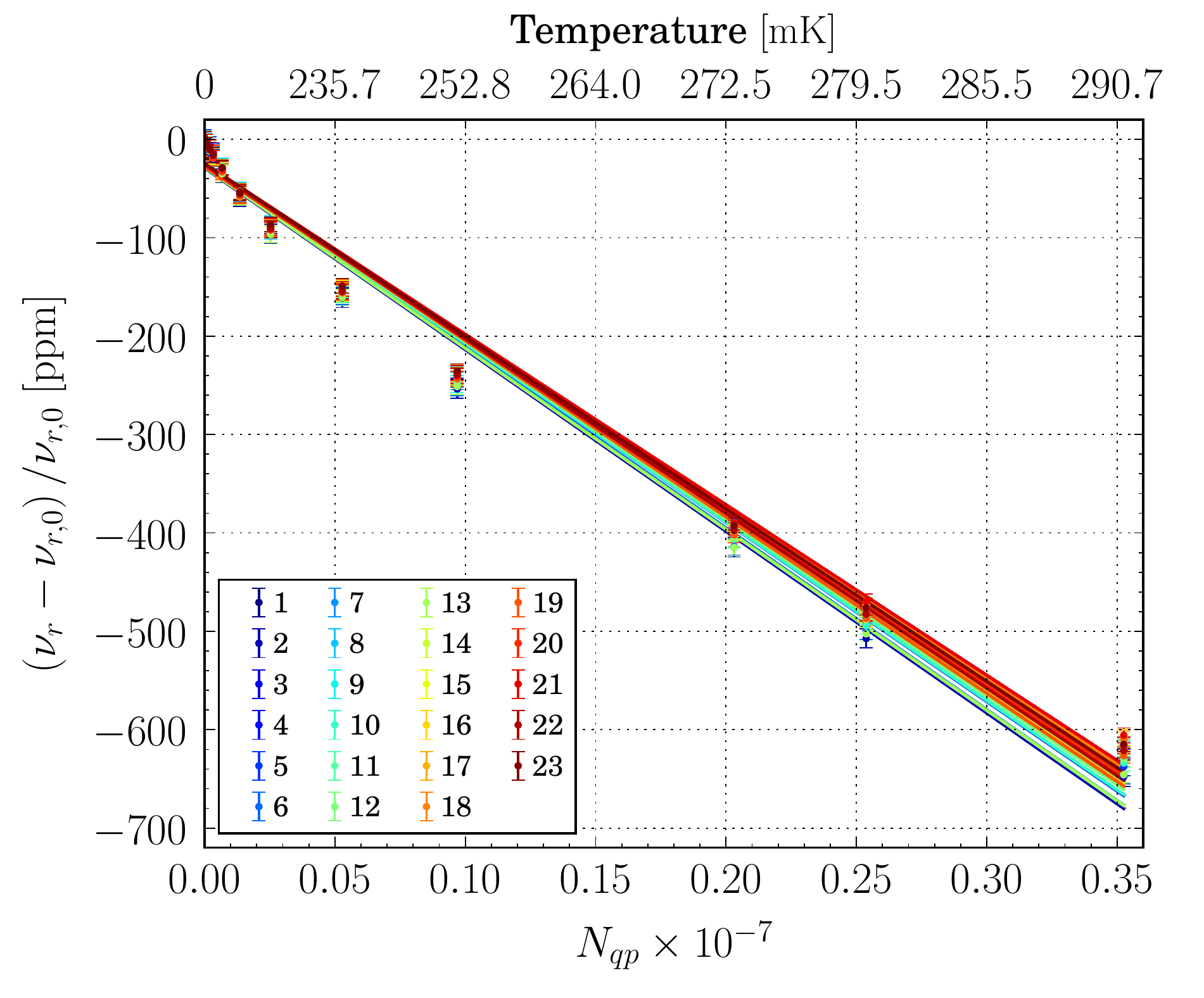}
\includegraphics[scale=0.395]{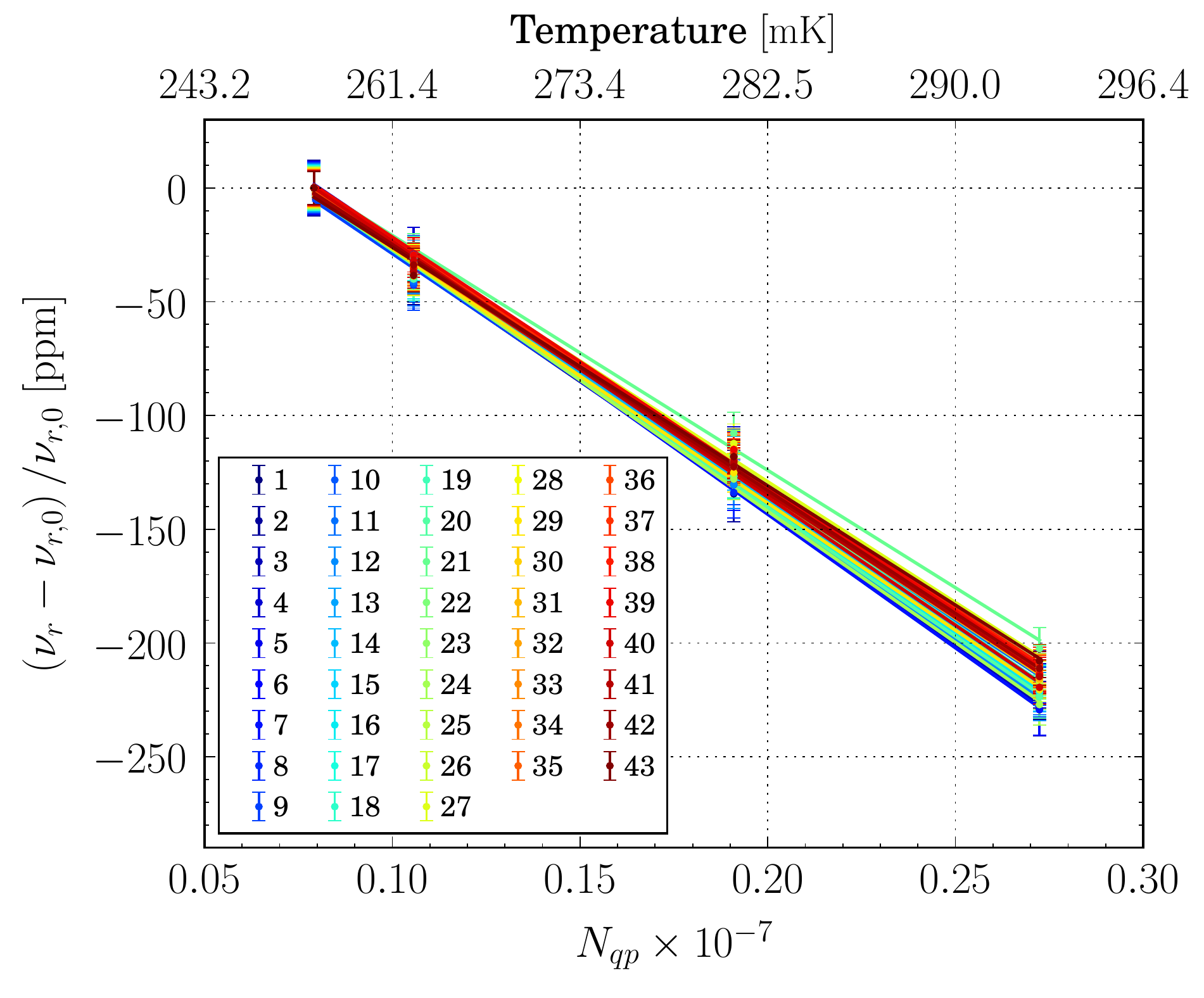}
\caption{\small Fractional frequency shift as a function of bath temperature for all the four arrays. Here, the temperatures varies from approximately 200 to \SI{300}{mK} (\emph{Fit Range $\#$2} in tab.~\ref{tab:fit_range}). The \emph{dots} with the \emph{error bars} are the measured data, and the \emph{solid lines} are the linear fit results. Different colors indicate different resonators. \emph{Top--left panel}: \SI{150}{GHz} array, fit range $\left[185;284\right]\SI{}{mK}$. \emph{Top--right panel}: \SI{250}{GHz} array, fit range $\left[168;265\right]\SI{}{mK}$. \emph{Bottom--left panel}: \SI{350}{GHz} array, fit range $\left[155;291\right]\SI{}{mK}$. \emph{Bottom--right panel}: \SI{460}{GHz} array, fit range $\left[255;293\right]\SI{}{mK}$.}
\phantomsection\label{fig:fit_range2}
\end{figure}

Fig.~\ref{fig:fit_range1} and \ref{fig:fit_range2} show the fit results in the \emph{Fit Range $\#$1} and \emph{Fit Range $\#$2} respectively, for all the detectors of the OLIMPO detector arrays. The electrical phase responsivity has been obtained through eq.~\eqref{eq:respon}, where we used the $Q$ values measured at the base temperatures. We used the values of $\tau_{qp}$ measured at \SI{300}{mK}, hence the $\mathcal{R}_{\vartheta}$ calculated here have to be considered lower limits, because the value of $\tau_{qp}$ at \SI{300}{mK} is theoretically lower than those at the base temperatures. Tab.~\ref{tab:responsivity} collects the electrical phase responsivity averaged, for each array, over all the detectors.

\begin{table}[!h]
	\centering
		\fontsize{10pt}{15pt}\selectfont{
		\begin{tabular}{c|c|c|c}
		\hline
		\hline
		\multicolumn{1}{c|}{\multirow{1}{*}{Channel}}&
		\multicolumn{3}{c}{\multirow{1}{*}{average $\mathcal{R}_{\vartheta}$ $\left[\SI{}{rad/pW}\right]$}}\\
		\cline{2-4}
		\multicolumn{1}{c|}{\multirow{1}{*}{$\left[\SI{}{GHz}\right]$}}&
		\multicolumn{1}{c|}{\multirow{1}{*}{Fit Range $\#$1}}&
		\multicolumn{1}{c|}{\multirow{1}{*}{Fit Range $\#$2}}&
		\multicolumn{1}{c}{\multirow{1}{*}{Fit Range $\#$3}}\\
		\hline
		\hline
150& $2.28\pm 0.26$& $2.71\pm 0.31$& $1.35\pm 0.10$\\
250& $3.22\pm0.30$& $6.37\pm0.42$& $1.49\pm 0.22$\\
350& $6.99\pm 0.63$& $10.3\pm 1.0$& $2.09\pm 0.10$\\
460& $5.52\pm 0.27$& $6.03\pm 0.30$& $2.14\pm 0.11$\\
\hline
		\hline
		\end{tabular}		
		}
	\caption{\small Array--average of the electrical phase responsivity in the three fit ranges.}
	\phantomsection\label{tab:responsivity}
\end{table}

\bibliographystyle{JHEP} 
\addcontentsline{toc}{section}{References}
\bibliography{bib_.bib}

\end{document}